\documentclass[11pt,a4paper,final]{article}
\usepackage{amsmath,amssymb}
\usepackage{epsfig,graphicx}
\usepackage{lscape}
\usepackage{caption}
\usepackage{subcaption}
\usepackage{multirow}
\usepackage[numbers,sort&compress]{natbib}

\begin{document}

\title{\bf Unifying ``soft" and ``hard" diffractive exclusive vector meson production and
deeply virtual Compton scattering}

 \author{
  S.~Fazio$^{a}$\footnote{{\bf e-mail}: sfazio@bnl.gov},
  R.~Fiore$^{b}$\footnote{{\bf e-mail}: roberto.fiore@cs.infn.it},
  L.~Jenkovszky$^{c}$\footnote{{\bf e-mail}: jenk@bitp.kiev.ua},
  A.~Salii$^{c}$\footnote{{\bf e-mail}: saliy.andriy@gmail.com}
  \\
 $^a$ \small{\em Brookhaven National Laboratory, Physics Department, 11973 Upton, NY - U.S.A.}\\
 $^b$ \small{\em Dipartimento di Fisica, Universit\'a della Calabria and}\\
 \small{\em Istituto Nazionale di Fisica Nucleare, Gruppo collegato di Cosenza}\\
 \small{\em I-87036 Arcavacata di Rende, Cosenza, Italy}\\
 $^c$ \small{\em Bogolyubov Institute for Theoretical Physics} \\
 \small{\em National Academy of Sciences of Ukraine, UA-03680 Kiev, Ukraine } \\}

\date{\today}
\maketitle

\begin{abstract}
A Pomeron model applicable to both ``soft" and ``hard" processes is suggested and tested against the high-energy data from virtual photon-induced reactions.
The Pomeron is universal, containing two terms, a ``soft" and a ``hard" one, whose relative weight varies with $\widetilde {Q^2}=Q^2+M_V^2$, where $Q^2$ is the virtuality of the incoming photon and $M_V$ is the mass of the produced vector particle. With a small number of adjustable parameters, the model fits all available data on vector meson production and deeply virtual Compton scattering from HERA. 
Furthermore, we attempt to apply the model to hadron-induced reactions, by using high-energy data from proton-proton scattering.
\end{abstract}

\section{Introduction}\label{sec:Intro}
According to perturbative QCD calculations, the Pomeron corresponds to the exchange of an infinite gluon ladder, producing an infinite set of moving Regge poles, the so-called BFKL Pomeron \cite{BFKL}, whose highest intercept $\alpha(0)$ is near $1.3\div 1.4$. Phenomenologically,
``soft" (low virtuality $Q^2$) and ``hard'' (high virtuality $Q^2$) diffractive ({\it i.e.} small squared momentum transfer $t$)
processes with Pomeron exchange are described by the exchange of two different objects in the $t$ channel, a ``soft'' and a ``hard'' Pomeron (or their QCD gluon images), (see, for instance, Refs.~\cite{BP, DDLN}). This implies the existence of two (or even more) scattering amplitudes, differing by the values of the parameters of the Pomeron trajectory, their intercept $\alpha(0)$ and slope $\alpha'(t=0)$, typically $(1.08\div 1.09)$ and $(0.25),$ respectively, for the ``soft'' Pomeron, and
 $(1.3\div 1.4),$ and $(0.1$ or even less$)$ for the ``hard" one, each attached to vertices of the relevant reaction or kinematical region.
    A simple ``unification" is to make theses parameters $Q^2$-dependant. This breaks Regge
factorization, by which Regge trajectories should not depend on $Q^2$.
\begin{figure}[ht]
  \vspace{-0.2cm}
  \begin{center}\includegraphics[trim = 0mm 8mm 10mm 14mm,clip,scale=0.6]{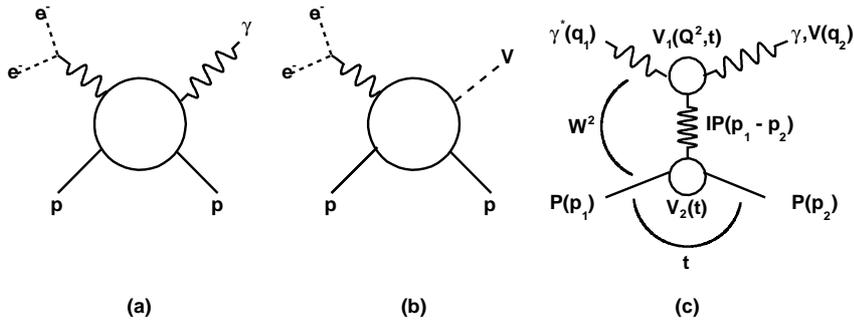}\end{center}
  \caption{ \label{fig:diagram} Diagrams of DVCS (a) and VMP (b); (c) DVCS (VMP) amplitude in a Regge-factorized form.}
\end{figure}

    In the present approach, initiated in Refs. \cite{Acta,Capua}, we postulate that

1. Regge factorization holds, {\it i.e.} the dependence on the virtuality of the external particle (virtual photon) enters only the relevant vertex, not the propagator;

2. there is only one Pomeron in nature and it is the same in all reactions. It may be 
complicated, e.g. having many, at least two, components.

The first postulate was applied, for example, in Refs.~\cite{Francesco, Capua, FazioPhysRev} to study the deeply virtual Compton scattering (DVCS) and the vector meson production (VMP). In Fig. \ref{fig:diagram}, where diagrams (a) and (b) represent the DVCS and the VMP, respectively, the $Q^2$ dependence enters only the upper vertex of the diagram (c).
The particular form of this dependence and its interplay with $t$ is not unique.

In Refs.~\cite{Capua,FazioPhysRev} the interplay between $t$ and $Q^2$ was realized by the introduction of a new variable, $z=t-\widetilde {Q^2}$, where $\widetilde {Q^2}$ is the familiar variable $\widetilde {Q^2} = Q^2 + M_V^2$, $M_V$ being the vector meson mass. The model (called also ``scaling model") is simple and fits the data on DVCS (in this case $M_V = 0$) and VMP, although the physical meaning of this new variable is not clear.

In a series of subsequent papers (see Refs.~\cite{Confer,Acta,FazioPhysRev}), $\widetilde {Q^2}$ was incorporated in a ``geometrical'' way reflecting the observed trend in the decrease of the forward slope as a function of  $\widetilde {Q^2}$. This geometrical approach, combined with the Regge-pole model was named ``Reggeometry''. A Reggeometric amplitude dominated by a single Pomeron  shows \cite{Acta} reasonable agreement with the HERA data on VMP and DVCS, when fitted separately to each reaction, i.e. with a large number of parameters adjusted to each particular reaction.

    As a further step, to reproduce the observed trend of hardening\footnote{In what follows we use the variable $\widetilde {Q^2}$ as a measure of
``hardness''.}, as $\widetilde{Q^2}$ increases, and following Donnachie and Landshoff \cite{L, DL}, a two-term amplitude, characterized by a two-component - ``soft'' + ``hard'' - Pomeron,  was suggested \cite{Acta}. We stress that the Pomeron is unique, but we construct it as a sum of two terms. Then, the amplitude is defined as
\begin{equation}
A(\widetilde {Q^2},s,t)=A_s (\widetilde {Q^2},s,t)+A_h(\widetilde {Q^2},s,t)
\label{two-term-amp}
\end{equation}
($s=W^2$ is the square of the c.m.s. energy),
such that the relative weight of the two terms changes with $\widetilde {Q^2}$ in a right way (see Fig.~\ref{fig:Rsh}), i.e. such that the ratio $r=A_h/A_s$ increases as the reaction becomes ``harder'' and v.v. It is interesting to note that this trend is not guaranteed ``automatically'': both the ``scaling'' model  \cite{Capua, FazioPhysRev} or the Reggeometric one \cite{Acta} show the opposite tendency, that may not be merely an accident and whose reason should be better understood. This ``wrong'' trend can and should be corrected, and in fact it was corrected \cite{L, DL} by means of additional $\widetilde {Q^2}$-dependent factors $H_i(\widetilde {Q^2}),\ i=s,h$ modifying the $\widetilde {Q^2}$
dependence of the amplitude,
in a such way as to provide increasing of the weight of the hard component with increasing $\widetilde {Q^2}$. To avoid conflict with unitarity, the rise with $\widetilde {Q^2}$ of the hard component is finite (or moderate), and it terminates at some saturation scale,
whose value is determined phenomenologically. In other words, the ``hard" component, invisible at small $\widetilde {Q^2}$, gradually takes over as $\widetilde {Q^2}$ increases. An explicit example of these functions will be given below.

This paper is organized as follows. In Sec.~\ref{sec:Single} we remind and update the single-component Reggeometric model of Ref.~\cite{Acta}.
In Sec.~\ref{sec:Two-components model} the model is extended to a two-component Pomeron: ``soft'' + ``hard''. A global fit to the HERA data on all VMP and DVCS, with a unique (and small!) number of parameters,  is presented. In Sec.~\ref{sec:Balance} the $\widetilde {Q^2}$-dependent balance between the two components is studied.
In Sec. \ref{sec:pp} we attempt to unify virtual photon- and hadron-induced reactions taking high-energy $pp$ scattering as an example. Hadron-hadron elastic scattering is different from exclusive VMP and DVCS not only because the photon is different from a hadron (although they are related by vector meson dominance), but even more so by the transition between space- and time-like regions: while the virtual photon's ``mass square" $q^2$ is negative, that of the hadron is positive and that makes this attempt interesting!
Our main results and the open questions are summarized in Sec.~\ref{sec:conclusion}.


\section{Single-component Reggeometric Pomeron}
\label{sec:Single}
We start by reminding the properties and some representative results of the single-term Reggeometric model \cite{Acta}.

The invariant scattering amplitude is defined as
\begin{equation}\label{Amplitude1}
A(Q^2,s,t)=\widetilde He^{-\frac{i\pi\alpha(t)}{2}}\left(\frac{s}{s_0}\right)^{\alpha(t)} e^{2\left(\frac{a}{\widetilde{Q^2}}+\frac{b}{2m_N^2}\right)t},
\end{equation}
where
\begin{equation}
\alpha(t)=\alpha_0+\alpha't
\end{equation}
is the linear Pomeron trajectory, $s_0$ is a scale for the square of the total energy $s$, $a$ and $b$ are two parameters to be determined with the fitting procedure and $m_N$ is the nucleon mass. The coefficient $\widetilde H$ is a function providing the right behavior of elastic cross sections in $\widetilde{Q^2}$:
\begin{equation}
\widetilde H\equiv \widetilde H(\widetilde{Q^2})=\frac{\widetilde{A_0}}{\left(1+\frac{\widetilde{Q^2}}{{Q_0^2}}\right)^{n_s}},
\label{eq:norm}
\end {equation}
where $\widetilde{A_0}$ is a normalization factor, $Q_0^2$ is a scale for the virtuality and $n_s$ is a real positive number.

In this model we use an effective Pomeron, which can be ``soft'' or ``hard'', depending on the reaction and/or kinematical region defining its ``hardness''. In other words, the  values of the parameters $\alpha_0$ and $\alpha'$ must be fitted to each set of the data. Apart from $\alpha_0$ and $\alpha'$, the model contains five more sets of free parameters, different in each reaction, as shown in Table~\ref{tab:one_term}.
The exponent in the exponential factor in Eq.~(\ref{Amplitude1}) reflects the geometrical nature of the model: $a/\widetilde {Q^2}$  and $b/2m_N^2$ correspond to the ``sizes" of upper and lower vertices in Fig. \ref{fig:diagram}c.

By using the {Eq.~\ref{eq:norm}) wkth the norm
\begin{equation}\label{eq:dcsdt_from_Amplitude}
\frac{d\sigma_{el}}{dt}=\frac{\pi}{s^2}|A(Q^2,s,t)|^2,
\end{equation}
the differential and integrated elastic cross sections become,
\begin{equation}\label{eq:dcsdt_1}
\frac{d\sigma_{el}}{dt}=\frac{A_0^2}{\left(1+\frac{\widetilde{Q^2}}{{Q_0^2}}\right)^{2n}}\left(\frac{s}{s_0}\right)^{2(\alpha(t)-1)}e^{4\left(\frac{a}{\widetilde{Q^2}}+\frac{b}{2m_N^2}\right)t}
\end{equation}
and
\begin{equation}\label{eq:cs_1}
\sigma_{el}=\frac{A_0^2} {\left(1+\frac{\widetilde{Q^2}}{{Q_0^2}}\right)^{2n}}
\frac{\left(\frac{s}{s_0}\right)^{2(\alpha_0-1)}}
{4\left(\frac{a}{\widetilde{Q^2}}+\frac{b}{2m_N^2}\right)+2\alpha'\ln\left(\frac{s}{s_0}\right)},
\end{equation}
where
$$A_0=-\frac{\sqrt{\pi}}{s_0}\widetilde{A_0}.$$

Eqs.~(\ref{eq:dcsdt_1}) and (\ref{eq:cs_1}) (for simplicity we set $s_0 = 1$ GeV$^2$) were fitted to the HERA data obtained the by ZEUS and H1 Collaborations on exclusive diffractive VMP~\cite{z1, z4, z19, h15, h9, h6, z9, z13, h16, z24, z16, h3, z15, z26, z8, h14, h19, z28, z27, z18, z6, i'69, h17, z21, z1d, z25, h22, z29, h2} and DVCS~\cite{z5, h4, h7, z10, h13}.

In the present paper we have updated and extended the fits shown earlier in Ref.~\cite{Acta}, the results being collected in Table \ref{tab:one_term}, where the parameters with indefinite error bars were fixed at the fitting stage. The ``mass parameter'' for DVCS was set to $M=0$ GeV, therefore in this case $\widetilde{Q^2}=Q^2$. Each type of reaction was fitted separately. The representative fits, for $J/\psi$ and $\rho^{0}$ production, are shown in Figs.~\ref{fig:onetermJpsi} and \ref{fig:oneterm_rho}, respectively.
As it can be seen from the right plot of Fig.~\ref{fig:oneterm_rho}, the single-term model fails to fit both the high- and low-$|t|$ regions properly, especially when soft (photoproduction or low $Q^2$) and hard (electroproduction or high $Q^2$) regions are considered.
One of the problems of the single-term Reggeometric Pomeron model, Eq.~\eqref{Amplitude1}, is that the fitted parameters in this model acquire particular values for each reaction, 
which is one of the reasons for its extension to two terms  (next Section). 
 %
\begin{figure}[ht]
  \vspace{-0.2cm}
  \centering
  \includegraphics[trim = 0mm 0mm 1mm 0mm,clip,scale=0.34]{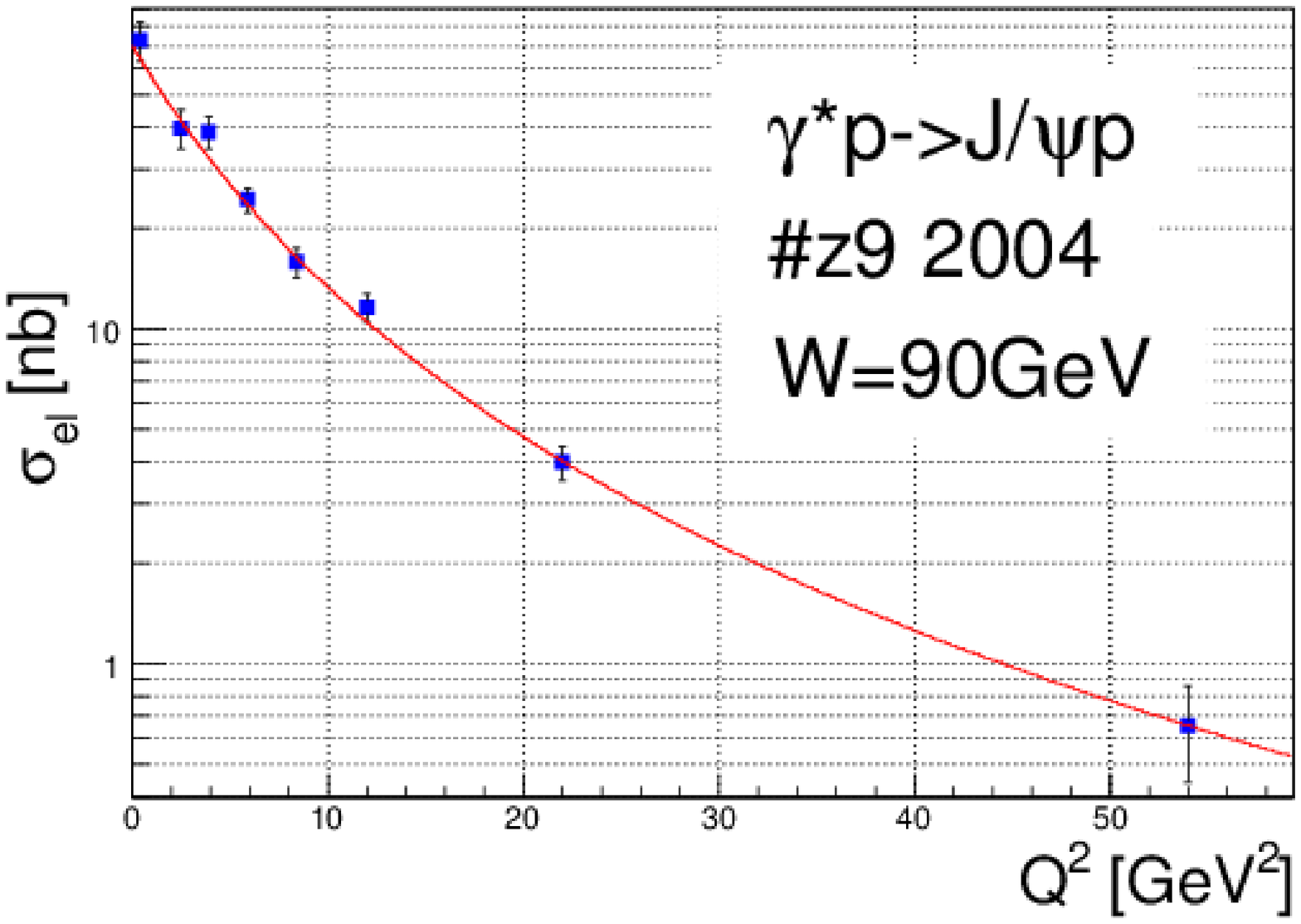}
  \includegraphics[trim = 1mm 0mm 1mm 0mm,clip,scale=0.72]{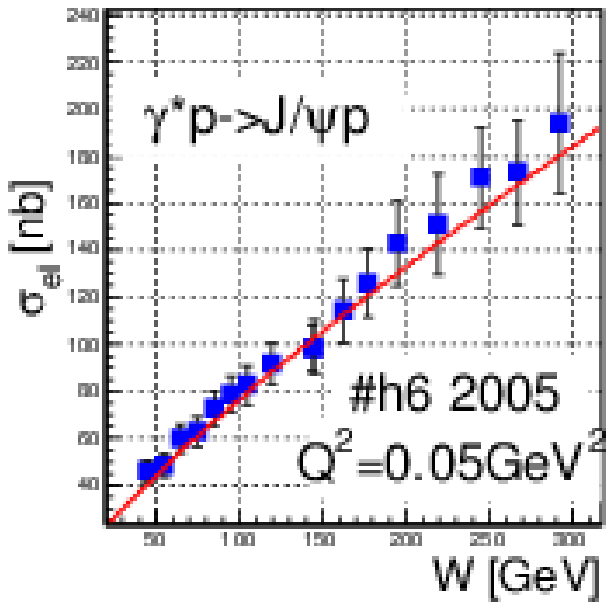}
  \includegraphics[trim = 1mm 2mm 1mm 0mm,clip,scale=0.72]{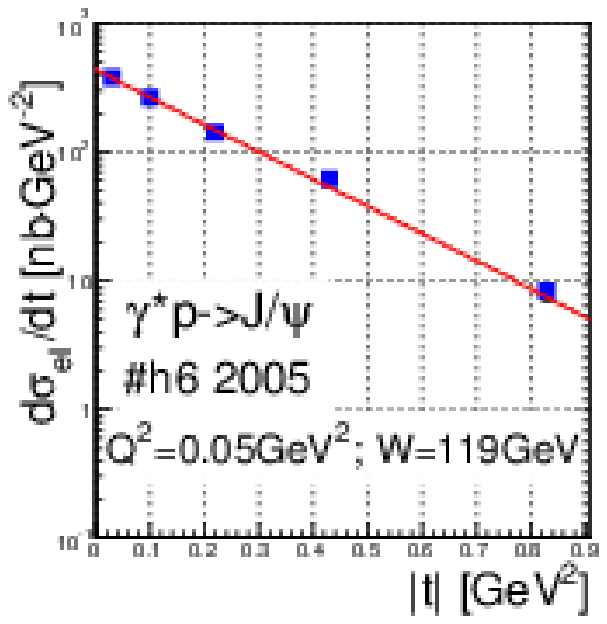}
  \caption{ \label{fig:onetermJpsi}Representative fits of Eqs.~\eqref{eq:dcsdt_1} and \eqref{eq:cs_1} to the data on $J/\psi$ production. The values of the fitted parameters are compiled in Table~\ref{tab:one_term}.}
\end{figure}

\begin{figure}[!ht]
  \vspace{-0.2cm}
  \centering
  \includegraphics[trim = 3mm 0mm 15mm 0mm,clip,scale=0.41]{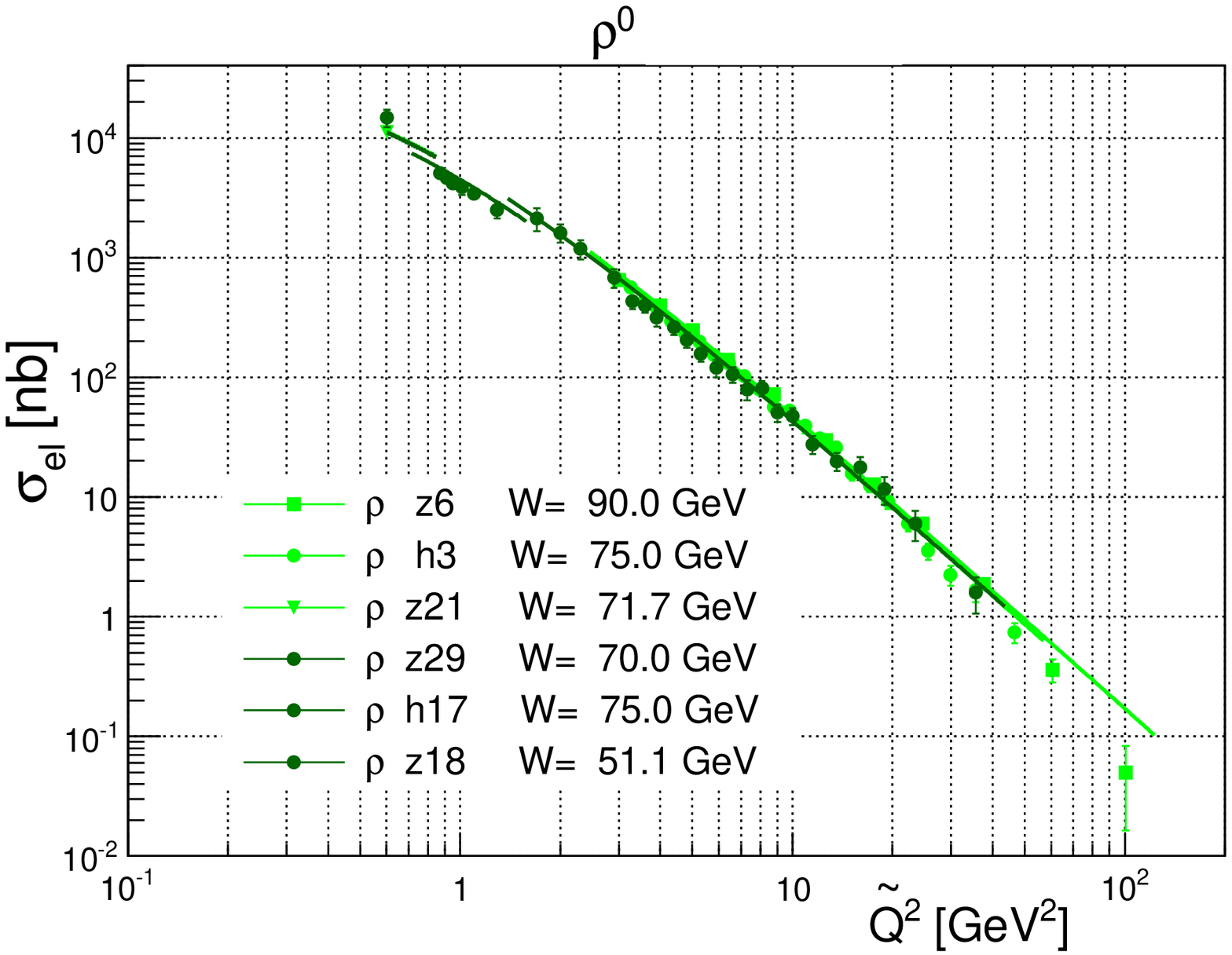}
  \includegraphics[trim = 2mm 0mm 5mm 0mm,clip,scale=0.41]{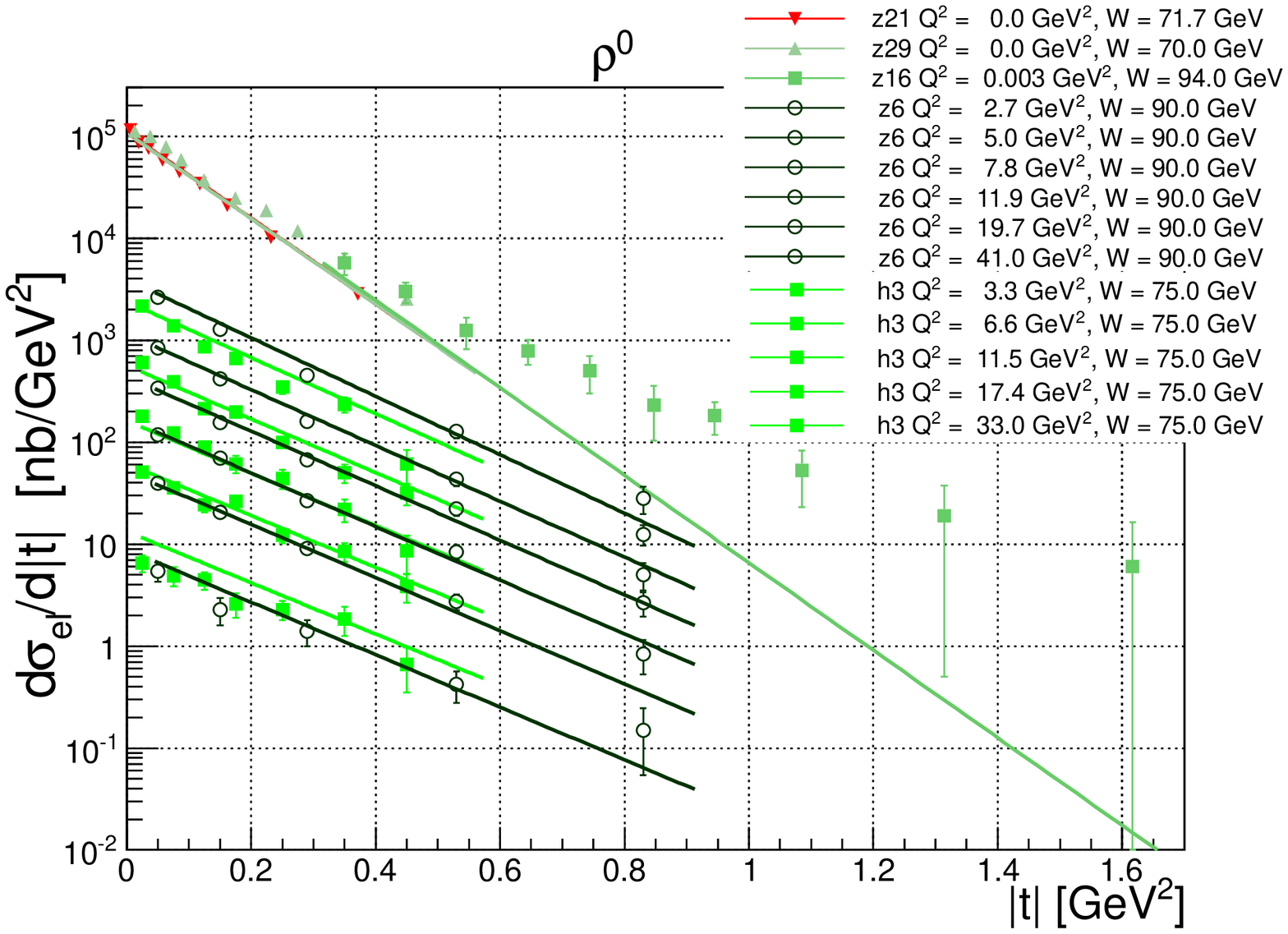}
  \caption{ \label{fig:oneterm_rho}Representative fits of Eqs.~\eqref{eq:dcsdt_1} and \eqref{eq:cs_1} to the data on $\rho^{0}$ electroproduction. The values of the fitted parameters are compiled in Table~\ref{tab:one_term}. }
\end{figure}

\begin{table}[!b]
  \centering
  \footnotesize
  \caption{Values of the parameters in Eqs.~(\ref{eq:dcsdt_1}), (\ref{eq:cs_1}) fitted to data on VMP and DVCS at HERA.
  The parameters with indefinite error bars were fixed at the fitting stage.
  \label{tab:one_term}}
  \begin{tabular}{c||c|c|c|c|c|c|c|c}
     \hline
                 &$A_0$ $\left[\frac{\sqrt\text{nb}}{\text{GeV}}\right]$
                 &$\widetilde{Q^2_0}$ $\left[\text{GeV}^2\right]$&   $n$
                 &$\alpha_{0}$& $\alpha'$  $\left[\frac{1}{\text{GeV}^{2}}\right]$
                 &$a$&$b$&${\tilde\chi}^2$ \\ \hline
        $\rho^0$&344$\pm$376&0.29$\pm$0.14      &1.24$\pm$0.07 &1.16$\pm$0.14& 0.21$\pm$0.53& 0.60$\pm$0.33&0.9$\pm$4.3 &2.74\\ 
        $\phi$  &58$\pm$112 &0.89$\pm$1.40      &1.30$\pm$0.28 &1.14$\pm$0.19& 0.17$\pm$0.78& 0.0$\pm$19.8 &1.34$\pm$5.09&1.22\\
        $J/\psi$& 30$\pm$31 &2.3$\pm$2.2        &1.45$\pm$0.32 &1.21$\pm$0.09& 0.077$\pm$0.072& 1.72       &1.16        &0.27\\  
$\varUpsilon$(1S)&37$\pm$100&0.93$\pm$1.75      &1.45$\pm$0.53 &1.29$\pm$0.25& 0.006$\pm$0.6& 1.90         &1.03        &0.4\\  
          $DVCS$&14.5$\pm$41.3&0.28$\pm$0.98     &0.90$\pm$0.18 &1.23$\pm$0.14& 0.04$\pm$0.71& 1.6          &1.9$\pm$2.5 &1.05\\ \hline
        \end{tabular}
 \end{table}

The VMP results clearly show the hardening of Pomeron in the change of $\alpha_0$ and $\alpha'$ when going from light to heavy vector mesons.
As seen in the right plot of Fig.~\ref{fig:oneterm_rho} of the $\rho^{0}$ case, a single exponent of the type $Ae^{Bt}$ is not sufficient to reproduce the differential cross section above $|t|>0.5$ in the electroproduction and especially in photoproduction regions.
This is the reason why it is so difficult to describe $\rho^{0}$ production in the whole kinematic range within the single-term Pomeron model.
These two phenomena (``hardening" of Pomeron trajectory and problems with $\rho^0$ production) motivate the introduction of a two-component Pomeron.

 It is also interesting to note that the effective Pomeron trajectory for DVCS ($\alpha_0 = 1.23,$ $\alpha'=0.04$, see Table~\ref{tab:one_term}) is  typically a hard one, in contradiction to expectations, that it should be soft at low-$Q^2$.  


\section{Two-component Reggeometric Pomeron}
\label{sec:Two-components model}
\subsection{Amplitude with two, ``soft'' and ``hard'', components}
Now we introduce the universal, ``soft'' and ``hard'', Pomeron model.
Using the Reggeometric ansatz of Eq.~\eqref{Amplitude1}, we write the amplitude as a sum of two parts, corresponding to the ``soft'' and ``hard'' components of a universal, unique  Pomeron:
  \begin{equation}\label{eq:Amplitude_hs_prime}
 A(Q^2,s,t)=
      \widetilde{H_s}\,e^{-i\frac{\pi}{2}\alpha_s(t)}\left(\frac{s}{s_{0s}}\right)^{\alpha_s(t)} e^{2\left(\frac{a_{s}}{\widetilde{Q^2}}+\frac{b_{s}}{2m_N^2}\right)t}
     +\widetilde{H_h}\,e^{-i\frac{\pi}{2}\alpha_h(t)}\left(\frac{s}{s_{0h}}\right)^{\alpha_h(t)} e^{2\left(\frac{a_{h}}{\widetilde{Q^2}}+\frac{b_{h}}{2m_N^2}\right)t}.
  \end{equation}
Here  $s_{0s}$ and $s_{0h}$
are squared energy scales, and $a_i$ and $b_i$, with $i = s,h$, are parameters to be determined with the fitting procedure. The two coefficients $\widetilde{H_s}$ and $\widetilde{H_h}$ are functions similar to those defined in Ref. \cite{DL}:
\begin{equation}\label{eq:HsHh_tilde}
    \widetilde{H_s}\equiv\widetilde{H_s}(\widetilde{Q^2})=\frac{\widetilde{A_s}}{{\Bigl(1+\frac{\widetilde{Q^2}}{{Q_s^2}}\Bigr)}^{n_s}}, ~~~~~~\quad
    \widetilde{H_h}\equiv\widetilde{H_h}(\widetilde{Q^2})=\frac{\widetilde{A_h} \Bigl(\frac{\widetilde{Q^2}}{Q_h^2}\Bigr)}{{\Bigl(1+\frac{\widetilde{Q^2}}{{Q_h^2}}\Bigr)}^{n_h+1}},
\end{equation}
where $\widetilde{A_s}$ and $\widetilde{A_h}$ are normalization factors, $Q_s^2$ and $Q_h^2$ are scales for the virtuality, $n_s$ and $n_h$ are real positive numbers.
Each component of Eq.~\eqref{eq:Amplitude_hs_prime} has its own, ``soft" or ``hard", Regge (here Pomeron) trajectory:
$$\alpha_s(t)=\alpha_{0s}+\alpha_s't, ~~~~~~~~~~~\quad  \alpha_h(t)=\alpha_{0h}+\alpha_h't.$$
As an input we use the parameters suggested by Donnachie and Landshoff \cite{DL_tr}, so that
$$\alpha_s(t) = 1.08 + 0.25t,~~~~~~~~~~\quad \alpha_h(t) = 1.40 + 0.1t.$$

The ``Pomeron"  amplitude (\ref{eq:Amplitude_hs_prime}) is unique, valid for all diffractive
reactions, its ``softness'' or ``hardness'' depending on the relative $\widetilde{Q^2}$-dependent weight of the two components, governed by the relevant factors $\widetilde H_s(
\widetilde Q^2)$ and $\widetilde {H_h}(\widetilde {Q^2)}$.

 Fitting Eq.~\eqref{eq:Amplitude_hs_prime} to the data, we have found that the parameters assume rather large errors and, in particular, the parameters $a_{s,h}$ are close to $0$. Thus, in order  to reduce the number of free parameters, we simplified the model, by fixing $a_{s,h}=0$ and substituting the exponent $2\left(\frac{a_{s,h}}{\widetilde{Q^2}}+\frac{b_{s,h}}{2m_N^2}\right)$ with $b_{s,h}$ in Eq.~\eqref{eq:Amplitude_hs_prime}. The proper variation with $\widetilde {Q^2}$ will be provided by the factors $\widetilde{H_s}(\widetilde{Q^2})$ and $\widetilde{H_h}(\widetilde{Q^2})$.

Consequently, the scattering amplitude assumes the form
  \begin{equation}\label{eq:Amplitude_hs}
 A(s,t,Q^2,M_V^2)=
      \widetilde{H_s}\,e^{-i\frac{\pi}{2}\alpha_s(t)}\left(\frac{s}{s_{0s}}\right)^{\alpha_s(t)} e^{b_st}
     +\widetilde{H_h}\,e^{-i\frac{\pi}{2}\alpha_h(t)}\left(\frac{s}{s_{0h}}\right)^{\alpha_h(t)} e^{b_ht}.
  \end{equation}

 The ``Reggeometric'' combination $2\left(\frac{a_{s,h}}{\widetilde{Q^2}}+\frac{b_{s,h}}{2m_N^2}\right)$ was important for the description of the slope $B(Q^2)$ within the single-term Pomeron model (see previous Section), but in the case of two terms the $Q^2$-dependence of $B$ can be reproduced without this extra combination, since each term in the amplitude \eqref{eq:Amplitude_hs} has its own $Q^2$-dependent factor $\widetilde{H_{\ }}_{\!\!s,h}(Q^2)$.

  By using the amplitude (\ref{eq:Amplitude_hs}) and Eq.~\eqref{eq:dcsdt_from_Amplitude}, we calculate the differential and elastic cross sections, by setting for simplicity $s_{0s} = s_{0h} = s_0$, to obtain
\begin{equation}\label{eq:dcsdt(h+s)}
\frac{d\sigma_{el}}{{dt}}=H_s^2e^{2\{L(\alpha_s(t)-1)+{b_s}t\}}+H_h^2e^{2\{L(\alpha_h(t)-1)+{b_h}t\}}
\end{equation}
$$+2H_sH_he^{\{L(\alpha_s(t)-1)+L(\alpha_h(t)-1)+({b_s}+{b_h})t\}}\cos\Bigl(\frac{\pi}{2}(\alpha_s(t)-\alpha_h(t))\Bigr),$$
\begin{equation}\label{eq:cs(h+s)}
     \sigma_{el}=\frac{H_s^2e^{2\{L(\alpha_{0s}-1)\}}}{2(\alpha_s'L+{b_s})}
                +\frac{H_h^2e^{2\{L(\alpha_{0h}-1)\}}}{2(\alpha_h'L+{b_h})}
                +2H_sH_he^{L(\alpha_{0s}-1)+L(\alpha_{0h}-1)}
                 \frac{\mathfrak{B}\cos\phi_0+\mathfrak{L}\sin\phi_0} {\mathfrak{B}^2 + \mathfrak{L}^2}.
\end{equation}

In these two equations we used the notations

  \begin{equation}\label{eq:Denotation}
  \begin{array}{l}
    L=\ln\left({s/s_{0}}\right),
    \\\phi_0=\frac{\pi}{2}(\alpha_{0s}-\alpha_{0h}),
  \end{array}\qquad
  \begin{array}{l}
    \mathfrak{B}=L\alpha_s' + L\alpha_h'+({b_s}+{b_h}),
    \\\mathfrak{L}=\frac{\pi}{2}(\alpha_s'-\alpha_h'),
  \end{array}
  \nonumber
  \end{equation}
\begin{equation}
    H_s(\widetilde{Q^2})=\frac{A_s}{{\Bigl(1+\frac{\widetilde{Q^2}}{{Q_s^2}}\Bigr)}^{n_s}}, \quad
    H_h(\widetilde{Q^2})=\frac{A_h\Bigl(\frac{\widetilde{Q^2}}{{Q_h^2}}\Bigr)}{{\Bigl(1+\frac{\widetilde{Q^2}}{{Q_h^2}}\Bigr)}^{n_h+1}}, \nonumber
\end{equation}
with
  $$A_{s,h}=-\frac{\sqrt{\pi}}{s_{0}}\widetilde{A_{\ }}_{\!\!s,h}.$$

Finally, we notice that amplitude~(\ref{eq:Amplitude_hs}) can be rewritten in the form
  $$
   A(s,t,Q^2,{M_v}^2)= \widetilde{A_s}e^{-i\frac{\pi}{2}\alpha_s(t)}\left(\frac{s}{s_{0}}\right)^{\alpha_s(t)}
    e^{b_st - n_s\ln{\left(1+\frac{\widetilde{Q^2}}{\widetilde{Q_s^2}}\right)}}
  $$
  \begin{equation}
  +\widetilde{A_h}e^{-i\frac{\pi}{2}\alpha_h(t)}\left(\frac{s}{s_{0}}\right)^{\alpha_h(t)}
    e^{b_ht - (n_h+1)\ln{\left(1+\frac{\widetilde{Q^2}}{\widetilde{Q_h^2}}\right)}
    +\ln{\left(\frac{\widetilde{Q^2}}{\widetilde{Q_h^2}}\right)} },
    \label{eq:Amplitude_hs_modif}
    \end{equation}
  where  the two exponential factors $e^{b_st - n_s\ln{\left(1+\frac{\widetilde{Q^2}}{\widetilde{Q_s^2}}\right)}}$ and $e^{b_ht - (n_h+1)\ln{\left(1+\frac{\widetilde{Q^2}}{\widetilde{Q_h^2}}\right)}
    +\ln{\left(\frac{\widetilde{Q^2}}{\widetilde{Q_h^2}}\right)}}$ can be interpreted as the product of the form factors of upper and lower vertices (see Fig.~\ref{fig:diagram}c). Interestingly, the amplitude~\eqref{eq:Amplitude_hs_modif}  resembles the scattering amplitude in Ref.~\cite{Capua}.

\subsection{Fitting the two-component Pomeron to VMP and DVCS data from HERA}
\subsubsection{Normalization of the data from different reactions}
Before fitting our model to the available HERA experimental data on $d\sigma_{el}/dt(t)$ and $\sigma_{el}(Q^{2},W)$ of VMP and DVCS reactions, it is necessary to normalize these data such that they lie on the same surface, i.e. give the same values of the cross sections for the same values of $W$, $\widetilde{Q^2}$ (and $t$). 
We chose the $J/\psi$ production as a ``reference point''. The normalization procedure is not unique. For example, according to Ref. \cite{Nikolaiev} there are three sets of normalization parameters. For vector mesons in our calculations we used
\begin{equation}\label{eq:f}
 f_{\rho^0}:f_{\omega}:f_{\phi}:f_{J/\psi} = 0.68:0.068:0.155:1.
\end{equation}
Also, from the fits we set $f_{\varUpsilon}:f_{J/\psi}=0.75$.

Let us stress that we compared the data for different reactions at the same values of $\widetilde{Q^2}$
rather than $Q^2$.

After normalization, the differential and elastic cross sections lie on  the same surface, so that
$$
f_{\rho^0}\sigma_{\rho^0}=f_{\omega}\sigma_{\omega}=f_{\phi}\sigma_{\phi}=f_{J/\psi}\sigma_{J/\psi}=f_{\varUpsilon}\sigma_{\varUpsilon},
$$
where each $\sigma$ stands for all kinds of cross section. Just as an example, the fit of Eq.~(\ref{eq:cs(h+s)}) to the normalized data on cross section
$f_i\cdot\sigma_{el}(\widetilde{Q^2})$ relative to  ${\rho^0}$, ${\omega}$, ${\phi}$ and ${J/\psi}$ production is shown in Fig.~\ref{fig:cs(Q2)}.

\subsubsection{Fitting procedure}
 We performed a global fit of our model, using Eqs.~(\ref{eq:dcsdt(h+s)}) and (\ref{eq:cs(h+s)}), to all VMP (i.e. $J/\psi$, $\phi$, $\rho^0$ and $\varUpsilon$) and DVCS HERA data, with $W>30$GeV. Notice that
 in this energy range only diffractive events were selected at HERA and, consequently, the Pomeron is the only object exchanged in the $t$ channel.

The fitting strategy is based on the minimization of the quantity $\tilde\chi^2=\frac{1}{N}\sum_{i=1}^{N} {\tilde\chi_i^2}$,
where $N$ is the number of all reactions involved (i.e. $\rho$, $\phi$, $\omega$, $J/\psi$, $\varUpsilon$ and $\gamma$ production);
$\tilde\chi_i$ is the mean value of $\chi^2$ for different types of data for  selected class of reactions, defined as $\tilde\chi_i=\frac{1}{N_i}\sum_{k=1}^{N_i} {\tilde\chi_{k,i}^2}$,
where $\tilde\chi_{k,i}^2$  is $\chi_{k,i}^2/d.o.f.$ for i-th class of reactions and k-th type of data, i.e. those relative to $\sigma_{el}(Q^2)$, $\sigma_{el}(W)$ and ${d\sigma_{el}(t)}/{dt}$;
$N_i$ is number of different type of data for i-th class of reactions.

 Following relation \eqref{eq:f}, we fixed the normalization parameters at

\begin{equation}\label{eq:f_i}
f_{\rho}=0.680,\ f_{\phi}=0.155,\ f_{\omega}=0.068,\ f_{J/\psi}=1,\ f_{\varUpsilon}=0.750
\end{equation}
and set $s_{0}$ equal to $1$\;GeV$^2$.


DVCS and VMP are similar in the sence that in both reactions a vector particle is produced. However there are differences between the two because of the vanishing rest mass of the produced real photon. The unified description of these two types of related reactions does not work by simply setting $M_{\gamma}=0.$  From fits we found $M_{DVCS}^{eff}=1.8$~GeV and a normalization factor $f_{DVCS}=0.091$ follows.


The results of the fit are shown in Fig.~\ref{fig:cs(Q2)} (for $\sigma_{el}(\widetilde{Q^2})$), in Figs.~\ref{fig:cs_rho1(W)}-\ref{fig:cs_Ups(W)} (for $\sigma_{el}(W)$ and Figs.~\ref{fig:dcsdt_rho}-\ref{fig:dcsdt_Jpsi.php1} (for $d\sigma_{el}(t)/dt$) for vector meson production, and in Figs.~\ref{fig:dcsdt.DVCS}-\ref{fig:csQ2.DVCS} for DVCS, with the values of the fitted parameters given in Table~\ref{tab:fit1(s+h)}.
The mean value of the total $\tilde\chi^2$ (see above its definition) is equal to 0.986. The mean values of $\tilde\chi^2$ of the fit for different observables (i.e. $\sigma_{el}(Q^2)$, $\sigma_{el}(W)$ or $d\sigma_{el}(t)/{dt})$ and different reactions (VMP or DVCS), together with the numbers of degrees of freedom (number of data points) and the global mean value $\tilde\chi_i^2$, are shown in Table~\ref{tab:fit1_chi}.
 Furthermore, in Table~\ref{tab:fixed_trajec} the  parameters of the two-component Pomeron model (Eqs.~(\ref{eq:dcsdt(h+s)}) and (\ref{eq:cs(h+s)})) fitted to the combined VMP and DVCS data are quoted, when Pomeron trajectories are fixed to $\alpha_{s}(t)=1.08+0.25t$ and $\alpha_{h}(t)=1.20+0.01t$.


Next, by using Eq.~\eqref{eq:dcsdt(h+s)} with the values of the parameters from Table~\ref{tab:fit1(s+h)} and the formula
  \begin{equation}\label{eq:B(h+s)}
   B(Q^2,W,t)=\frac{d}{dt}\ln{\frac{d\sigma_{el}}{dt}},
  \end{equation}
we calculate the forward slopes and compare them with the experimental data on VMP, including those for the $\Psi$(2S) production. To do so,
 the experimental data were grouped in four separate $t$ bins with the mean values of $0.12$, $0.25$, $0.5$ and $0.6$ GeV$^2$. The results of these calculations, showing the $\widetilde{Q^2}$ dependence, are presented in Fig.~\ref{fig:B(Q2)_in_t}. A compilation of all results is presented in  Figs.~\ref{fig:B(Q2)} and \ref{fig:BLog(Q2)}. Note that in Figs.~\ref{fig:B(Q2)_in_t} - \ref{fig:BLog(Q2)} the results on $\Psi$(2S) are also shown. The model reproduces correctly also the $W$ dependence of the slope $B$, as shown in Fig.~\ref{fig:B(W)} for
$\rho^0$ and $J/\psi$.

\begin{table}[!htb]
   \centering
   \caption{Parameters of the two-component Pomeron model (Eqs.~(\ref{eq:dcsdt(h+s)}) and (\ref{eq:cs(h+s)})) fitted to the combined VMP and DVCS data. The value of $\tilde\chi^2$ is equal to 0.986}.
   \label{tab:fit1(s+h)}
    \begin{tabular}{c|c c c c c c c}
      \hline
    &$A_{0s,h}$ $\left[\frac{\sqrt{\text{nb}}}{\text{GeV}}\right]$&$\widetilde{Q^2_{s,h}}$  $\left[\text{GeV}^2\right]$&$n_{s,h}$ & $\alpha_{0\,s,h}$&$\alpha'_{s,h}$ $\left[\frac{1}{\text{GeV}^2}\right]$&$b_{s,h}$  $\left[\frac{1}{\text{GeV}^2}\right]$\\ \hline
%
soft&2104$\pm$1749 &0.29$\pm$0.20 &1.63$\pm$0.40 &1.005$\pm$0.090 &0.32$\pm$ 0.57 &2.93$\pm$5.06\\
hard&  44$\pm$  22 &1.15$\pm$0.52 &1.34$\pm$0.16 &1.225$\pm$0.055 &0.0 $\pm$17    &2.22$\pm$3.09\\
\hline
   \end{tabular}
  \end{table}

 \begin{table}[!htb]
   \centering
   \caption{Values of $\tilde\chi^2$ of the fit and
   the numbers of degrees of freedom (number of data points) for different observables (i.e. $\sigma_{el}(W)$, $\sigma_{el}(Q^2)$ or $d\sigma_{el}(t)/dt$), and values of
   $\tilde\chi^2_i$ for different reactions (VMP or DVCS).}
   \label{tab:fit1_chi}
    \begin{tabular}{c|c c|c c|c c|c}
  \hline
   Meson&\multicolumn{2}{c|}{$\sigma_{el}(W)$}&\multicolumn{2}{c|}{$\sigma_{el}(Q^2)$}&\multicolumn{2}{c|}{$\frac{d\sigma_{el}}{dt}$}\\
   production&$\tilde\chi^2$&$N_{d.o.f.}$&$\tilde\chi^2$&$N_{d.o.f.}$&$\tilde\chi^2$&$N_{d.o.f.}$& $\tilde\chi_{i}^2$\\ \hline

$\varUpsilon$&  0.47&  4&  0.00&  1& 0.00&  1&   0.469\\
      $J\psi$&  0.47& 43&  0.47& 16& 2.37& 92&   1.105\\
     $\omega$&  0.10&  3&  0.09&  4& 0.33&  7&   0.174\\
       $\phi$&  1.19& 46&  1.42& 22& 1.10& 85&   1.238\\
       $\rho$&  1.49&112&  0.97& 64& 3.85& 94&   2.104\\
       $DVCS$&  1.83& 89&  2.20& 38& 1.41& 84&   1.815\\ \hline
\end{tabular}
  \end{table}

 \begin{table}[!htb]
   \centering
    \caption{ Parameters of the two-component Pomeron model (Eqs.~(\ref{eq:dcsdt(h+s)})
    and (\ref{eq:cs(h+s)})) fitted to the combined VMP and DVCS data, with fixed    parameters of the Pomeron trajectories $\alpha_{s}(t)=1.08+0.25t$ and $\alpha_{h}(t)=1.20+0.01t$.}
    \label{tab:fixed_trajec}
    \begin{tabular}{c|c c c c c c c}
      \hline
    &$A_{0s,h}$ $\left[\frac{\sqrt{\text{nb}}}{\text{GeV}}\right]$&$\widetilde{Q^2_{s,h}}$  $\left[\text{GeV}^2\right]$&$n_{s,h}$ & $\alpha_{0\,s,h}$&$\alpha'_{s,h}$ $\left[\frac{1}{\text{GeV}^2}\right]$&$b_{s,h}$  $\left[\frac{1}{\text{GeV}^2}\right]$\\ \hline
soft&807$\pm$1107&0.46$\pm$0.70&1.79$\pm$0.79&1.08&0.25&3.41$\pm$2.48\\
hard& 47.9$\pm$  46.9&1.30$\pm$1.12&1.33$\pm$0.26&1.20&0.01&2.15$\pm$1.14\\
\hline
    \end{tabular}
 \end{table}

\newpage
 \begin{landscape}
  \begin{table*}[ht]
    \centering
    \caption{ The experimental data used. (* stands for luminosity after triggers) \label{tab:Data1}}
     \begin{tabular}{cccc|cc|cc|c|ccc p{4.25cm}}
    \hline
      &cite      &year&                  &$W$&$\left<W\right>$  &$Q^2$        &$\left<Q^2\right>$&$|t|$  &type&Lumi $[\text{pb}^{-1}]$&Data& \\ \hline
    z1&\cite{z1} &2011&$\varUpsilon(1S)$ &$60\div220$&$90$      &$<1$         &$~10^{-3}$      &$0$-$5$   &php&$468$  &$[96$-$07]$&$\mu^+\mu^-$\\
    z4&\cite{z4} &2009&$\varUpsilon(1S)$ &$60\div220$&          &$<1$         &$~10^{-3}$      &          &php&$468$  &$[96$-$07]$&$\mu^+\mu^-$\\
   z19&\cite{z19}&1998&$\varUpsilon(1S)$ &$80\div160$&$120$     &$<1$         &$~5\cdot10^{-5}$&          &php&$43.2$ &$[95$-$97]$&$\mu^+\mu^-$\\
   h15&\cite{h15}&2000&$\varUpsilon(1S)$ &$70\div250$&$160$     &$<1$         &$0.011$         &$0\div1.2$&php&$27.5$ &$[94$-$97]$&$\mu^+\mu^-$\\
      &          &    &$J/\psi$          &$26\div285$&$160$     &$ $          &$0.05$          &$0\div1.2$&php&$20.5$ &$[96$-$97]$&$\mu^+\mu^-$, $e^+e^-$\\ \hline
    h9&\cite{h9} &2002&$\psi(2s)$, $J/\psi$&$40\div150$&$ $     &$<1$         &$0.055$         &$0\div1.0$&php&$77$   &$[96$-$00]$&$\mu^+\mu^-$, $e^+e^-$, $J/\psi+\pi^+\pi^-$\\
    h6&\cite{h6} &2005& $J/\psi$         &$40\div305$&$90$     &$<1$          &$0.05$         &$<1.2$    &php& $55$  &$[99$-$00]$ &$e^+e^-,$ $\mu^+\mu^-$ \\
      &          &    & $J/\psi$         &$40\div160$&$90$     &$2\div80$     &$8.9$          &$<1.2$    &   & $55$  &$[99$-$00]$ &\\ \hline
    z9&\cite{z9} &2004&$J/\psi$          &$30\div220$&$\sim\!90$&$0.15$ - $0.8$&$ $            &$<1$      &   &$69$   &$[98$-$00]$& $e^+e^-$\\
      &          &    &                  &           &          &$2$ - $100$  &$ $             &          &   &$83$   &           & $\mu^+\mu^-,$ $e^+e^-$\\ \hline
   z13&\cite{z13}&2002&$J/\psi$          &$20\div290$&$ $       &$<1$         &$5\cdot10^{-5}$ &$<1.25$   &php&$55.2$ &$[99$-$00]$& $e^+e^-$\\
      &          &    &                  &$20\div170$&$ $       &$ $          &$ $             &$<1.8 $   &php&$38$   &$[96$-$97]$& $\mu^+\mu^-$\\ \hline
   h16&\cite{h16}&1999&$J/\psi$          &$25\div180$&$96$      &$2$ - $80$   &$8$             &$<1.5$    &   &$27.3$ &$[95$-$97]$& $\mu^+\mu^-$, $e^+e^-$\\
   z24&\cite{z24}&1997&$J/\psi$          &$40\div140$&$ $       &$<4$         &$5\cdot10^{-5}$ &$<1$      &php&$2.7$  &$[94]$     & $e^+e^-$\\
      &          &    &                  &$ $        &$ $       &$ $          &$ $             &$ $       &   &$1.87$ &$ $        & $\mu^+\mu^-$\\ \hline
   z16&\cite{z16}&2000&$J/\psi$, $\phi$, $\rho$
                                         &$85\div105$&$94$      &$<0.01$      &$7\cdot10^{-6}$ &$<3$      &php&$1.98$ &$[95]$     & $J/\psi\rightarrow\mu^+\mu^-(e^+e^-);$\linebreak $\rho\rightarrow\pi^+\pi^-;$ $\phi\rightarrow K^+K^-$ \\
    h3&\cite{h3} &2010&$\phi$, $\rho$    &$35\div180$&$ $       &$2.5$ - $60$ &                &$<3$      &   &  $51$ &$[96$-$00]$& $\rho\rightarrow\pi^+\pi^-;$ $\phi\rightarrow K^+K^-$\\
   z15&\cite{z15}&2000&$\omega$, $\phi$  &$40\div120$&$70$      &$3$ - $20$   & $7$            &$<0.6$    &   &$37.7$ &$[96$-$97]$& $\pi^+\pi^-\pi^0;$ $\pi^0\rightarrow\gamma\gamma$\\
   z26&\cite{z26}&1999&$\omega$          &$70\div90$ &$80$      &$<4$         &$10^{-4}$       &$<0.6$    &php&$3.2;$ &$[94]$&$\pi^+\pi^-\pi^0;$ $\pi^0\rightarrow\gamma\gamma$\\ \hline
    z8&\cite{z8} &2005& $\phi$           &$35\div145$&$75$      &$2$ - $70$   & $5$            &$<0.6$    &   &$65.1$ &$[98$-$00]$& $K^+K^-$\\
   h14&\cite{h14}&2000& $\phi$           &$40\div130$&$75$      &$1$ - $5$    &$ $             &$<0.5$    &   &$0.125$&$[95]$     & $K^+K^-$\\
      &          &    &                  &$ $        &$ $       &$2.5$ - $15$ &                &$ $       &   &    $3$&$[96]$     & $K^+K^-$\\ \hline
   h19&\cite{h19}&1997& $\phi$           &$42\div134$&$100$     &$6$ - $20$   &                &$<0.6$    &   & $2.8$ &$[94]$     & $K^+K^-$\\
   z28&\cite{z28}&1996& $\phi$           &$60\div80$ &$70$      &$<4$         &$10^{-1}$       &$0.1\div0.5$&php&${}^*0.887$&$[94]$&$K^+K^-$\\
   z27&\cite{z27}&1996& $\phi$           &$42\div134$&$98$      &$7$ - $25$   &$12.3$          &$<0.6$    &   &$2.62$ &$[94]$     & $K^+K^-$\\ \hline
   \end{tabular}
 \end{table*}

 \newpage
\begin{table*}[rht]
  \centering
  \caption{ The experimental data used. (* stands for luminosity after triggers) \label{tab:Data2}}
   \begin{tabular}{cccc|cc|cc|c|ccc p{3cm}}
    \hline
      &cite      &year&                  &$W$ &$\left<W\right>$&$Q^2$      &$\left<Q^2\right>$&$|t|$&type&Lumi&Data& \\ \hline
   z18&\cite{z18}&1998& $J/\psi$         &$50\div150$&$97$     &$2$ - $40$    &$5.9$          &$<1$      &   &$ 6.0$ &$[95]$      &$e^+e^-,$ $\mu^+\mu^-$\\
      &          &    & $\rho$           &$20\div90$ &$47$     &$0.25$ - $0.85$&$0.47$        &$<0.6$    &php& $3.8$ &$[95]$      &$\pi^+\pi^-$ \\
      &          &    & $\rho$           &$32\div167$&$67$     &$3$ - $50$    &$6.2$          &$<0.6$    &   & $6.0$ &$[95]$      &$\pi^+\pi^-$ \\ \hline
    z6&\cite{z6} &2007& $\rho$           &$32\div180$&$90$     &$2$ - $160$   &               &$<1$      &   & $120$  &$[96$-$00]$ &$\pi^+\pi^-$ \\
   h1c&\cite{i'69}&2002&$\rho$           &$25\div70$ &$38.1$   &$<1$          &$10^{-4}$ &$0.073\div0.45$&php&$3$    &$[99]$      &$\pi^+\pi^-$\\
   h17&\cite{h17}&1999& $\rho$           &$30\div140$&$75$     &$1$ - $5$     &$ $            &$<0.5$    &   &$0.125$ &$[95]$      &$\pi^+\pi^-$ \\
      &          &    &                  &           &         &$2.5$ - $60$  &               &          &   & $3.87$ &$[96]$      & \\ \hline
   z21&\cite{z21}&1997& $\rho$           &$50\div100$&$71.7$   &$<4$          &$4\cdot10^{-6}$&$<0.5$    &php&${}^*2.17$&$[94]$  &$\pi^+\pi^-$ \\
   z1d&\cite{z1d}&1997&$\rho$&$50\div100$&$72$     &$<4$          &$10^{-5}$      &$<0.5$    &php&$0.691$&$[94]$      &$\pi^+\pi^-$\\
   z25&\cite{z25}&1996& $\rho$           &$50\div100$&$70$     &$<1$          &$10^{-4}$      &$0.073\div0.4$&php&$0.898$&$[94]$  &$\pi^+\pi^-$ \\
   h22&\cite{h22}&1996& $\rho$           &$40\div80$ &$55$     &$<0.5$        &$0.035 $       &$<0.5$    &php&$0.0198$&$[93$-$94]$&$\pi^+\pi^-$ \\
      &          &    & $\rho$           &$164\div212$&$187$   &$<0.01$       &$0.001$        &          &php&$0.0238$&$[93$-$94]$& \\
   z29&\cite{z29}&1995& $\rho$           &$60\div80$ &$70$     &$<4$          &$10^{-4}$      &$<0.5$    &php&$0.55$&$[93]$      &$\pi^+\pi^-$ \\ \hline
    h2&\cite{h2} &2009& $\gamma$         &$30\div140$&$82$     &$6.5\div80$   &$~10$          &$<1$      &   &$306$  &$[04$-$07]$ &\\
    z5&\cite{z5} &2008& $\gamma$         &$40\div170$&$104$    &$>1.5$        &$3.2$          &$0.08\div0.53$&&$61.1$&$[99$-$00]$ &\\
    h4&\cite{h4} &2007& $\gamma$         &$30\div140$&$82$     &$6.5\div80$   &$~8$           &$<1$      &   &$145$  &$[05$-$06]$ &\\
    h7&\cite{h7} &2005& $\gamma$         &$30\div140$&$82$     &$2\div80$     &$8$            &$<1$      &   &$46.5$ &$[96$-$00]$ &\\
   z10&\cite{z10}&2003& $\gamma$         &$40\div140$&$89$     &$5\div100$    &$9.6$          &$ $       &   &$111.7$&$[96$-$00]$ &\\
   h13&\cite{h13}&2001& $\gamma$         &$30\div120$&$75$     &$2\div20$     &$4.5$          &$<1$      &   &  $8$  &$ $         &\\
   \hline
   \end{tabular}
    \begin{tabular}[r]{cccc|c|c|c }
     \hline  \multicolumn{7}{c}{Low Energy (photoproduction)}\\ \hline
      &cite      &year&                   &$W$                &$|t|$         &  \\ \hline
    f1&\cite{f1} &1979& $\omega$          &$10.3\div18.4$     &$ $           & $ $\\
    f2&\cite{f2} &1979& $\rho,$ $\phi$    &$7.6\div18.4$      &$  $          &$\rho\rightarrow\pi^+\pi^-,$ $\phi\rightarrow K^+K^-$\\
    f3&\cite{f3} &1993& $J/\psi$          &$15.8\div26.5$     &$0\div1.5$    &$J/\psi\rightarrow\mu^+\mu^-$\\
   sl1&\cite{sl1}&1971& $\rho$            &$2.15\div4.0$      &$0.06\div0.8$ &$\rho\rightarrow\pi^+\pi^-$      \\
    c1&\cite{c1} &1982& $\rho,$ $\omega$  &$6.2\div9.2$       &$0.06\div1$   &$\rho\rightarrow\pi^+\pi^-,$ $\omega\rightarrow \pi^+\pi^0\pi^-$\\
    c2&\cite{c2} &1983& $\phi,$ $\omega$  &$6.2\div11.5$      &$0\div1$      &$\phi,\omega\rightarrow\pi^+\pi^-\pi^0$\\
   z2d&\cite{z2d}&1997&$\omega,$ $\phi,$ $\rho$&$9.2\div17.2$ &$ $           &\\
   \hline
    \end{tabular}
\end{table*}

\end{landscape}

\newpage
\begin{figure*}[!ht]
  \centering
   \includegraphics[trim = 0mm 0mm 12mm 0mm,clip, scale=0.71]{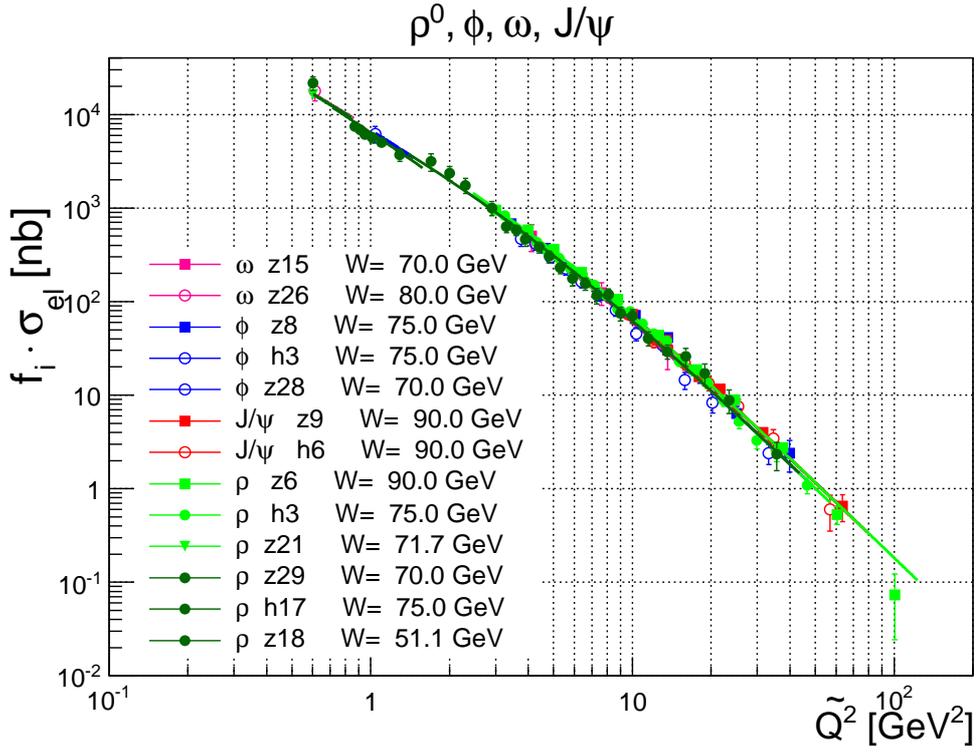}
  \caption{ \label{fig:cs(Q2)} Fit of Eq.~(\ref{eq:cs(h+s)}) to the data on the normalized elastic cross section $f_i\cdot\sigma_{el}(Q^2)$ for $\rho^0$, $\phi$, $\omega$ and $J/\psi$, for different values of $W$.
  Here $f_i$ is the  normalization factor (see Eq.~\eqref{eq:f_i}). }
\end{figure*}

\newpage
\begin{figure}[!ht] \centering
 \includegraphics[trim = 0mm 3mm 5mm 0mm,clip, scale=0.71]{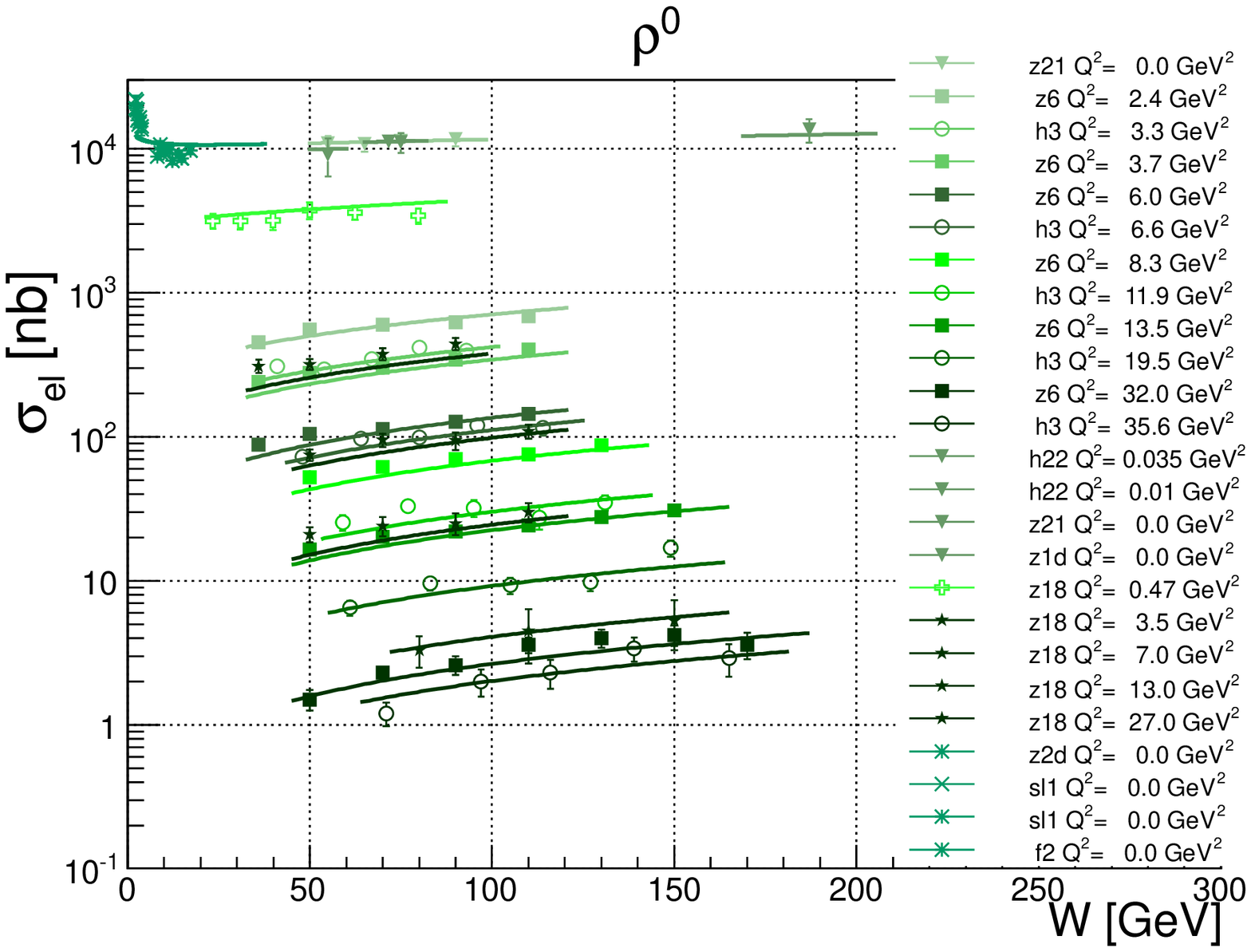} 
 \caption{ \label{fig:cs_rho1(W)} Fit of Eq.~(\ref{eq:cs(h+s)}) to the data on the elastic cross section $\sigma_{el}(W)$ for $\rho^0$, for different values of $Q^2$.}
\end{figure}

\begin{figure}[!ht] \centering
 \includegraphics[trim = 0mm 3mm 12mm 2mm,clip, scale=0.71]{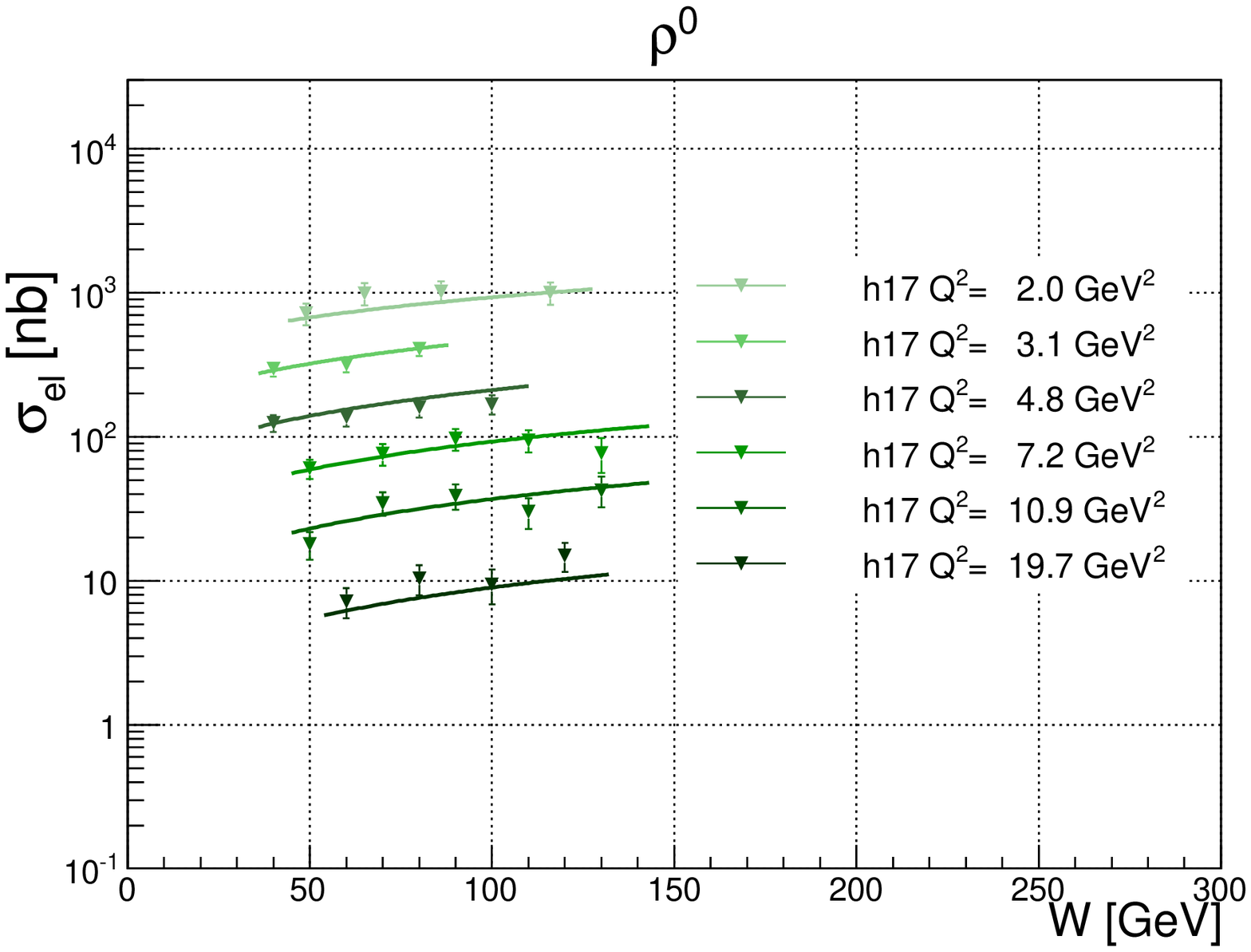} 
 \caption{ \label{fig:cs_rho2(W)} Fit of Eq.~(\ref{eq:cs(h+s)}) to the data on the elastic cross section $\sigma_{el}(W)$  for $\rho^0$, for different values of $Q^2$.}
\end{figure}

\newpage
\begin{figure}[!hb] \centering
 \includegraphics[trim = 0mm 3mm 10mm 2mm,clip, scale=0.71]{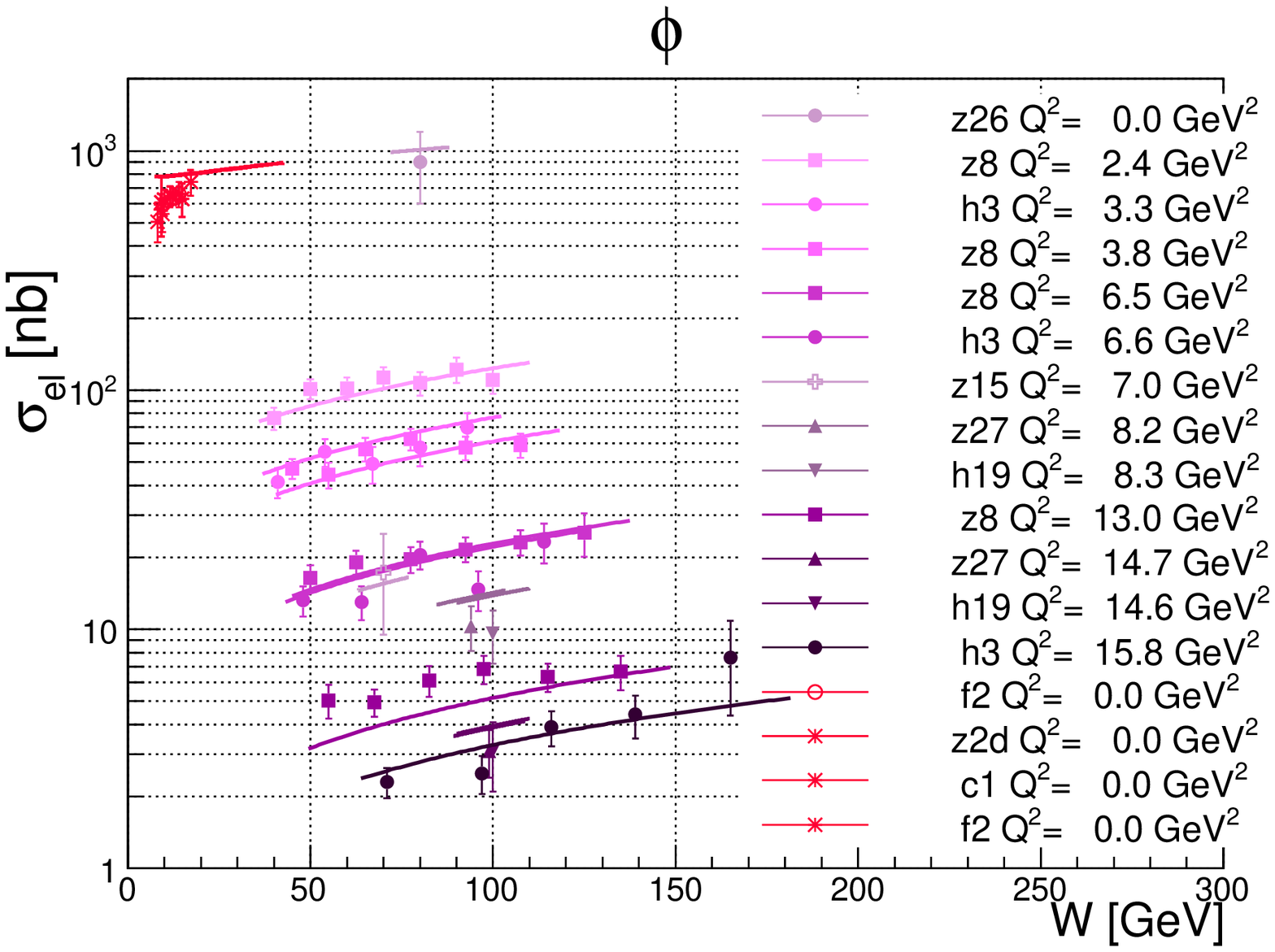}
 \caption{ \label{fig:cs_phi(W)} Fit of Eq.~(\ref{eq:cs(h+s)}) to the data on the elastic cross section $\sigma_{el}(W)$  for $\phi$, for different values of $Q^2$.}
\end{figure}

\begin{figure}[!hb] \centering
 \includegraphics[trim = 0mm 3mm 12mm 2mm,clip, scale=0.71]{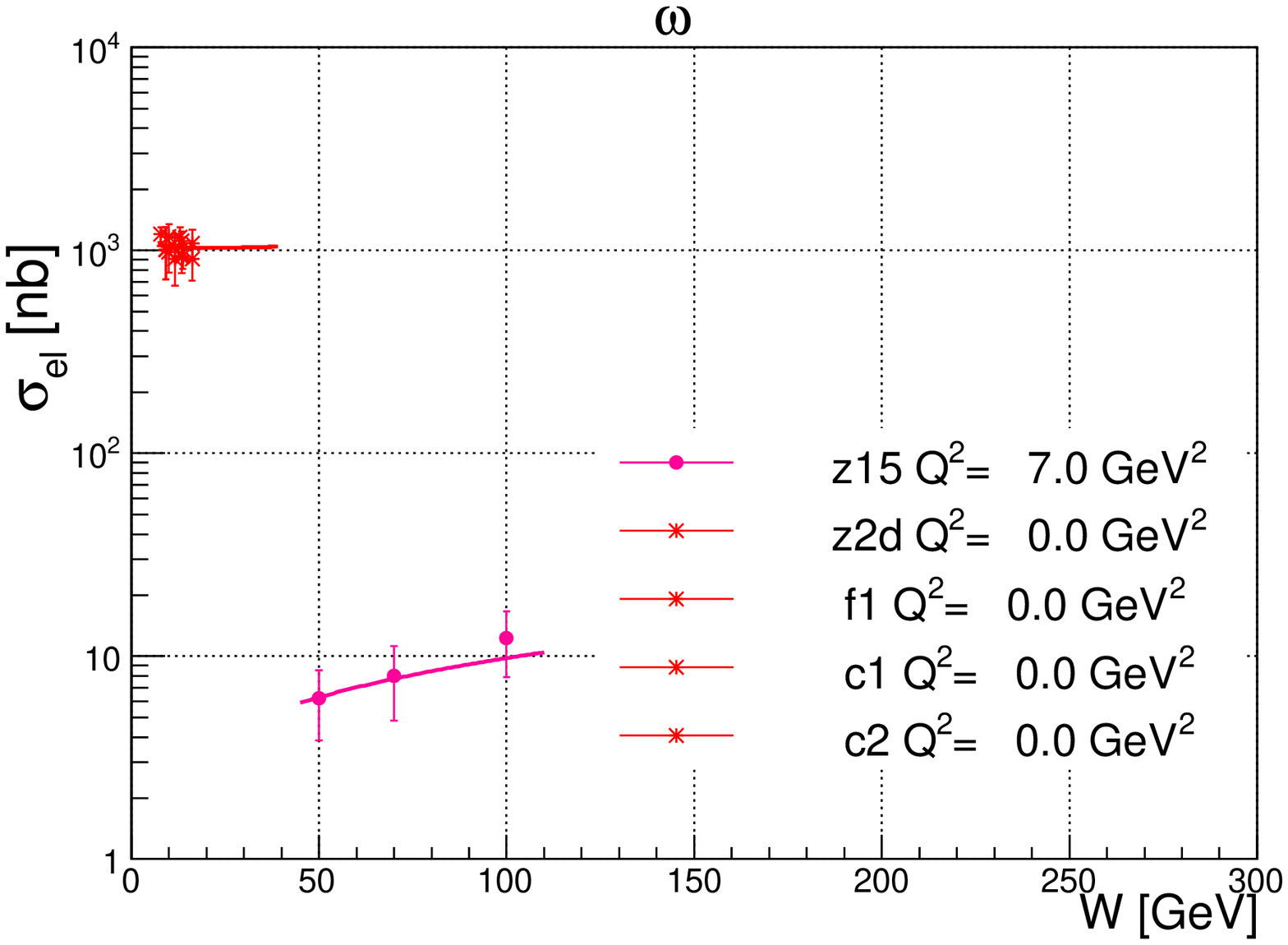}
 \caption{ \label{fig:cs_omega(W)} Fit of Eq.~(\ref{eq:cs(h+s)}) to the data on the elastic cross section $\sigma_{el}(W)$  for $\omega$, for different values of $Q^2$.}
\end{figure}
\newpage

\begin{figure}[!ht]  \centering
  \includegraphics[trim = 0mm 3mm 10mm 2mm,clip, scale=0.71]{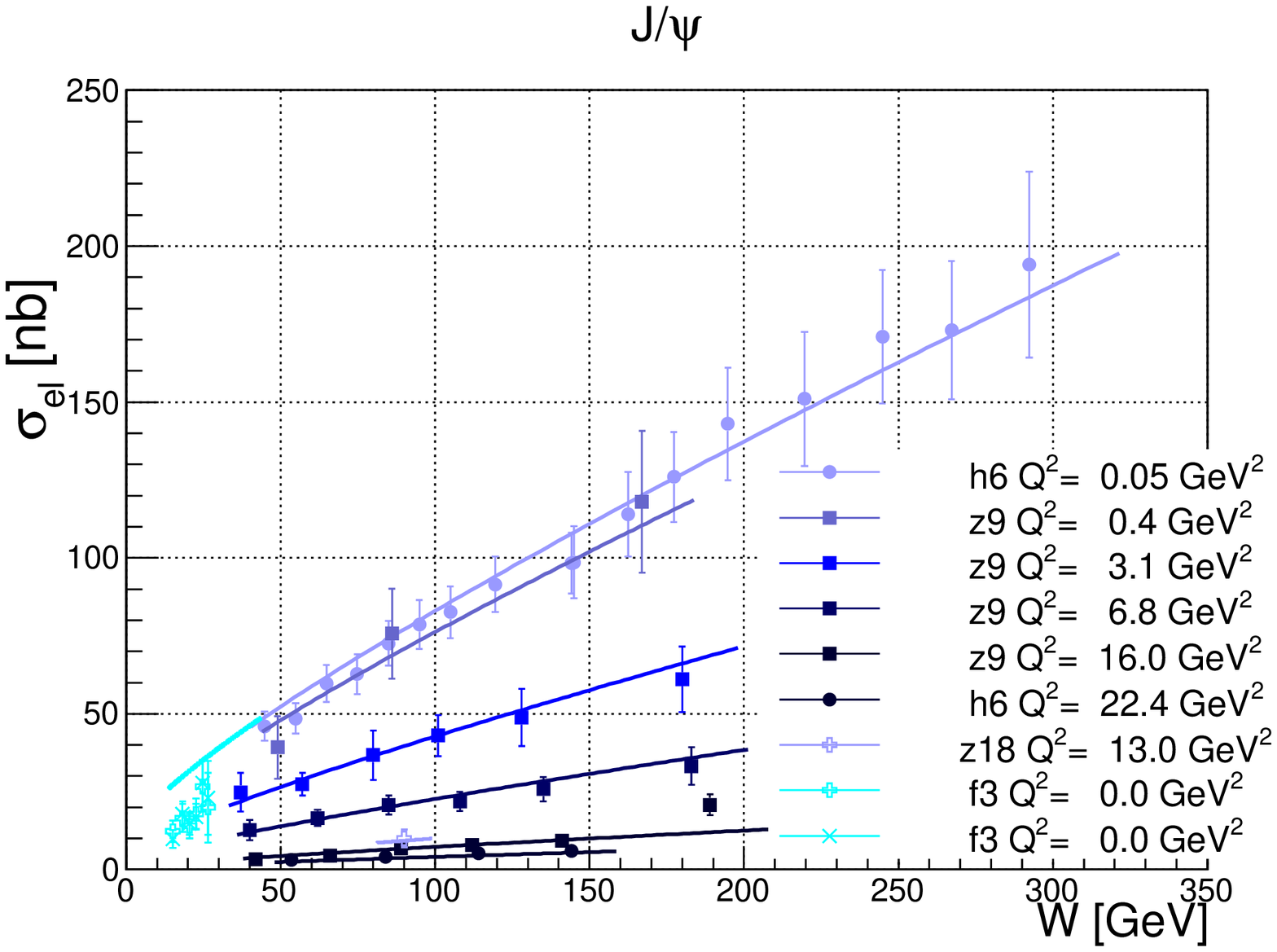}
  \caption{ \label{fig:cs_Jpsi(W)} Fit of Eq.~(\ref{eq:cs(h+s)}) to the data on the elastic cross section $\sigma_{el}(W)$  for $J/\psi$, for different values of $Q^2$.}
\end{figure}

\begin{figure}[!ht]  \centering
  \includegraphics[trim = 0mm 3mm 12mm 2mm,clip, scale=0.71]{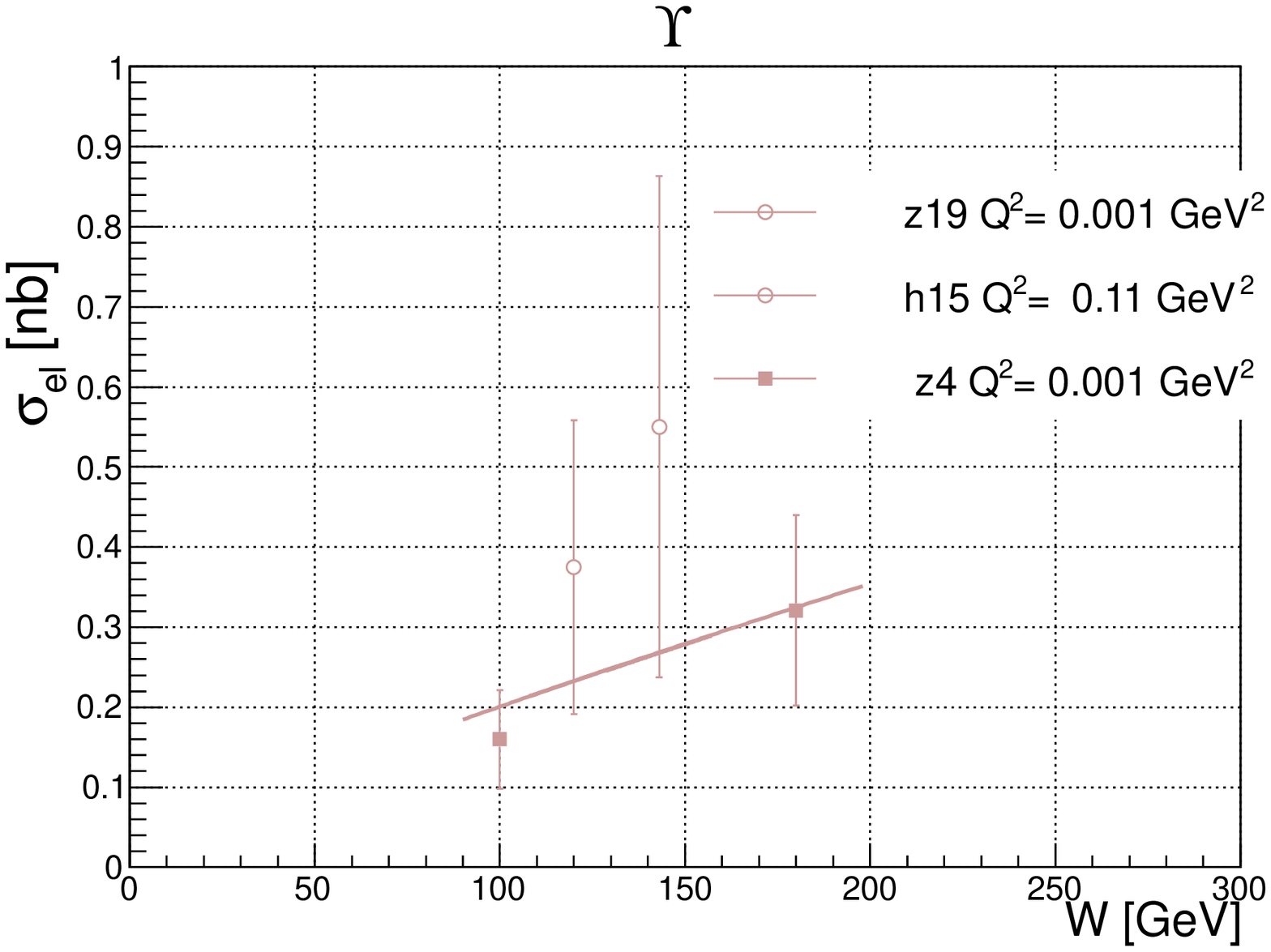}
  \caption{ \label{fig:cs_Ups(W)} Fit of Eq.~(\ref{eq:cs(h+s)}) to the data on the elastic cross section $\sigma_{el}(W)$  for $\varUpsilon$, for different values of $Q^2$.}
\end{figure}
\newpage

\begin{figure}[!ht]\centering
 \includegraphics[trim = 0mm 0mm 0mm 0mm,clip, scale=0.71]{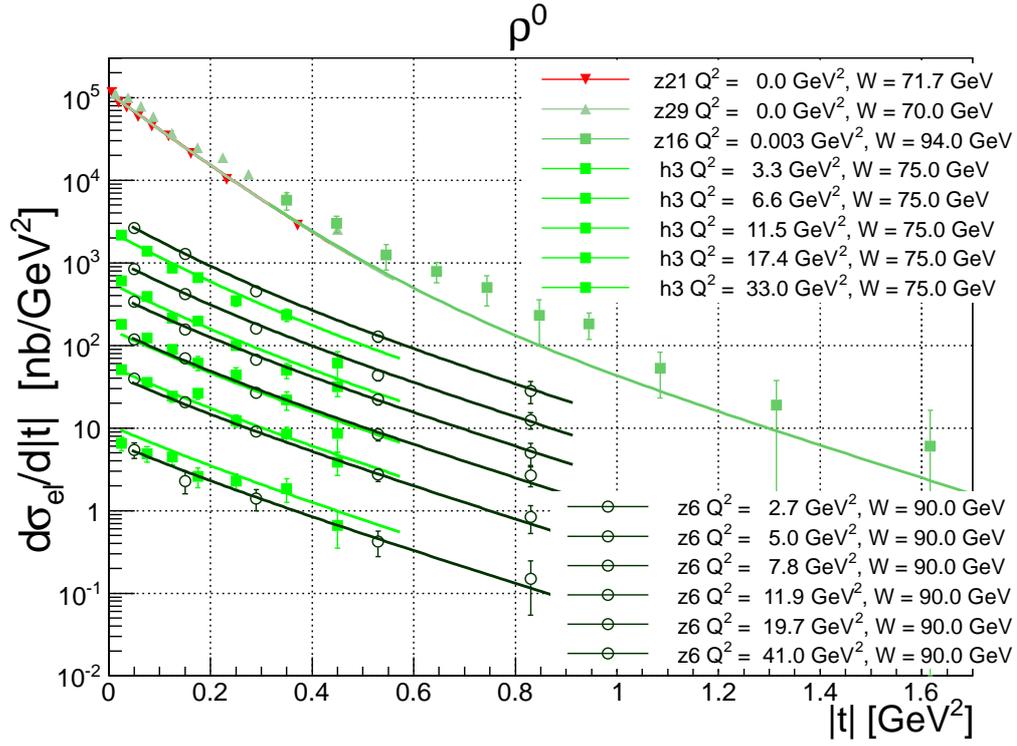}
 \caption{ \label{fig:dcsdt_rho} Fit of Eq.~(\ref{eq:dcsdt(h+s)}) to the data on the differential elastic cross section $d\sigma_{el}/dt$ for $\rho^0$, for different values of $Q^2$ and $W$.}
\end{figure}

\begin{figure}[!ht]\centering
 \includegraphics[trim = 0mm 0mm 5mm 0mm,clip, scale=0.71]{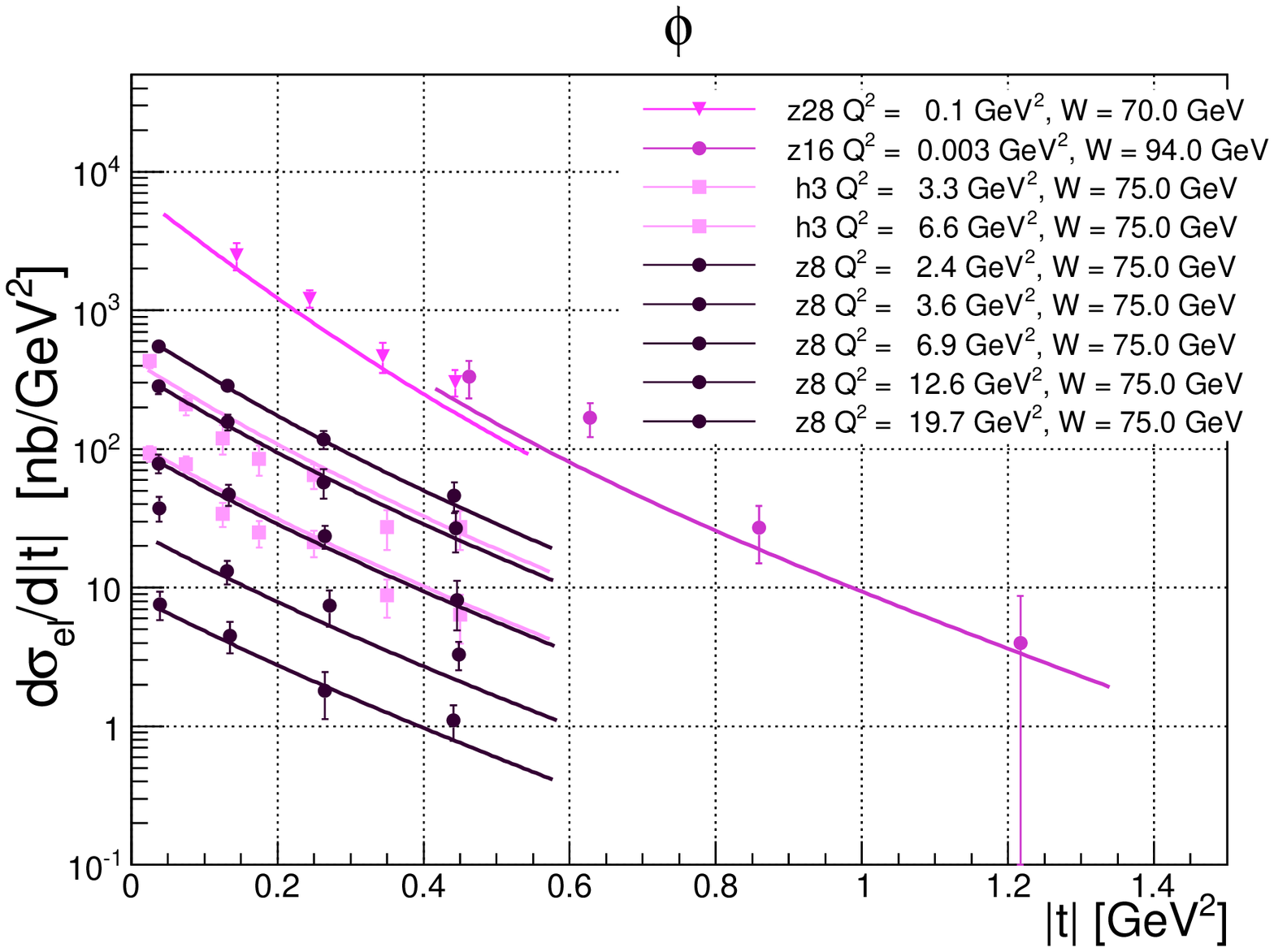}
 \caption{ \label{fig:dcsdt_phi} Fit of Eq.~(\ref{eq:dcsdt(h+s)}) to the data on the elastic differential cross section $d\sigma_{el}/dt$ for $\phi$, for different values of $Q^2$ and $W$.}
\end{figure}
\newpage

\begin{figure}[!ht]
\centering
\centering
\includegraphics[trim = 0mm 0mm 5mm 0mm,clip, scale=0.71]{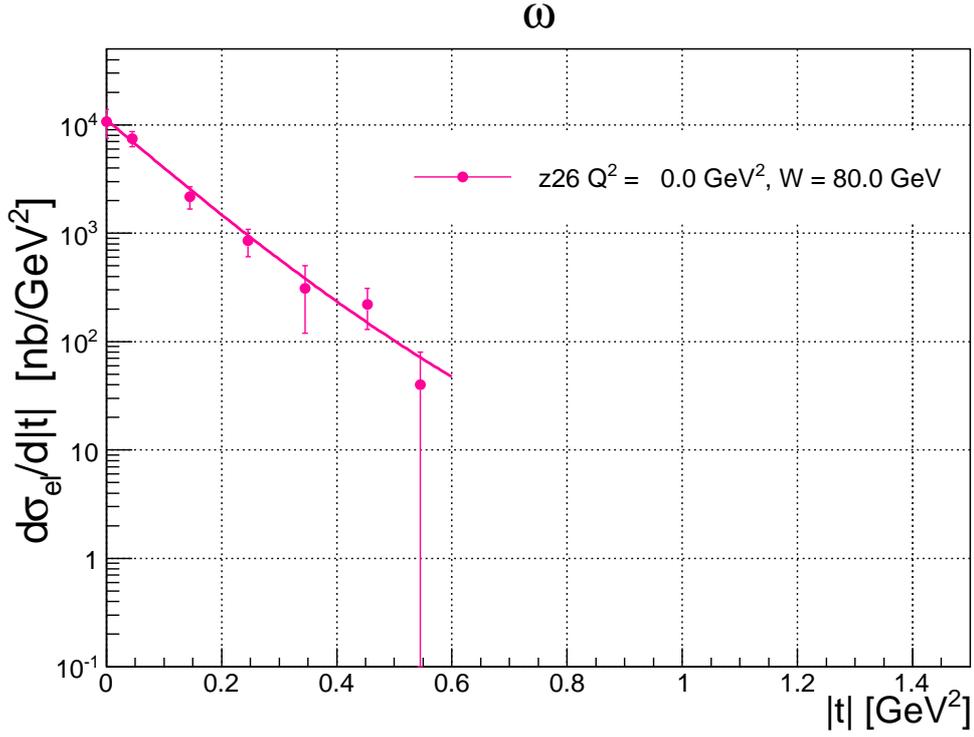}
\caption{ \label{fig:dcsdt_omega} Fit of Eq.~(\ref{eq:dcsdt(h+s)}) to the data on the elastic differential cross section $d\sigma_{el}/dt$ for $\omega$, for $Q^2 = 0.00$ GeV$^2$ and $W = 80$ GeV.}
\end{figure}

\begin{figure}[!ht]
\centering
 \centering
 \includegraphics[trim = 0mm 0mm 5mm 0mm,clip, scale=0.71]{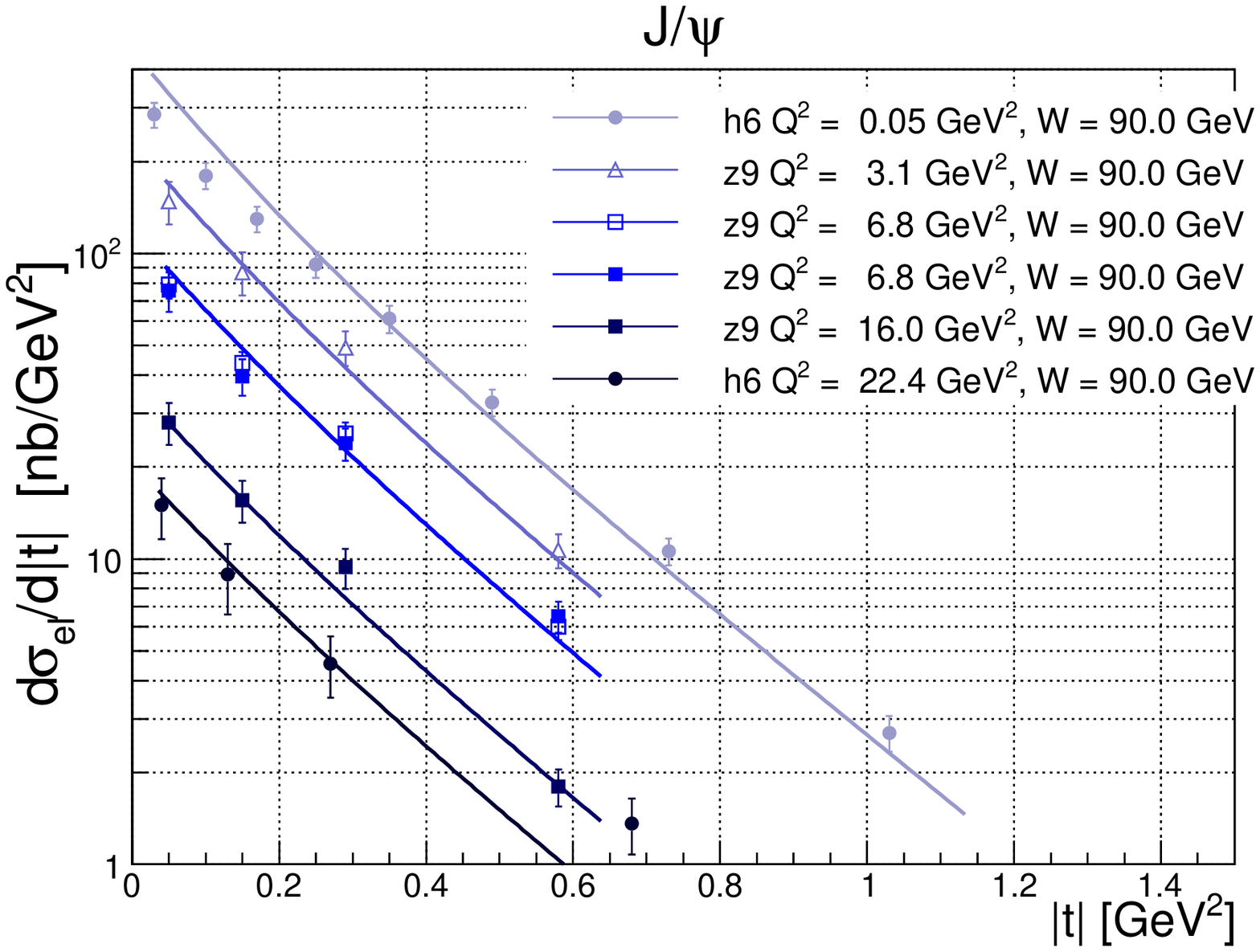}
 \caption{ \label{fig:dcsdt_Jpsi} Fit of Eq.~(\ref{eq:dcsdt(h+s)}) to the data on the elastic differential cross section $d\sigma_{el}/dt$ for $J/\psi$, for different values of $Q^2$ and $W = 90$ GeV.}
\end{figure}

\newpage
\begin{figure}[!ht]
 \centering
 $J/\psi$\\
 \includegraphics[trim = 0mm 0mm  0mm 0mm,clip, scale=0.75]{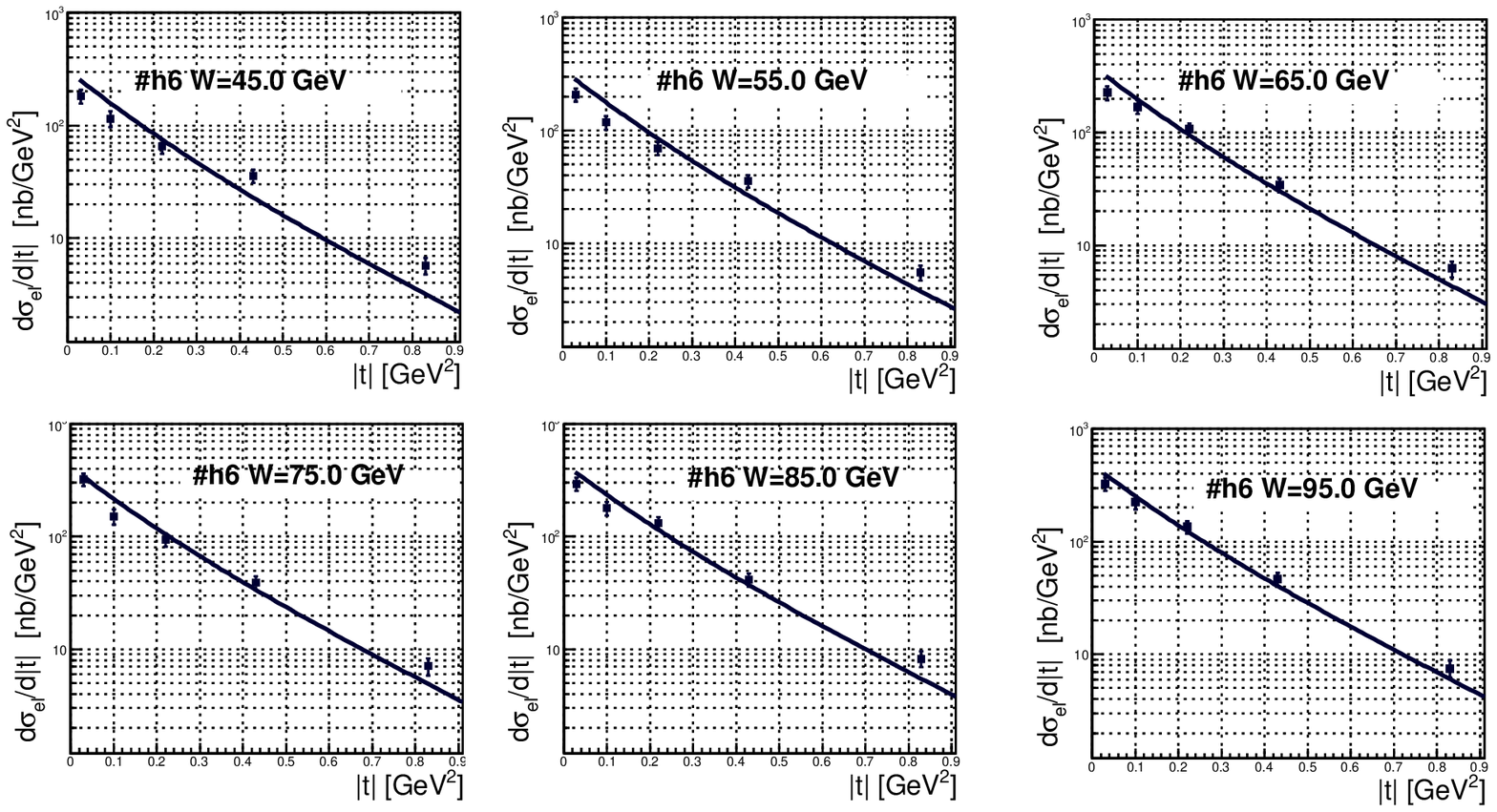}\\
 \includegraphics[trim = 0mm 0mm  0mm 0mm,clip, scale=0.75]{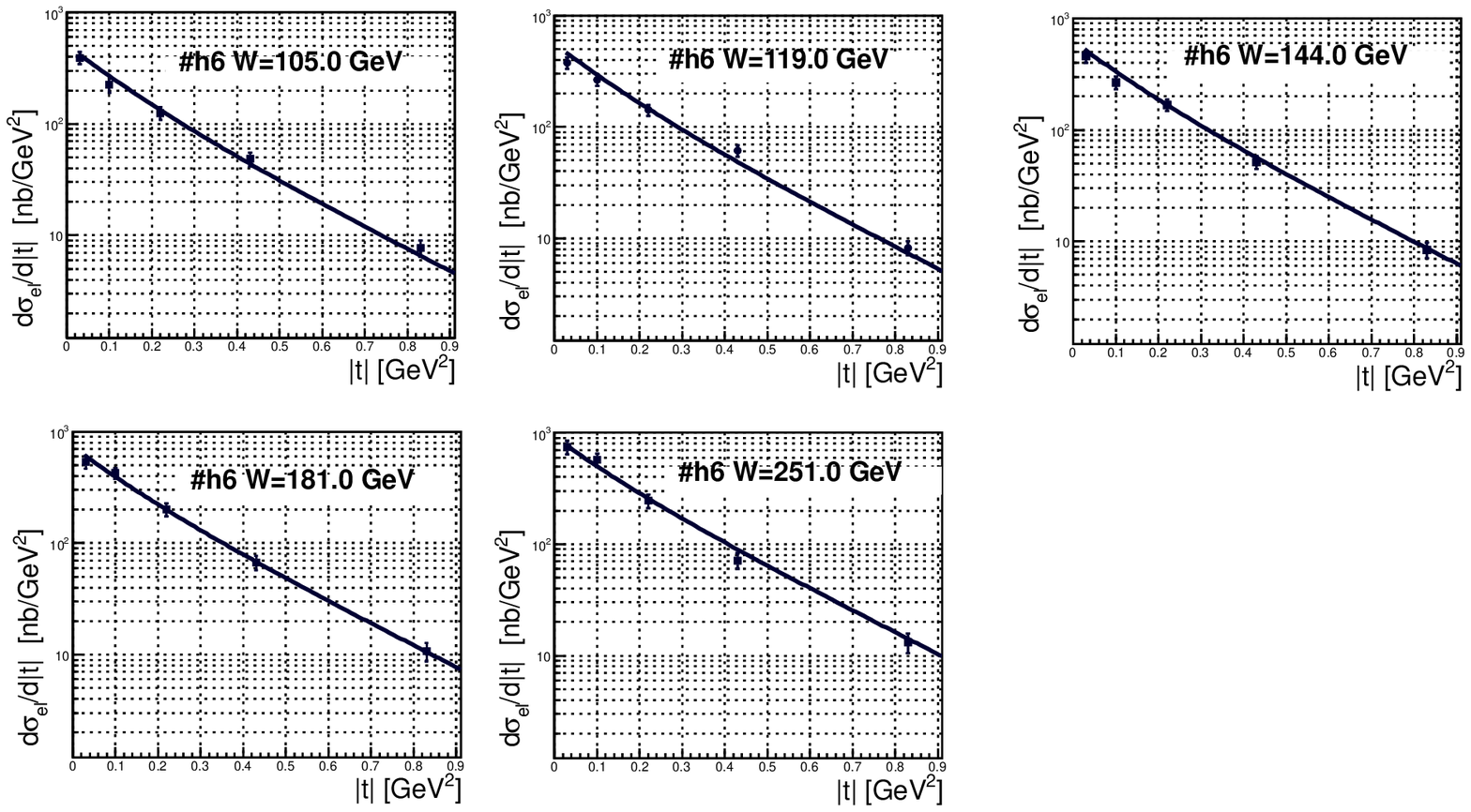}
 \caption{ \label{fig:dcsdt_Jpsi.php1} Fit of Eq.~(\ref{eq:dcsdt(h+s)}) to the data on the elastic differential cross section $d\sigma_{el}/dt$ for $J/\psi$ at photoproduction ($Q^2=0.05$ GeV$^2$), for different values of $W$.}
\end{figure}

 \newpage
\begin{figure}[!ht]  \centering
  \includegraphics[trim = 2mm 0mm 16mm 12mm,clip, scale=0.43]{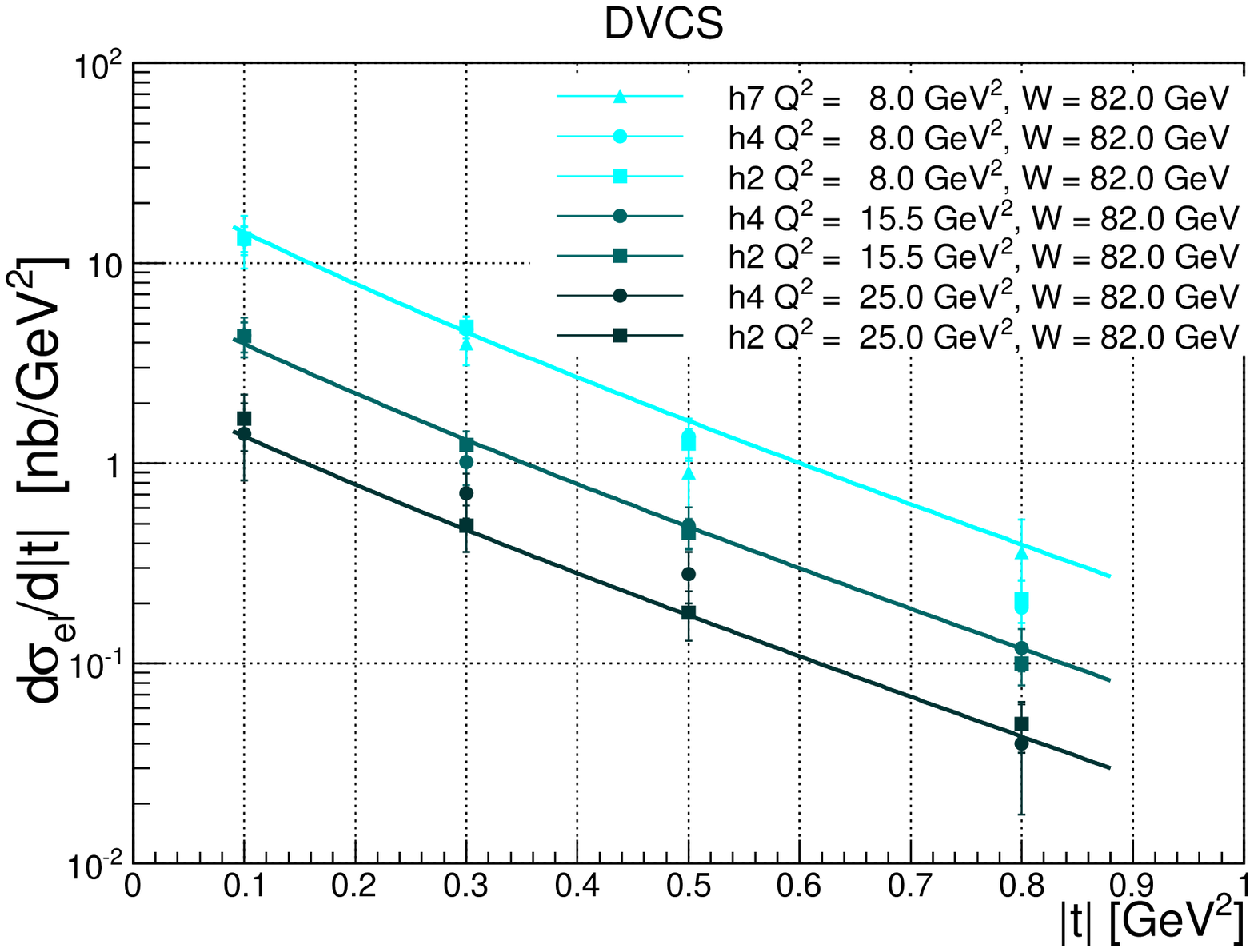}
  \includegraphics[trim = 2mm 0mm 16mm 12mm,clip, scale=0.43]{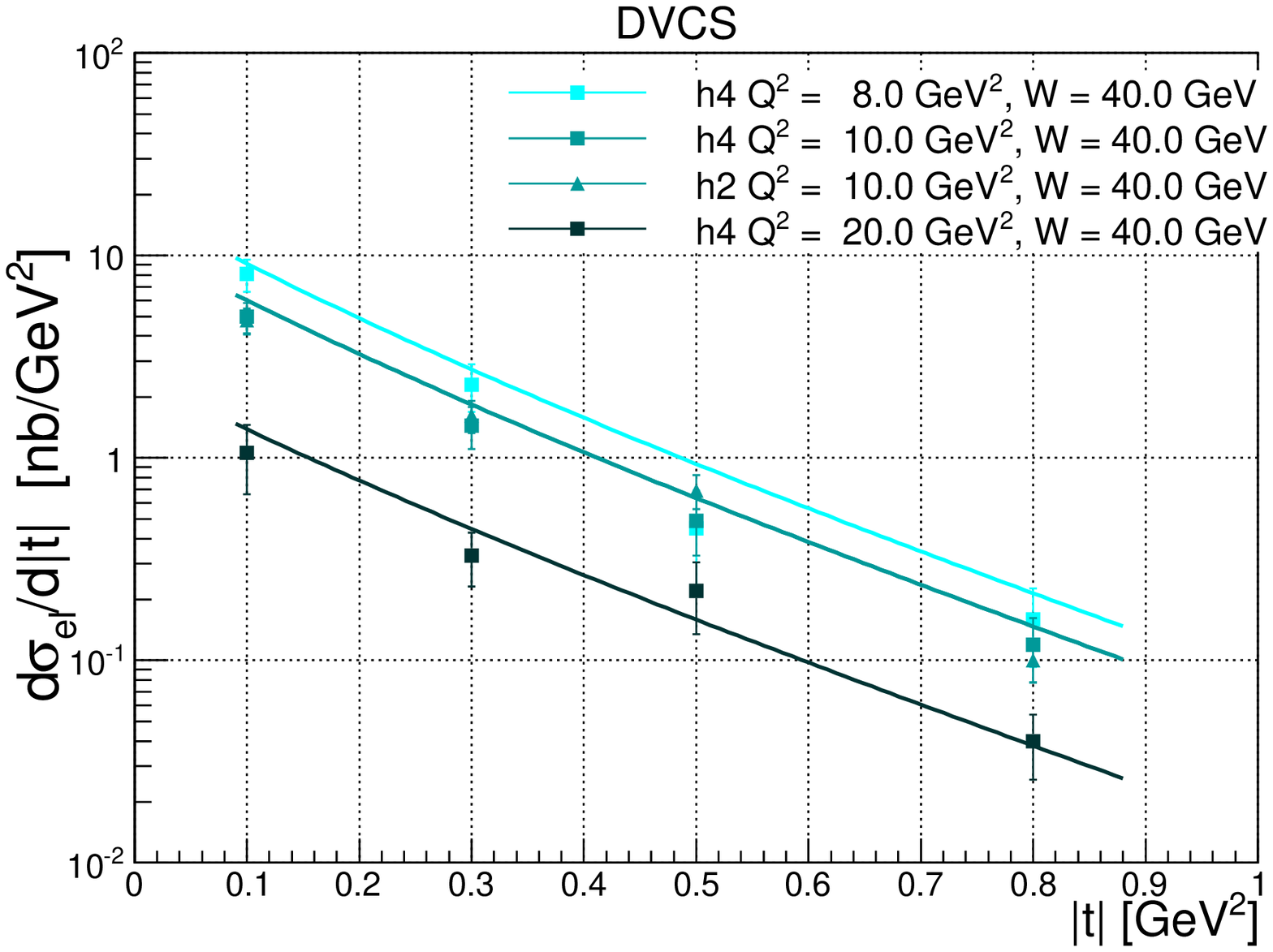}\\
  \includegraphics[trim = 2mm 0mm 16mm 12mm,clip, scale=0.43]{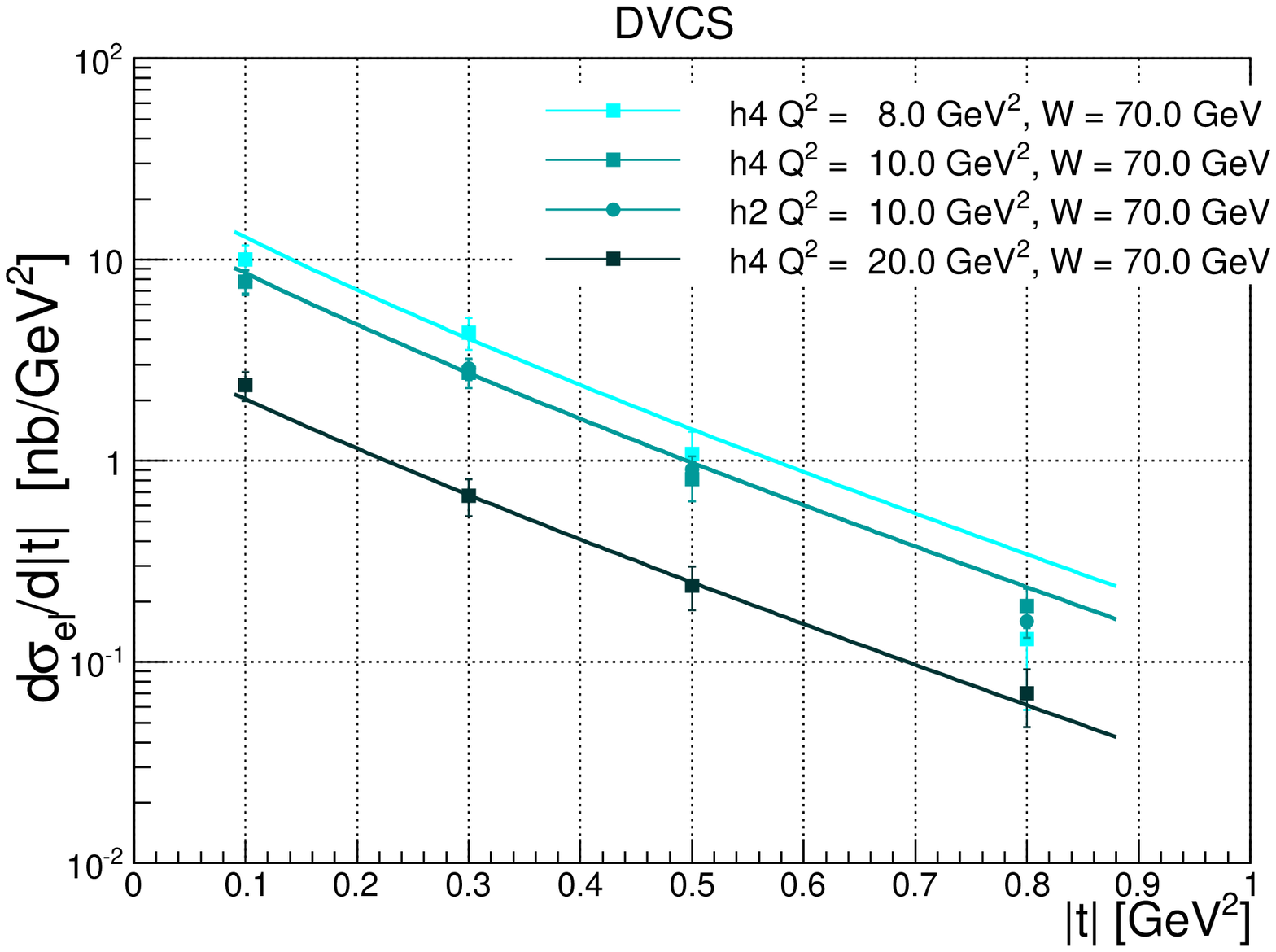}
  \includegraphics[trim = 2mm 0mm 16mm 12mm,clip, scale=0.43]{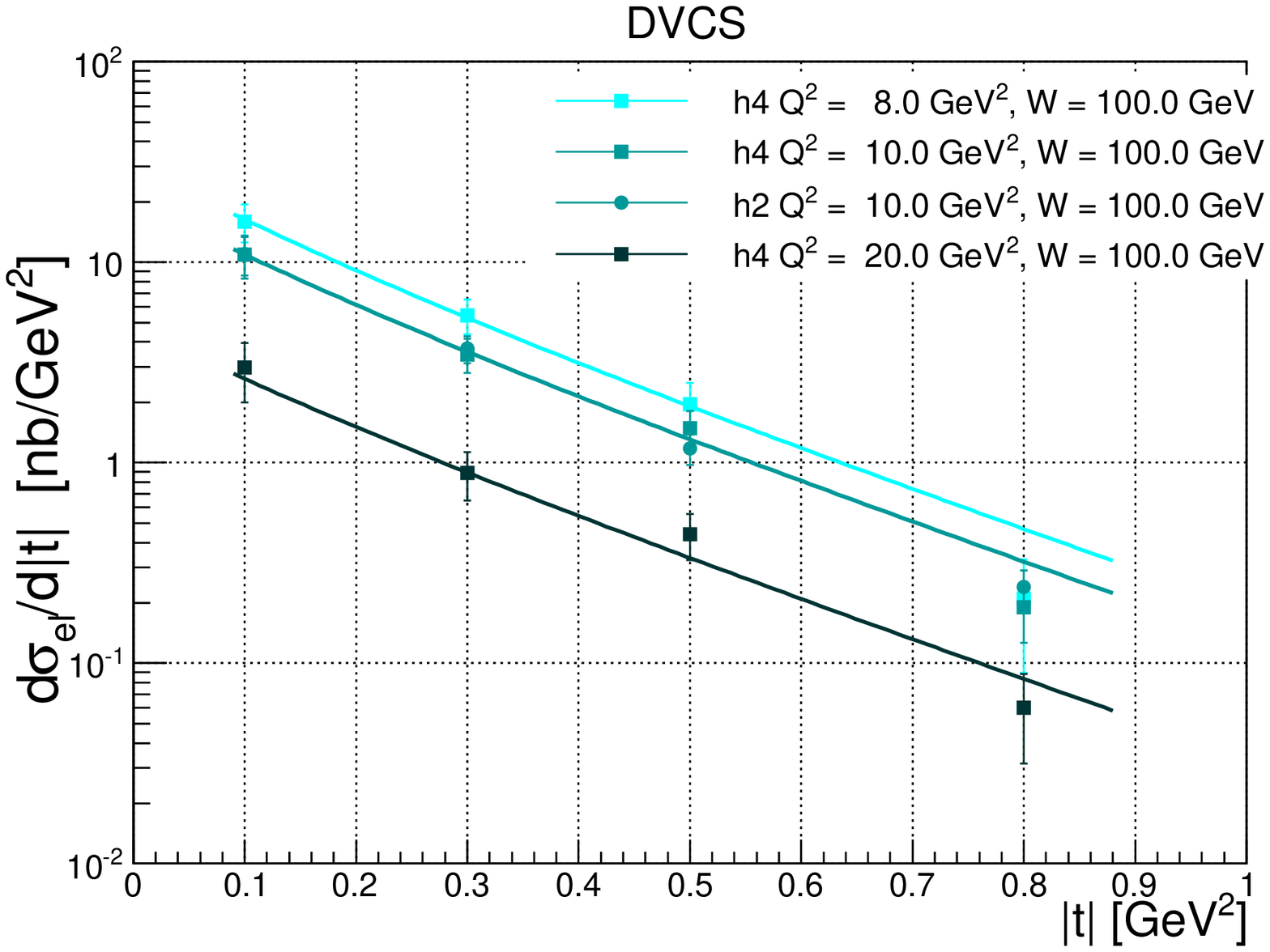}
 \vspace{-0.4cm}\caption{ \label{fig:dcsdt.DVCS} Fit of Eq.~(\ref{eq:dcsdt(h+s)}) to the the data on the elastic differential cross section $d\sigma_{el}/dt$ for DVCS.}
 \end{figure}

\begin{figure}[!ht]  \centering
  \includegraphics[trim = 4mm 0mm 16mm 12mm,clip, scale=0.43]{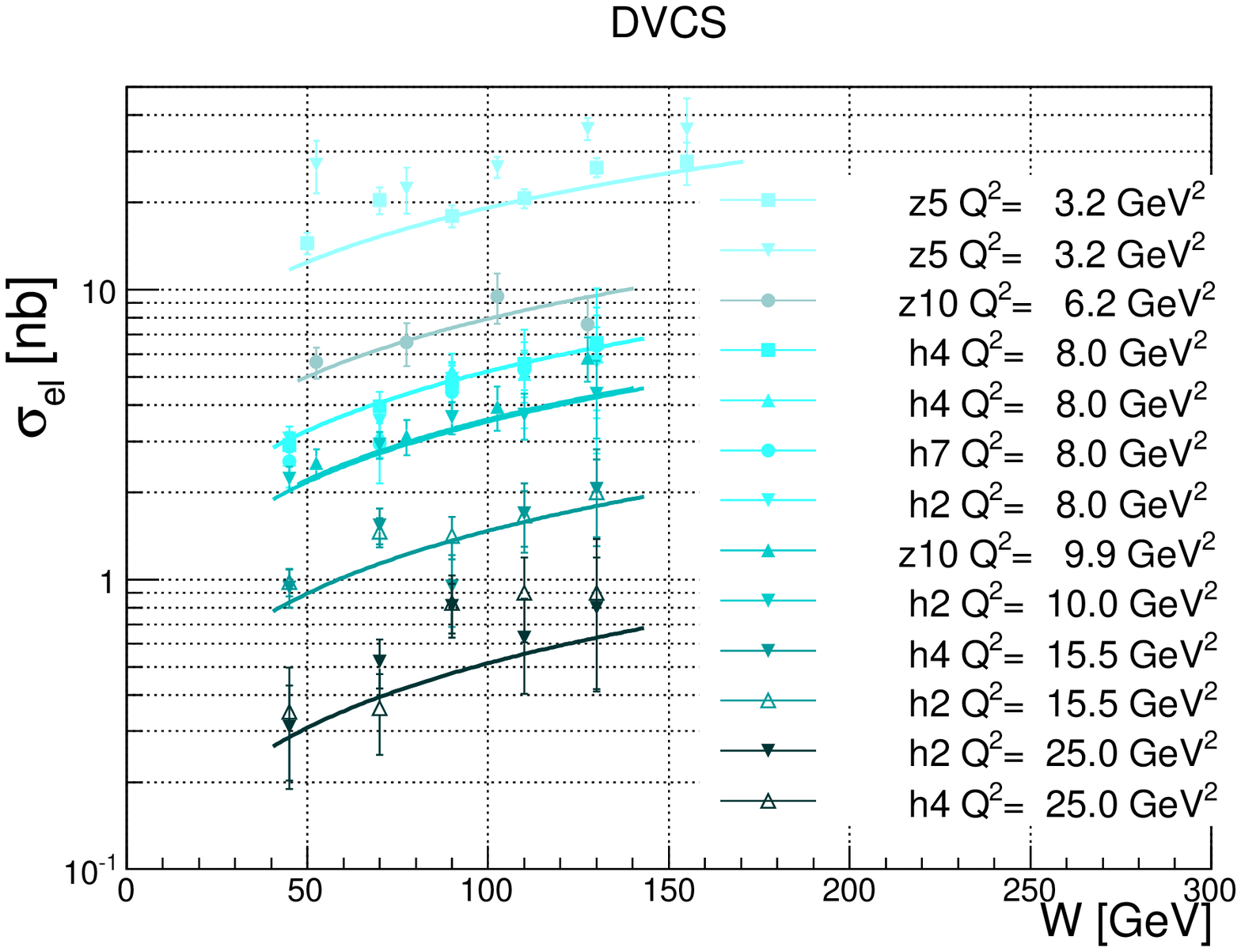}
  \includegraphics[trim = 4mm 0mm 16mm 12mm,clip, scale=0.43]{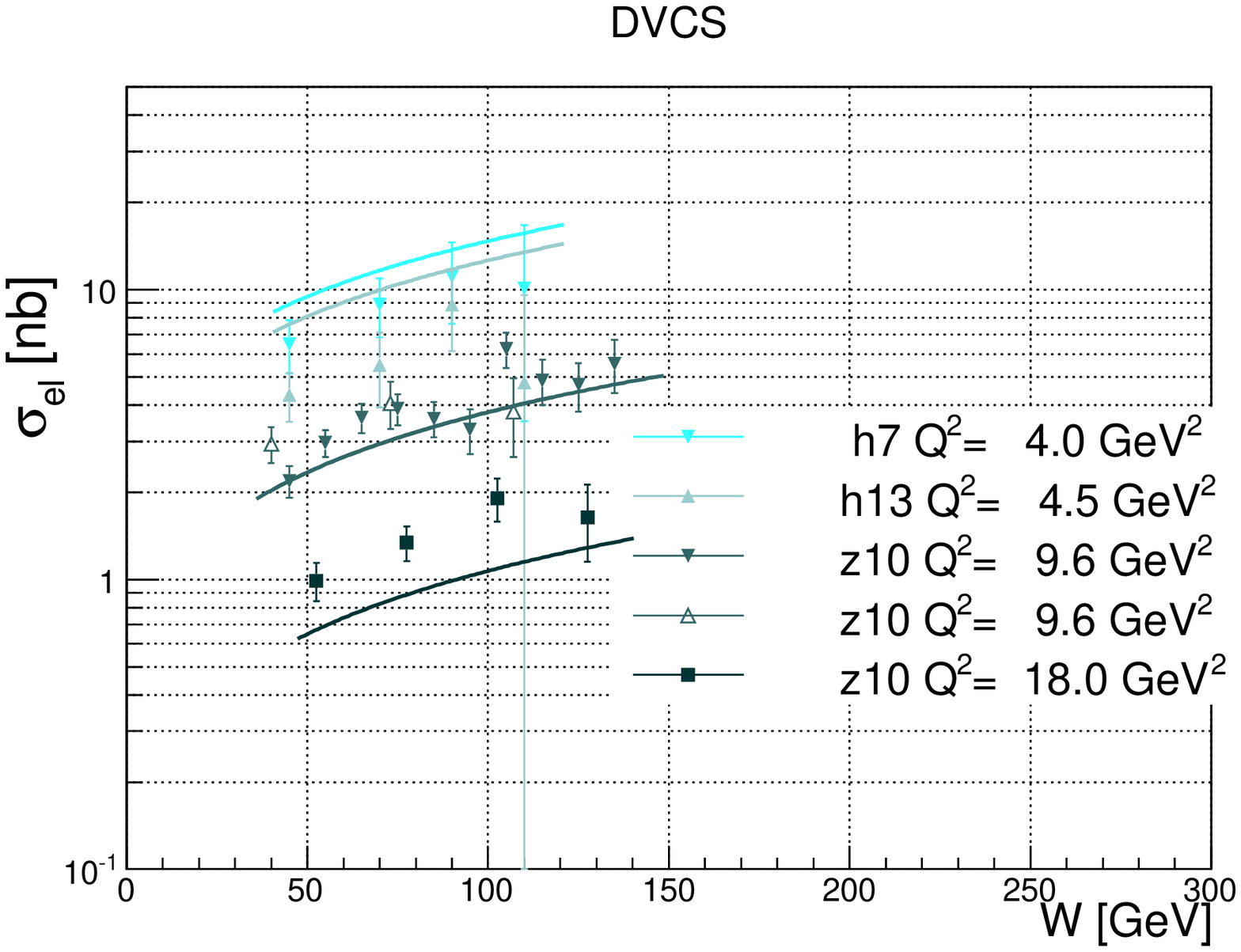}
  \caption{ \label{fig:csW.DVCS1} Fit of Eq.~(\ref{eq:cs(h+s)}) to the data on the elastic cross section $\sigma_{el}(W)$ for DVCS.}
\end{figure}

 \begin{figure}[!ht]  \centering
  \includegraphics[trim = 4mm 0mm 16mm 12mm,clip, scale=0.43]{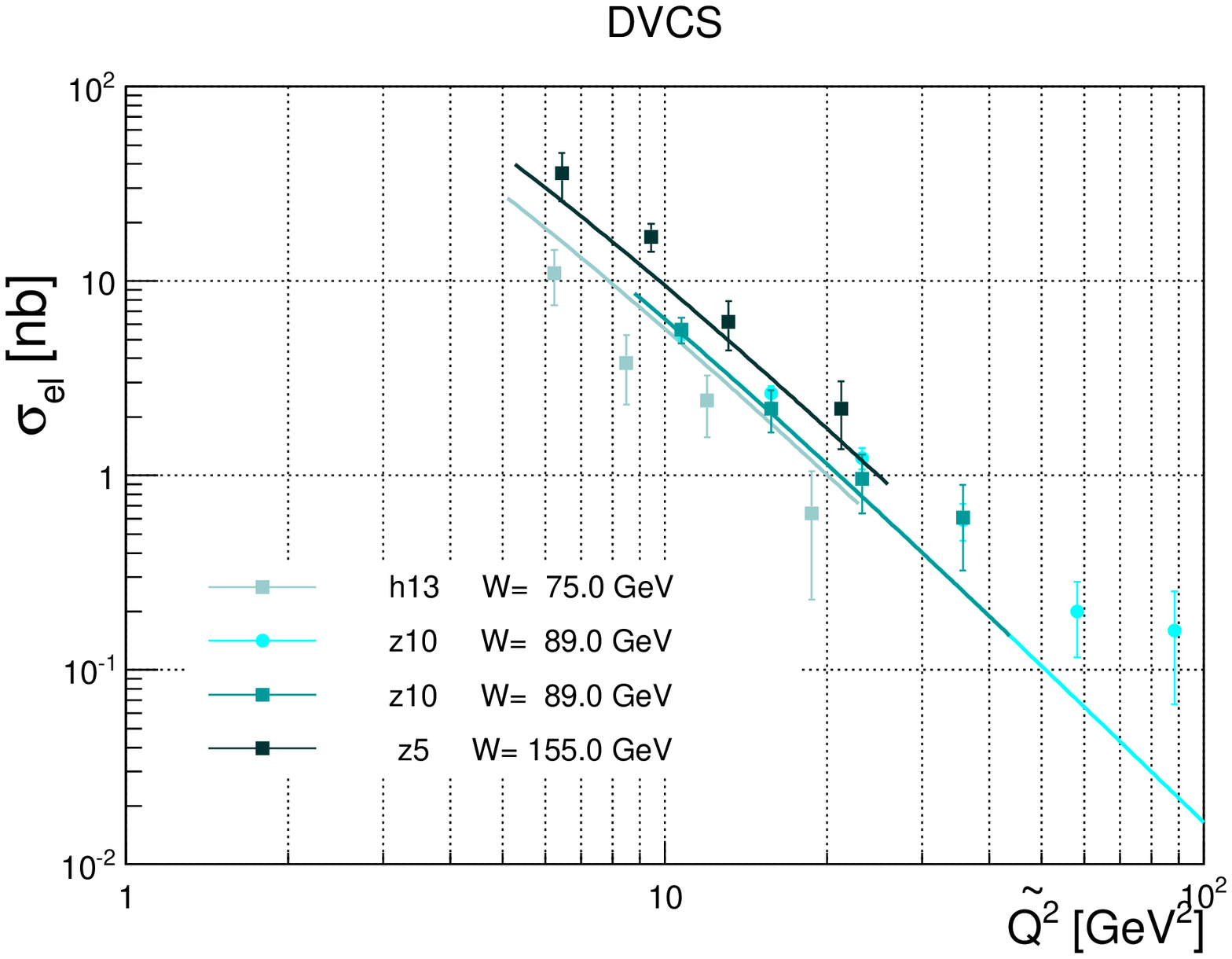}
  \includegraphics[trim = 4mm 0mm 16mm 12mm,clip, scale=0.43]{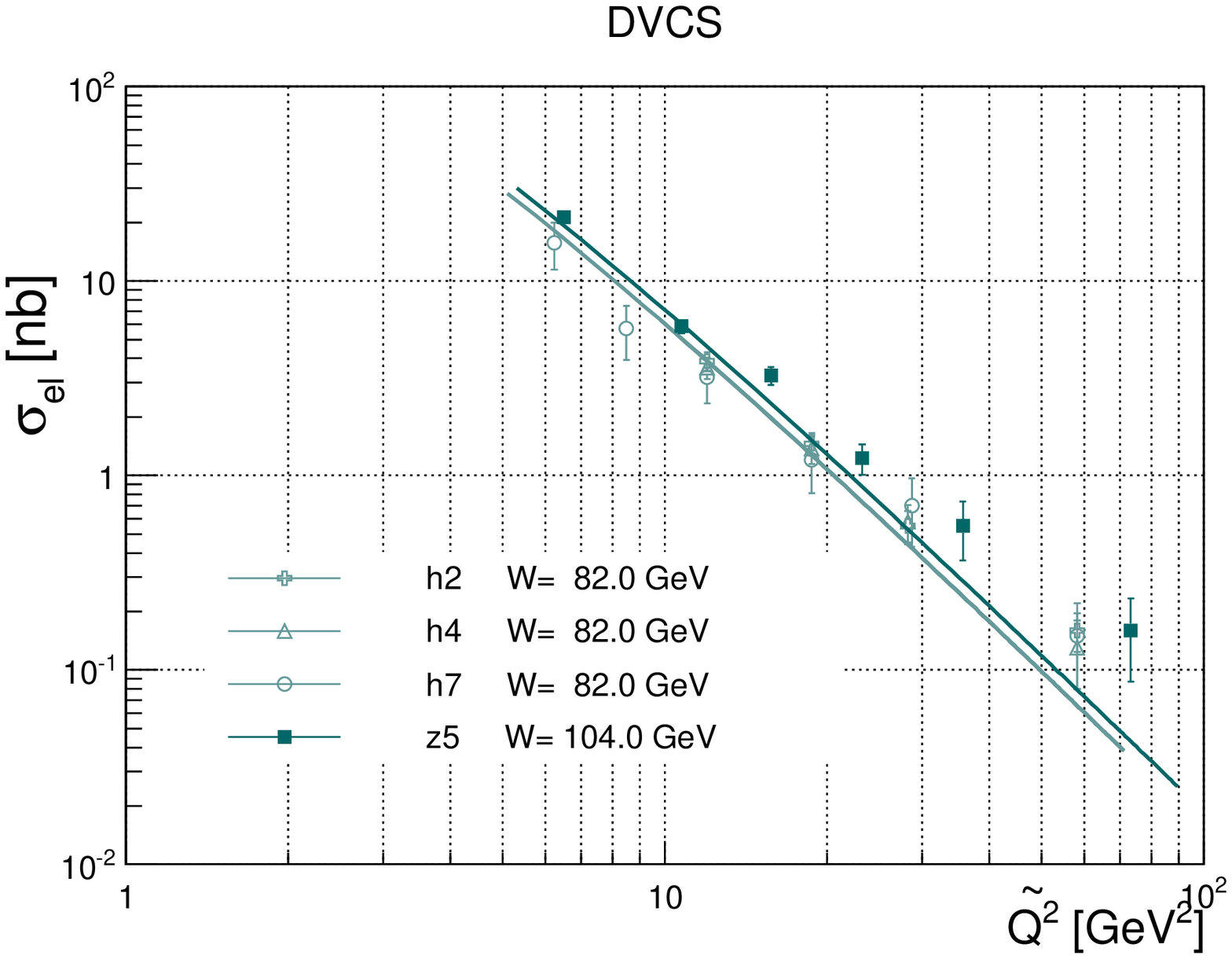}
  \caption{ \label{fig:csQ2.DVCS} Fit of Eq.~(\ref{eq:cs(h+s)}) to the data on the elastic cross section $\sigma_{el}(\widetilde{Q^2})$ for DVCS.}
\end{figure}

As seen from Figs.~\ref{fig:cs(Q2)} - \ref{fig:csQ2.DVCS} and Tables~\ref{tab:fit1(s+h)} - \ref{tab:fixed_trajec}, the overall fit to the large number of the diffractive data by the two-component Pomeron amplitude (Eq.~\ref{eq:Amplitude_hs}) is impressive,
apart from some peculiar 
cases. In particular, two points need to be better understood. They are

\begin{itemize}
\item
 compatibility of VMP and DVCS (the problem of the vanishing ``photon mass"),
\item
the problem of the description of $J/\psi$ production in the region of low $t$ and $Q^2$.
\end{itemize}

The number of the fitted parameters of the two-component Pomeron model (Eq.~\eqref{eq:f_i})  is 12 (Table~\ref{tab:fit1(s+h)}), with five additional normalization factors%
\footnote{Here we are taking into account $f_{DVCS}$, but excluding $f_{J/\psi}=1$ since it is the base line of our normalization.}
 for six vector particle productions ($\rho^0$, $\phi$, $\omega$, $J/\psi$, $\varUpsilon$ and $\gamma$). If we fix the Pomeron trajectories $\alpha_s(t)$ and $\alpha_h(t)$ (see Table~\ref{tab:fixed_trajec}), the number of free parameters reduces to $8$.
In the case of a single-component Pomeron (see Sec.~\ref{sec:Single}) the number of parameters for five different types reactions was much larger: $7\times 5=35$ (see Table~\ref{tab:one_term}).

\begin{figure}[!ht] \centering
 \includegraphics[trim = 0mm 0mm 10mm 0mm,clip, scale=0.40]{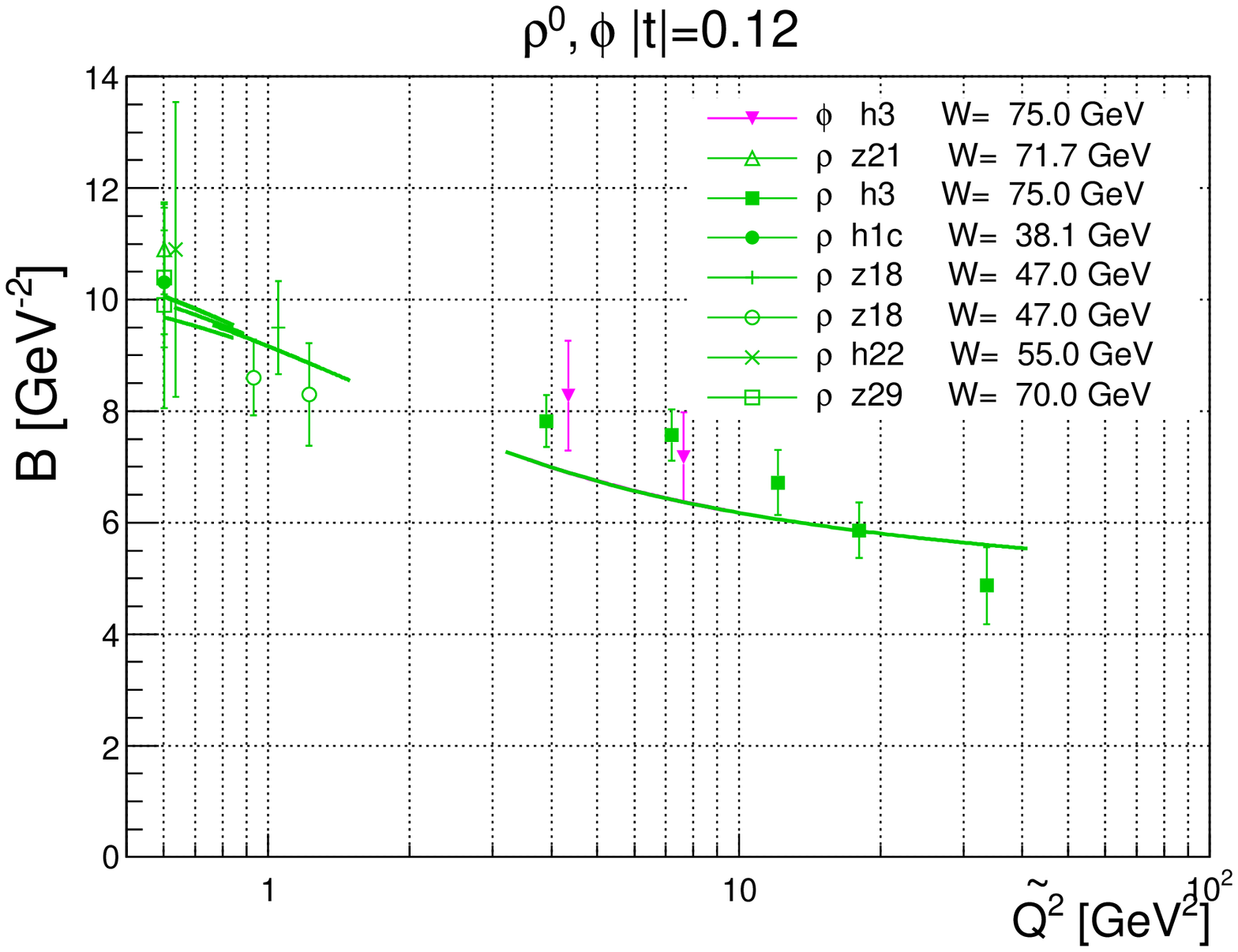}
 \includegraphics[trim = 0mm 0mm 10mm 0mm,clip, scale=0.40]{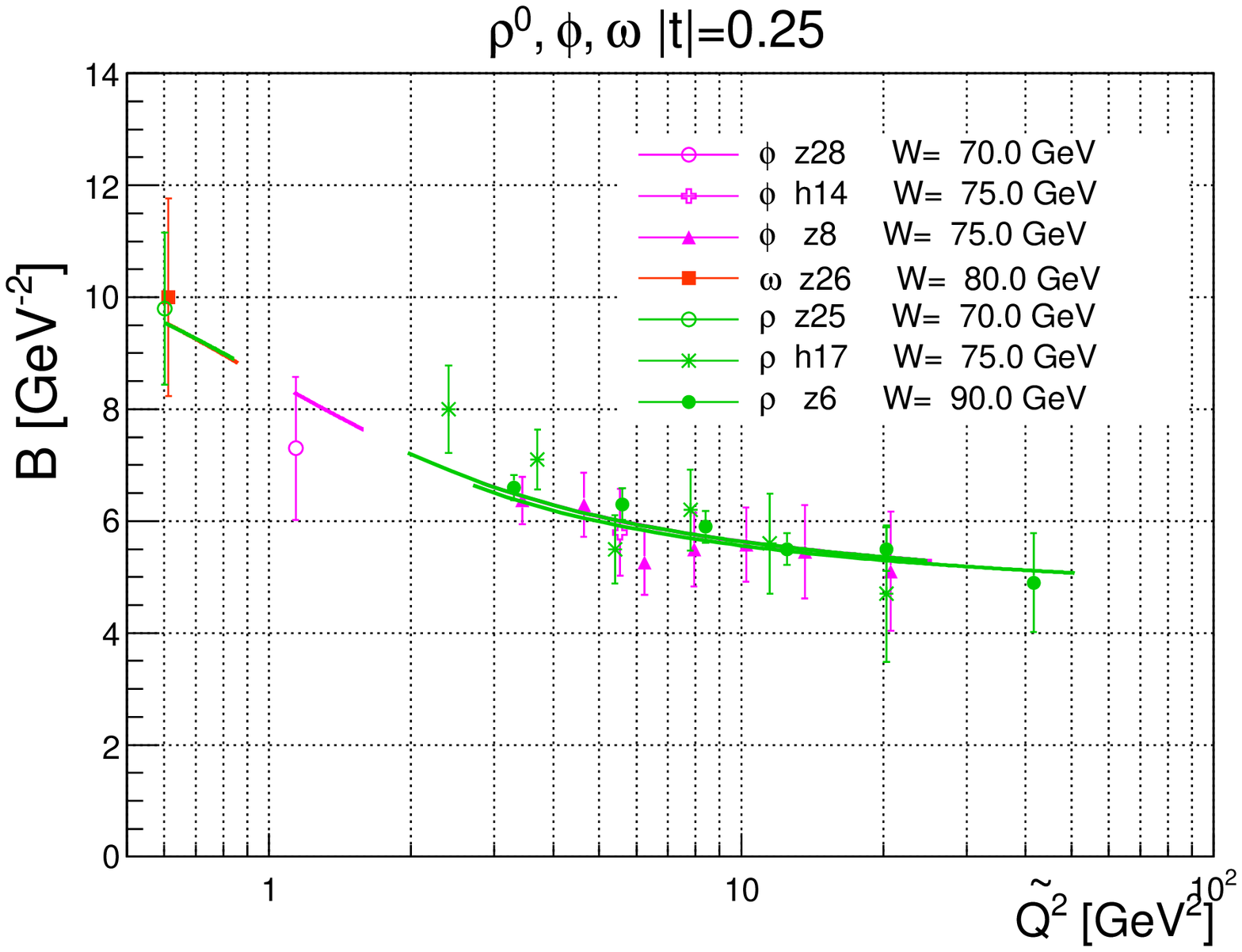}\\
 \includegraphics[trim = 0mm 0mm 10mm 0mm,clip, scale=0.40]{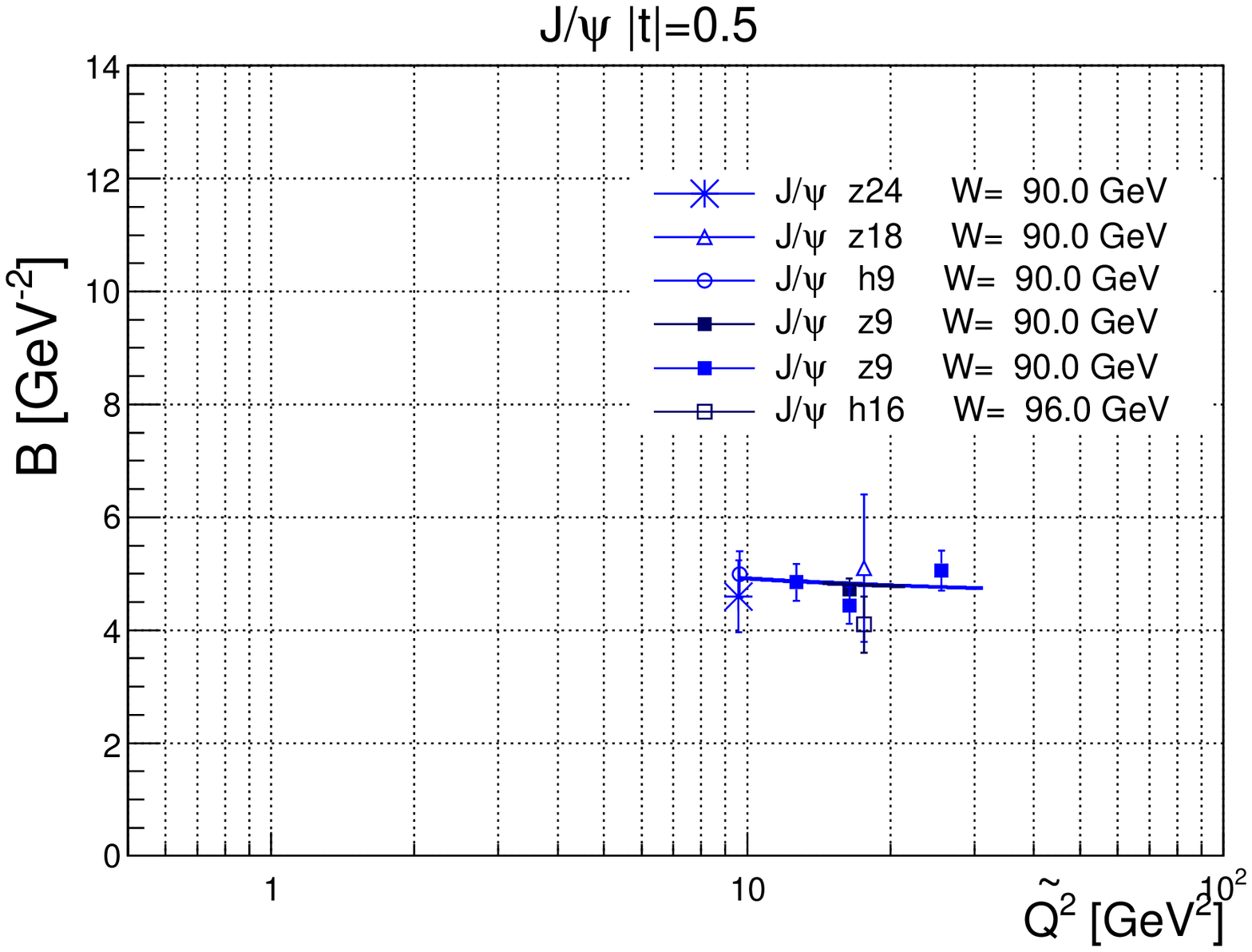}
 \includegraphics[trim = 0mm 0mm 10mm 0mm,clip, scale=0.40]{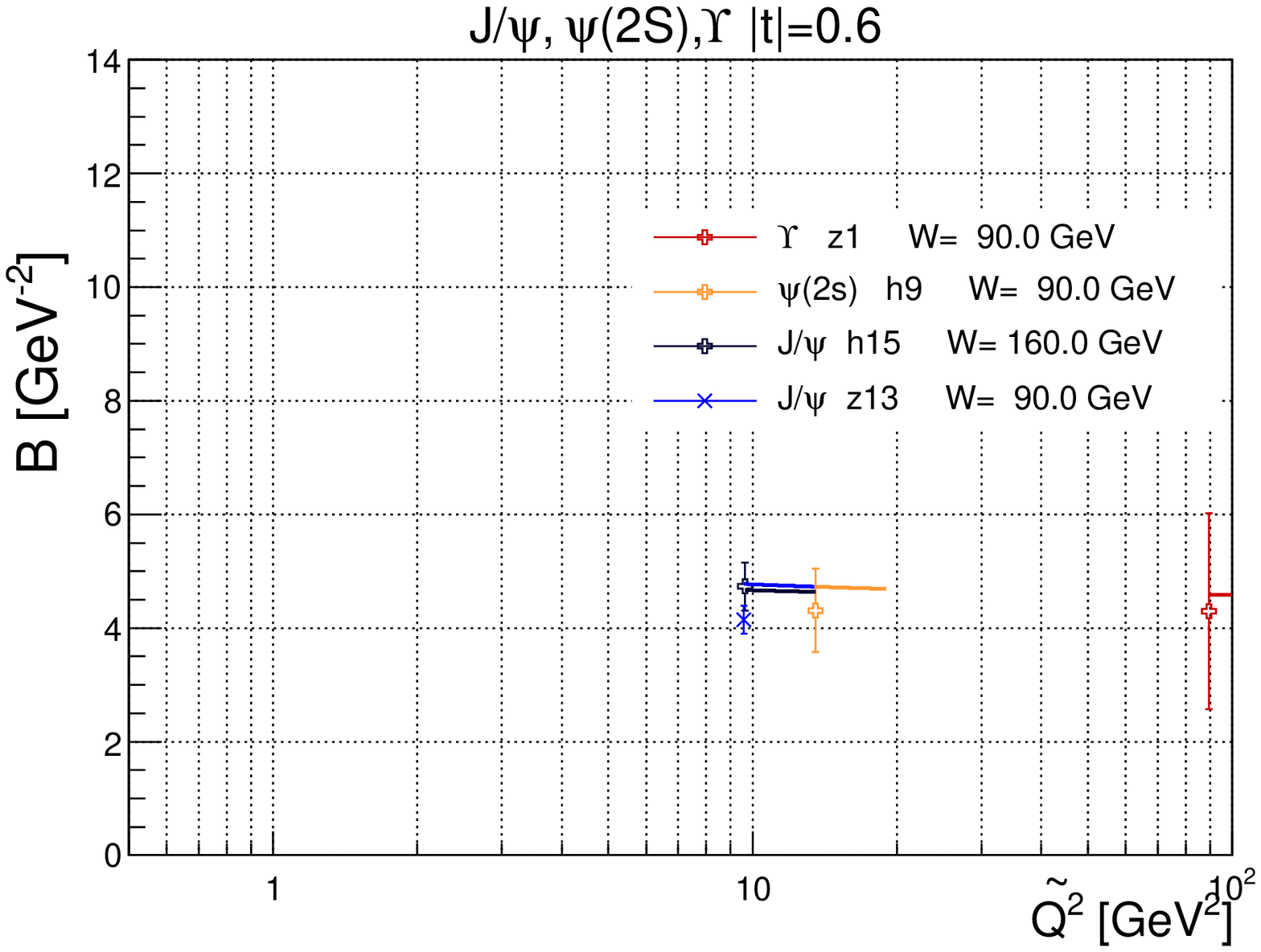}
\caption{ \label{fig:B(Q2)_in_t} Experimental data on the slope $B$ as a function of $\widetilde{Q^2}$ for $\rho^0, \phi, J/\psi$, $\varUpsilon$ and $\Psi$(2S) at $|t|=0.12,$ $0.25,$ $0.5,$ $0.6$ GeV$^{-2}$,  and our theoretical predictions coming from Eq.~(\ref{eq:B(h+s)}).}
\end{figure}

\begin{figure}[!ht]
\centering
  \includegraphics[trim = 0mm 0mm 10mm 3mm,clip, scale=0.70]{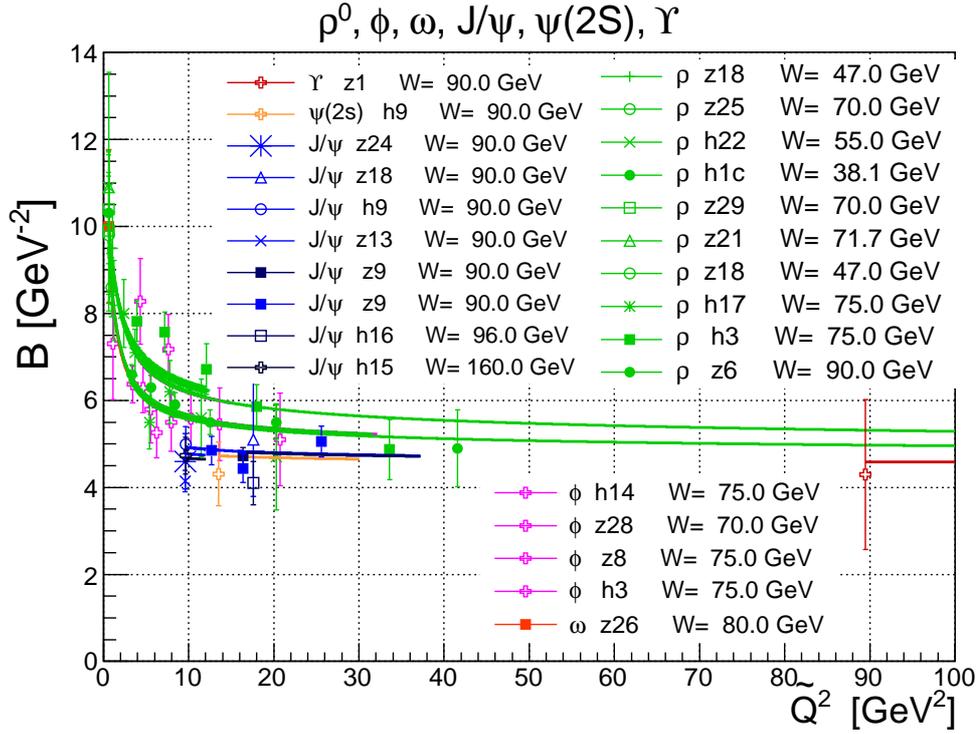}
  \caption{ \label{fig:B(Q2)} Experimental data on the slope $B$ as a function of $\widetilde{Q^2}$ for $\rho^0, \phi, J/\psi$, $\varUpsilon$ and $\Psi$(2S), and our theoretical predictions coming from Eq.~(\ref{eq:B(h+s)}).}
\end{figure}

\begin{figure}[!hb]  \centering
  \includegraphics[trim = 0mm 0mm 10mm 5mm,clip, scale=0.70]{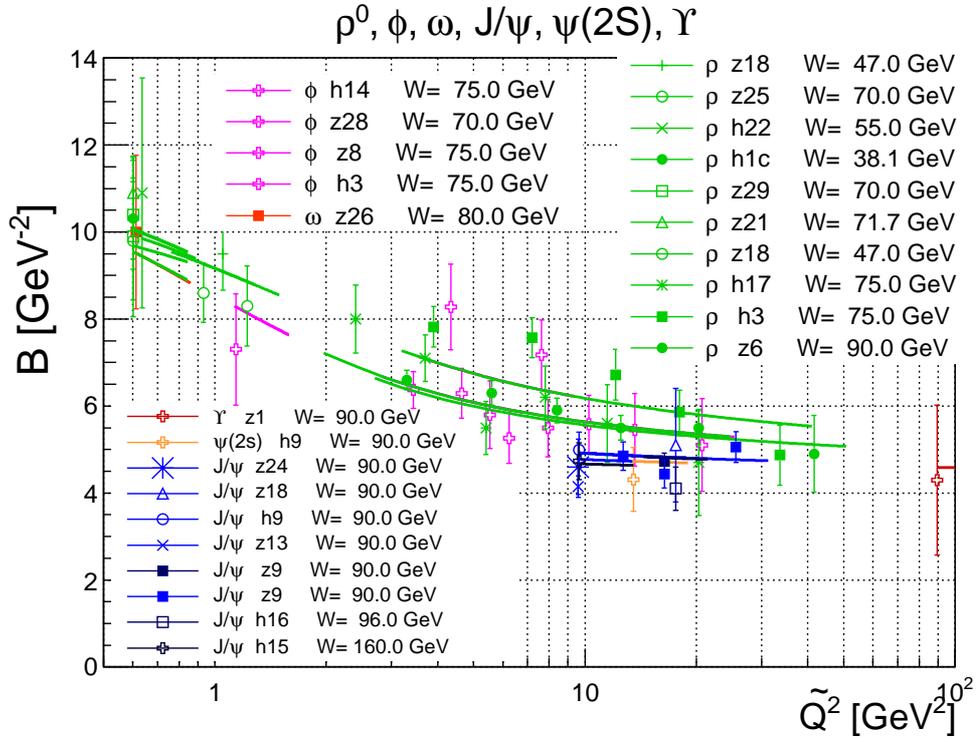}
  \caption{ \label{fig:BLog(Q2)} Experimental data on the slope $B$ as a function of $\widetilde{Q^2}$ for $\rho^0, \phi, J/\psi$, $\varUpsilon$ and $\Psi$(2S), and our theoretical predictions coming  from Eq.~(\ref{eq:B(h+s)}). The plot is the same as in Fig. \ref{fig:B(Q2)}, here with a logarithmic x-axis.}
\end{figure}
\newpage

\begin{figure}[!ht] \centering
 \includegraphics[trim = 0mm 3mm 10mm 0mm,clip, scale=0.40]{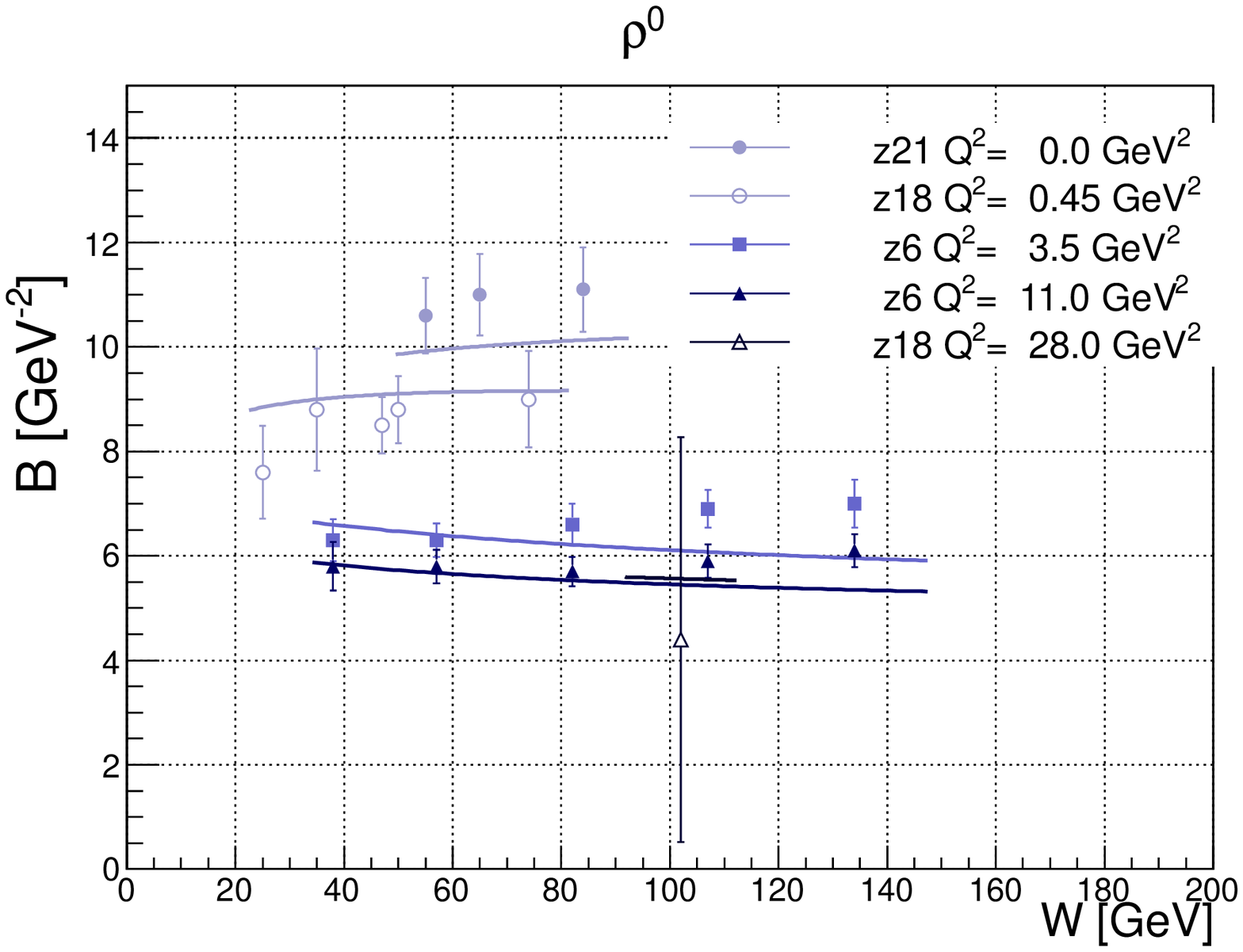}
 \includegraphics[trim = 0mm 3mm 10mm 0mm,clip, scale=0.40]{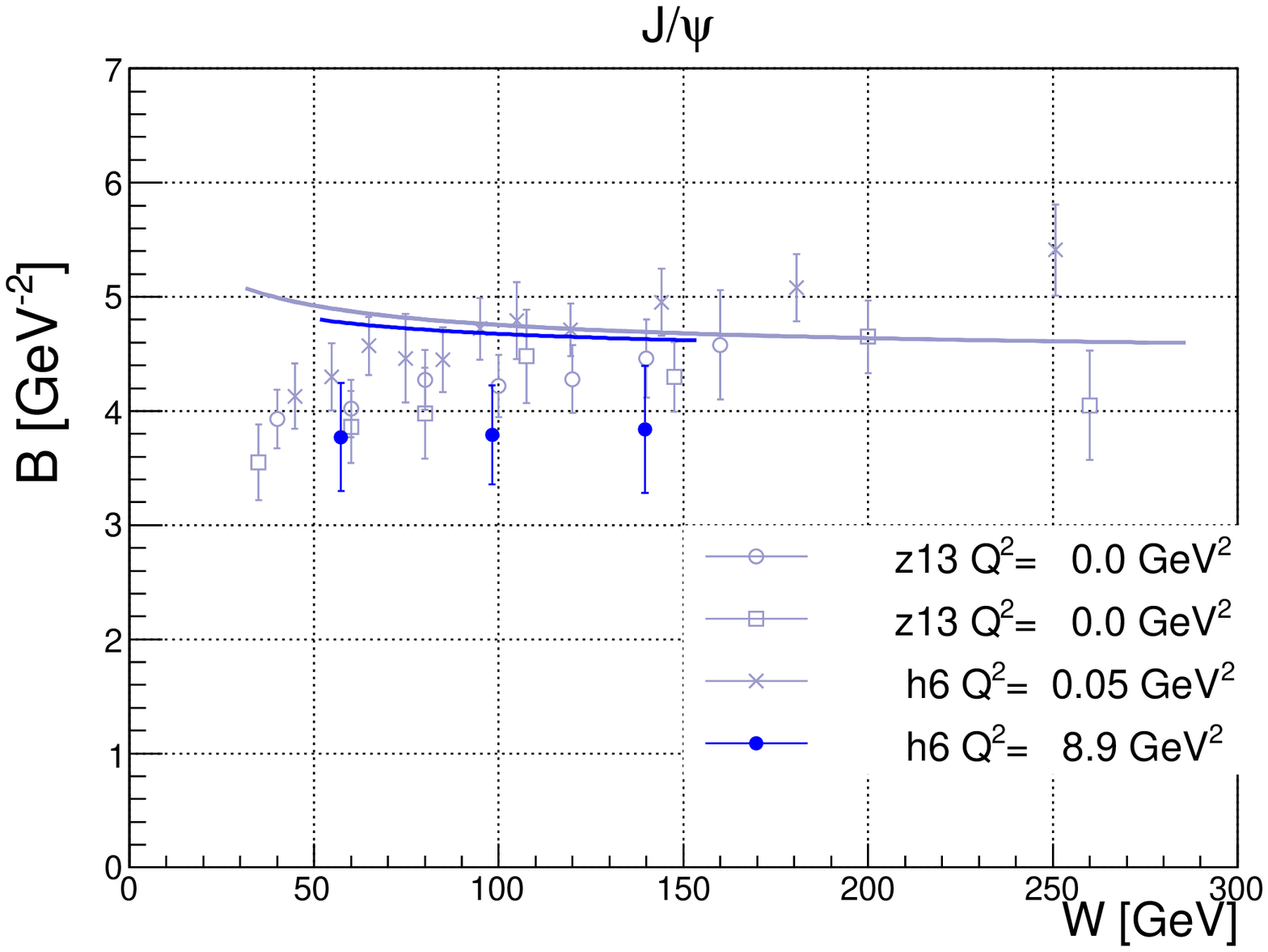}\\
\caption{ \label{fig:B(W)} Experimental data on the slope for $\rho^0$ and $J/\psi$ as functions of $W$,  and our theoretical predictions from Eq.~(\ref{eq:B(h+s)}).}
\end{figure}

\section{Balancing between the  ``soft" and ``hard" dynamics}\label{sec:Balance}
In this section we illustrate
the important and delicate interplay between the ``soft" and ``hard" components of our unique Pomeron.
Since the amplitude consists of two parts, according to the definition ~(\ref{two-term-amp}), it can be written as
\begin{equation}
A(Q^2,s,t)=A_s(Q^2,s,t)+A_h(Q^2,s,t).
\label{Ampl-2}
\end{equation}
As a consequence, the differential and elastic cross sections contain also an interference term between ``soft" and ``hard" parts, so that they read
\begin{equation}
\frac{d\sigma_{el}}{dt}=\frac{d\sigma_{s,el}}{dt}+\frac{d\sigma_{h,el}}{dt}+\frac{d\sigma_{interf,el}}{dt}
\label{dsigma_2}
\end{equation}
and
\begin{equation}
\sigma_{el}=\sigma_{s,el}+\sigma_{h,el}+\sigma_{interf,el},
\label{sigma_2}
\end{equation}
according to Eqs.~\eqref{eq:dcsdt(h+s)} and \eqref{eq:cs(h+s)}, respectively.

\begin{figure*}[!ht]
  \centering
   \includegraphics[trim = 0mm 0mm 0mm 0cm,clip, scale=0.42]{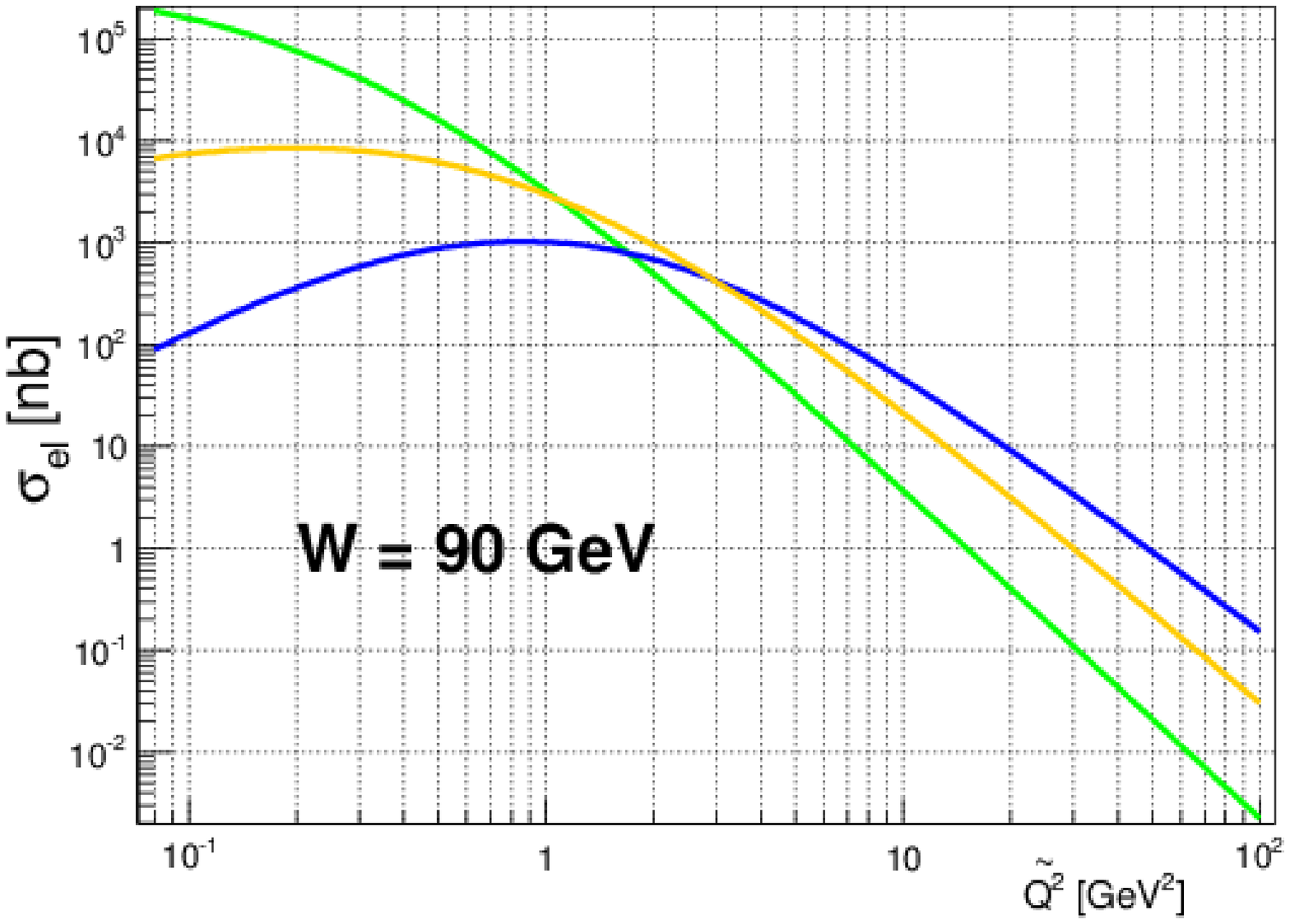}
   \includegraphics[trim = 0mm 0mm 0mm 0cm,clip, scale=0.43]{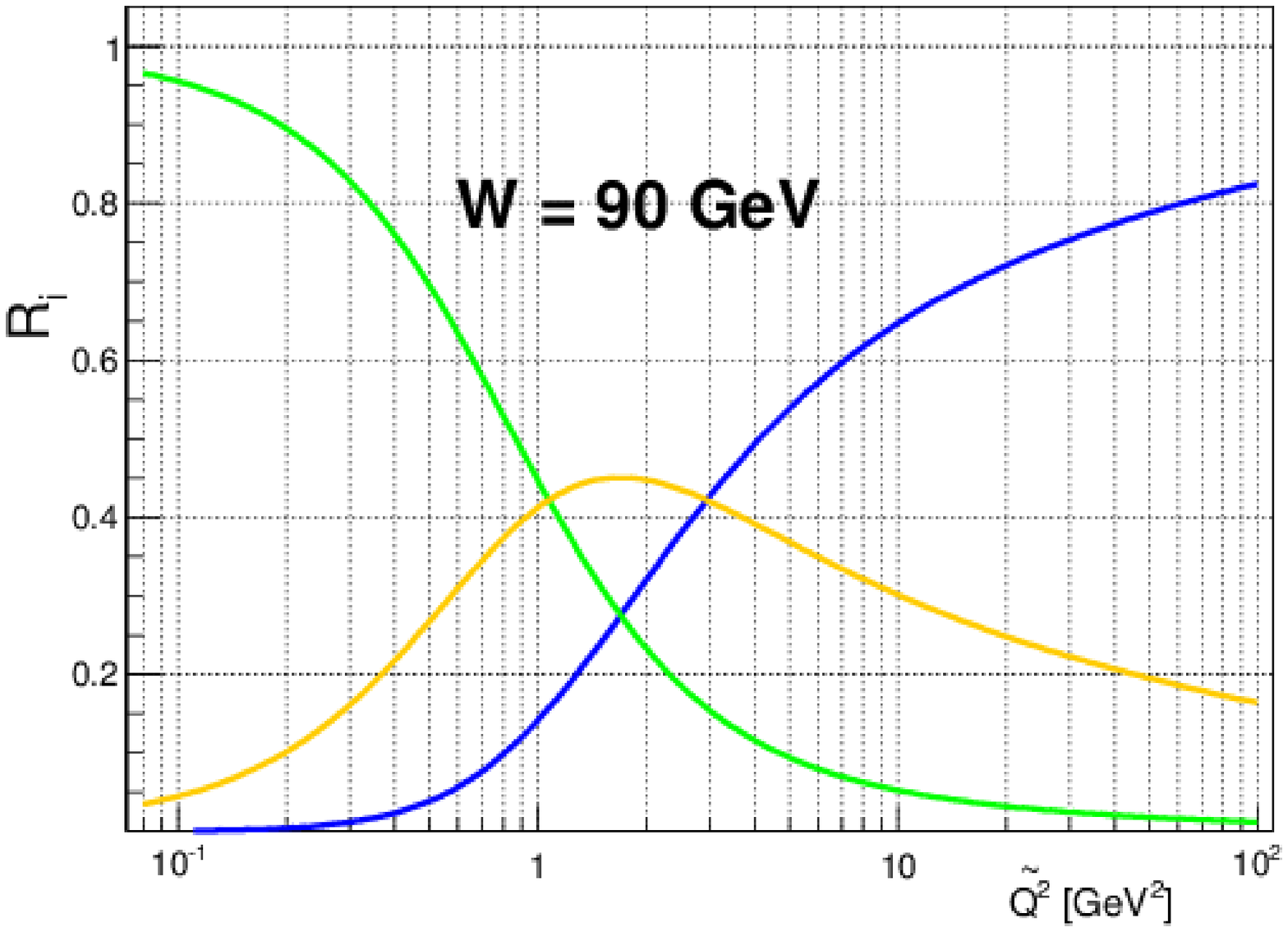}
  \caption{\label{fig:Rsh} Interplay between soft (green line), hard (blue line) and interference (yellow line) components of the cross section $\sigma_{i,el}$ (left plot) and $R_i(\widetilde{Q^2}, t)$ (right plot) as functions of $\widetilde{Q^2}$, for $W=70$ GeV.}
\end{figure*}

Given Eqs.~(\ref{dsigma_2}) and (\ref{sigma_2}), we can define the following ratios for each component:

\begin{equation}
R_i(\widetilde{Q^2}, W, t)=\frac{ \frac{d\sigma_{i,el}}{dt} }{ \frac{d\sigma_{el}}{dt} }
\label{ratio_dsigma}
\end{equation}
and

\begin{equation}
R_i(\widetilde{Q^2}, W)=\frac{\sigma_{i,el}}{\sigma_{el}},
\label{ratio_sigma}
\end{equation}
where $i$ stands for $\{s, h, interf\}$.


Fig.~\ref{fig:Rsh} shows the interplay between the components for both $\sigma_{i,el}$ and $R_i(\widetilde{Q^2}, t)$, as functions of $\widetilde {Q^2}$, for $W$ = 70  GeV.
In Fig.~\ref{fig:Rsh_surf} both plots show that not only $\widetilde{Q^2}$ is the parameter defining softness or hardness of the processs, but such is also the  combination of $\widetilde{Q^2}$ and $t$, similar to the variable $z=t-Q^2$ introduced in Ref.~\cite{Capua}.
 On the whole, it can be seen from the plots that the soft component dominates in the region of low $\widetilde{Q^2}$ and $t$, while the hard compontent dominates in the region of high $\widetilde{Q^2}$ and $t$. 

\begin{figure*}[!ht]
  \centering
   \includegraphics[trim = 2mm 1mm 0mm 0cm,clip, scale=0.38]{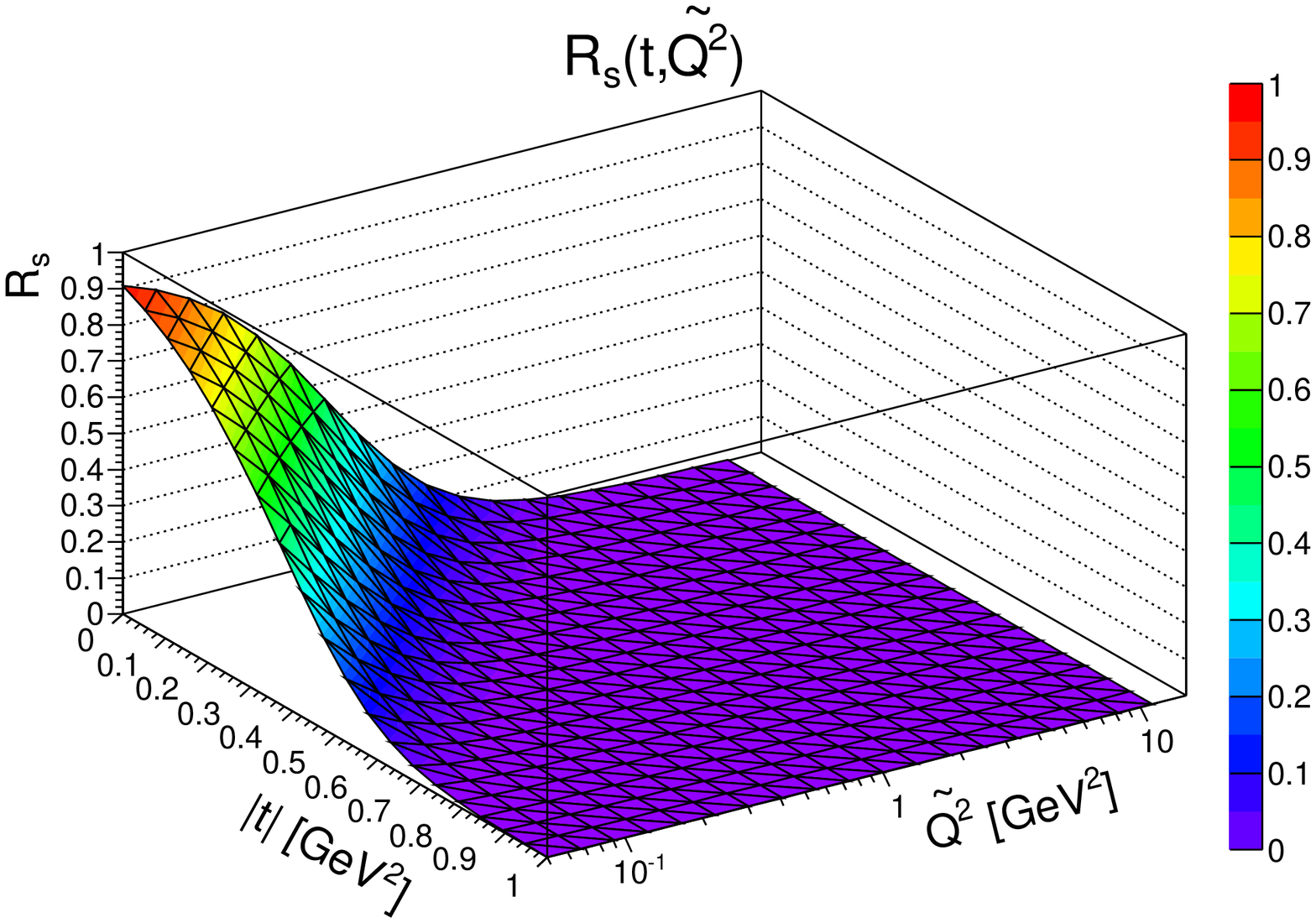}
   \includegraphics[trim = 0mm -5mm 0mm 13mm,clip, scale=0.34]{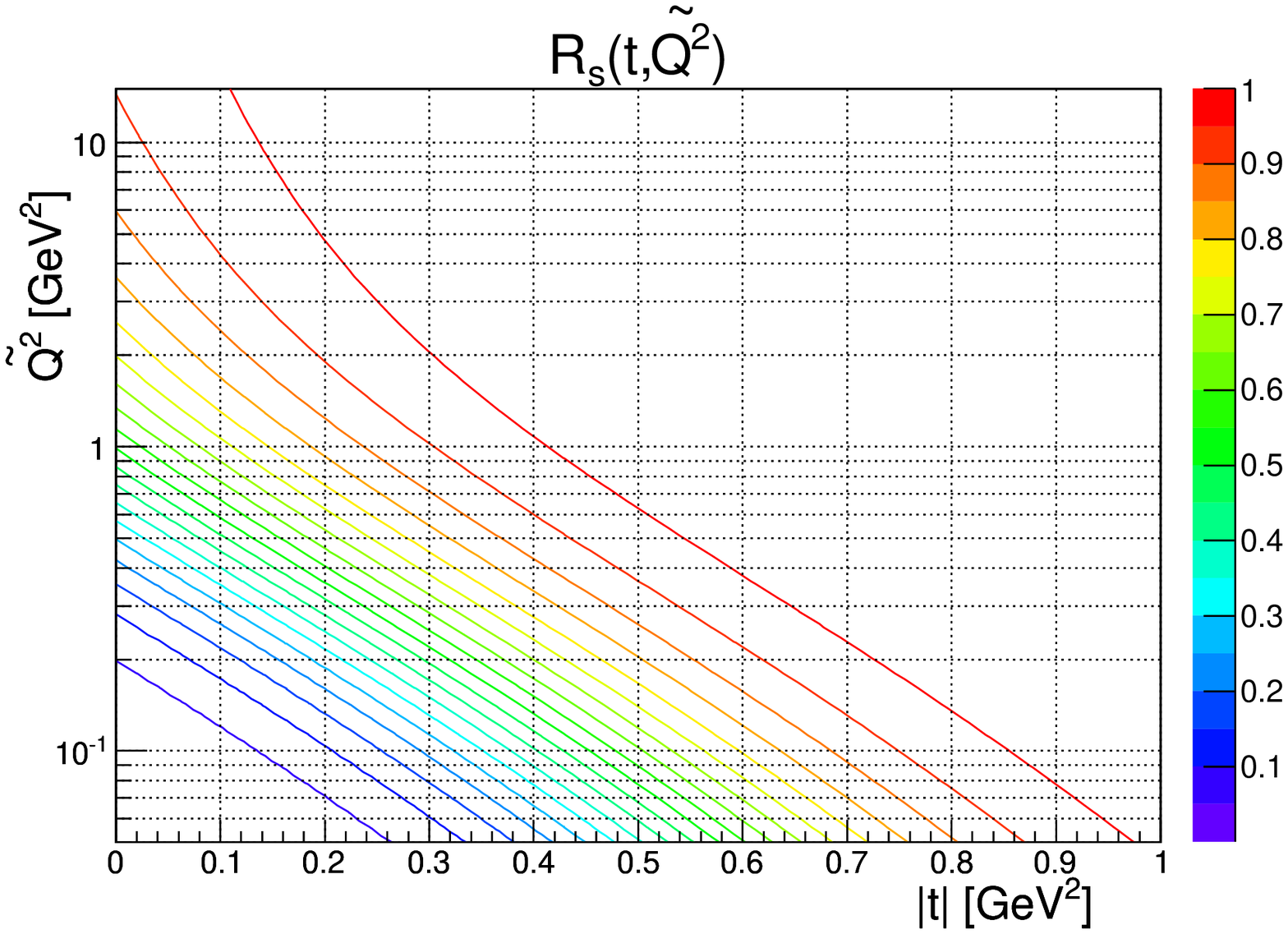}\\
   \includegraphics[trim = 3mm 1mm 0mm 0cm,clip, scale=0.38]{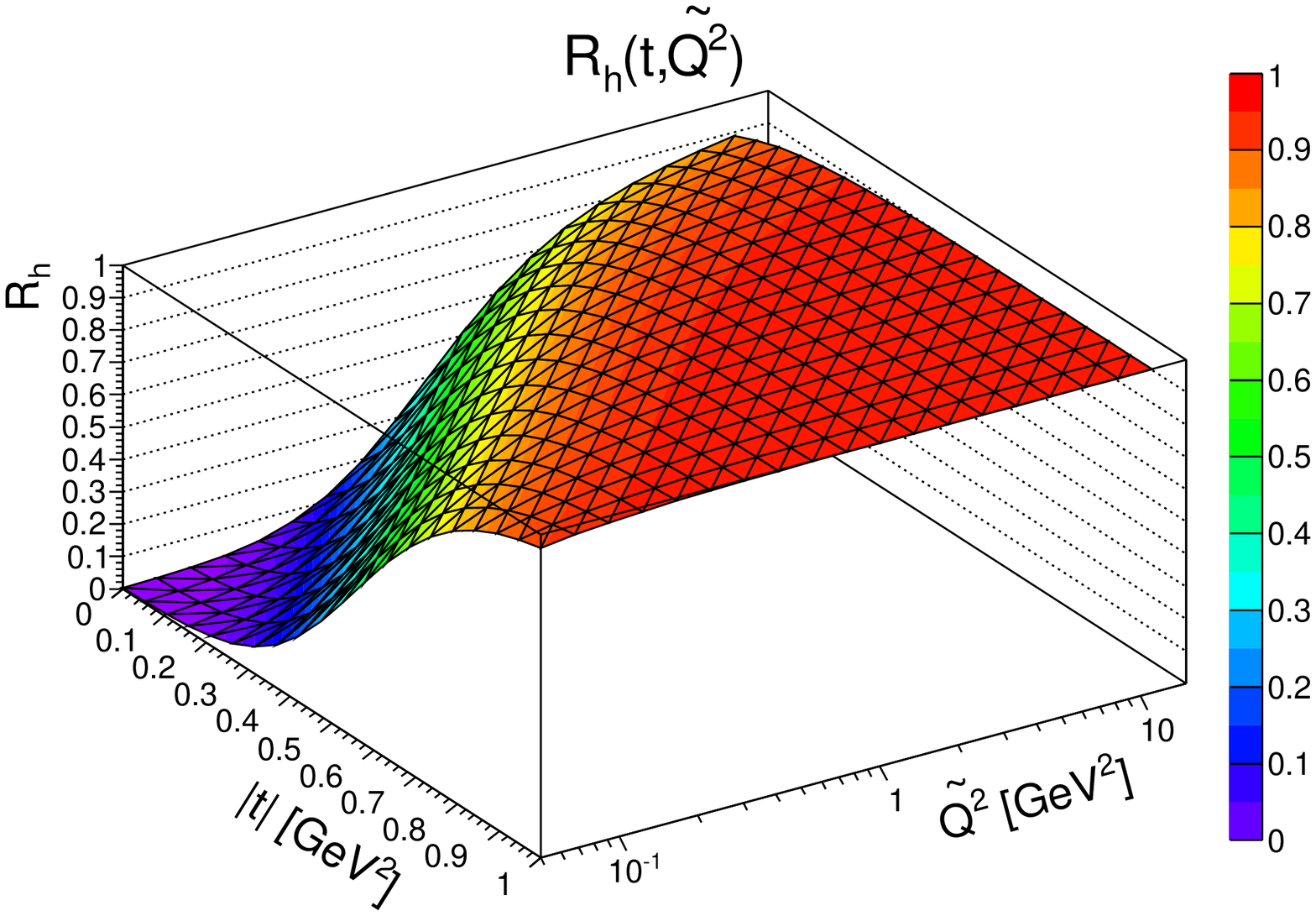}
   \includegraphics[trim = 0mm -5mm 0mm 13mm,clip, scale=0.34]{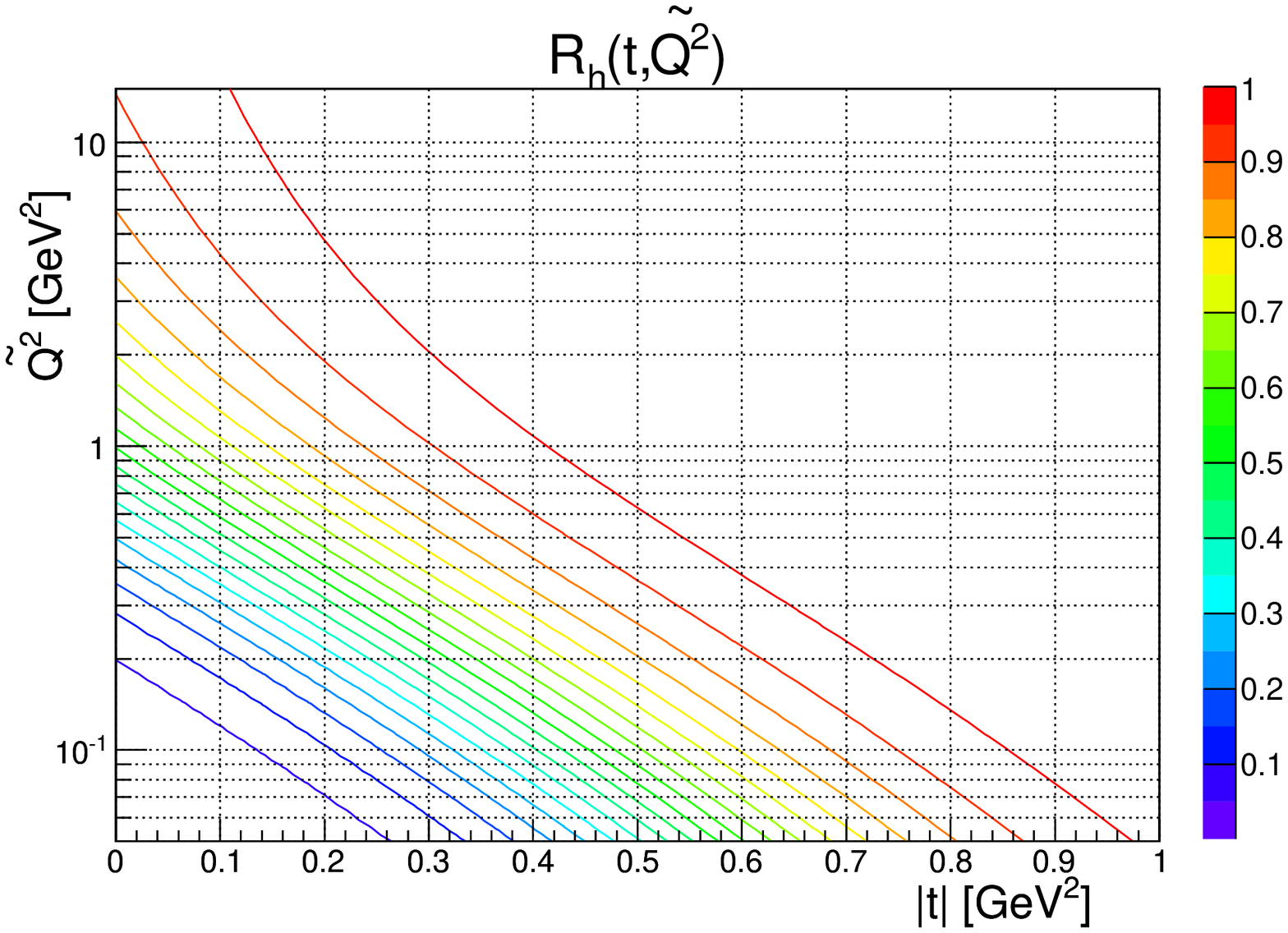}\\
   \includegraphics[trim = 2mm 1mm 0mm 0cm,clip, scale=0.38]{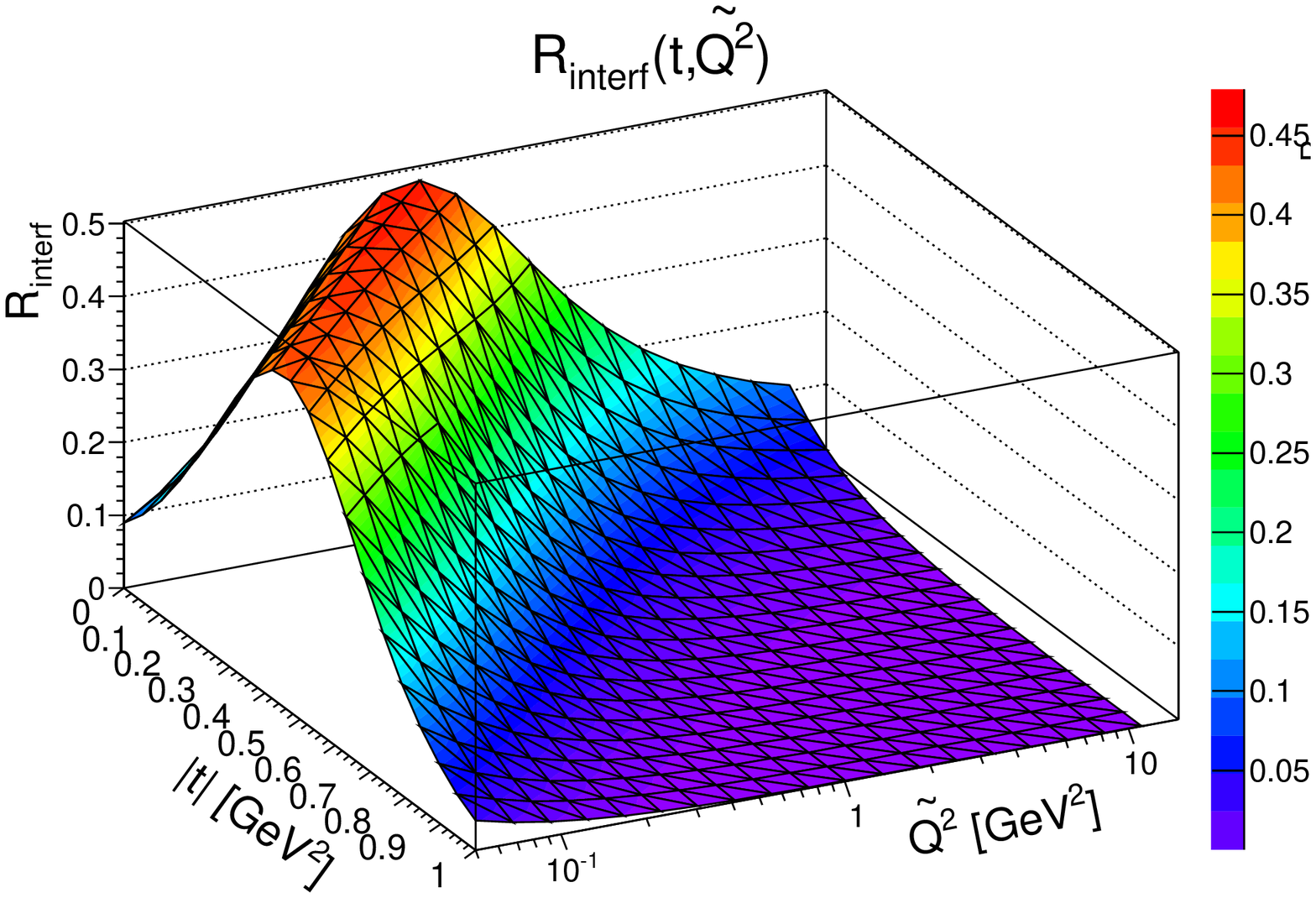}
   \includegraphics[trim = 0mm -5mm 0mm 13mm,clip, scale=0.34]{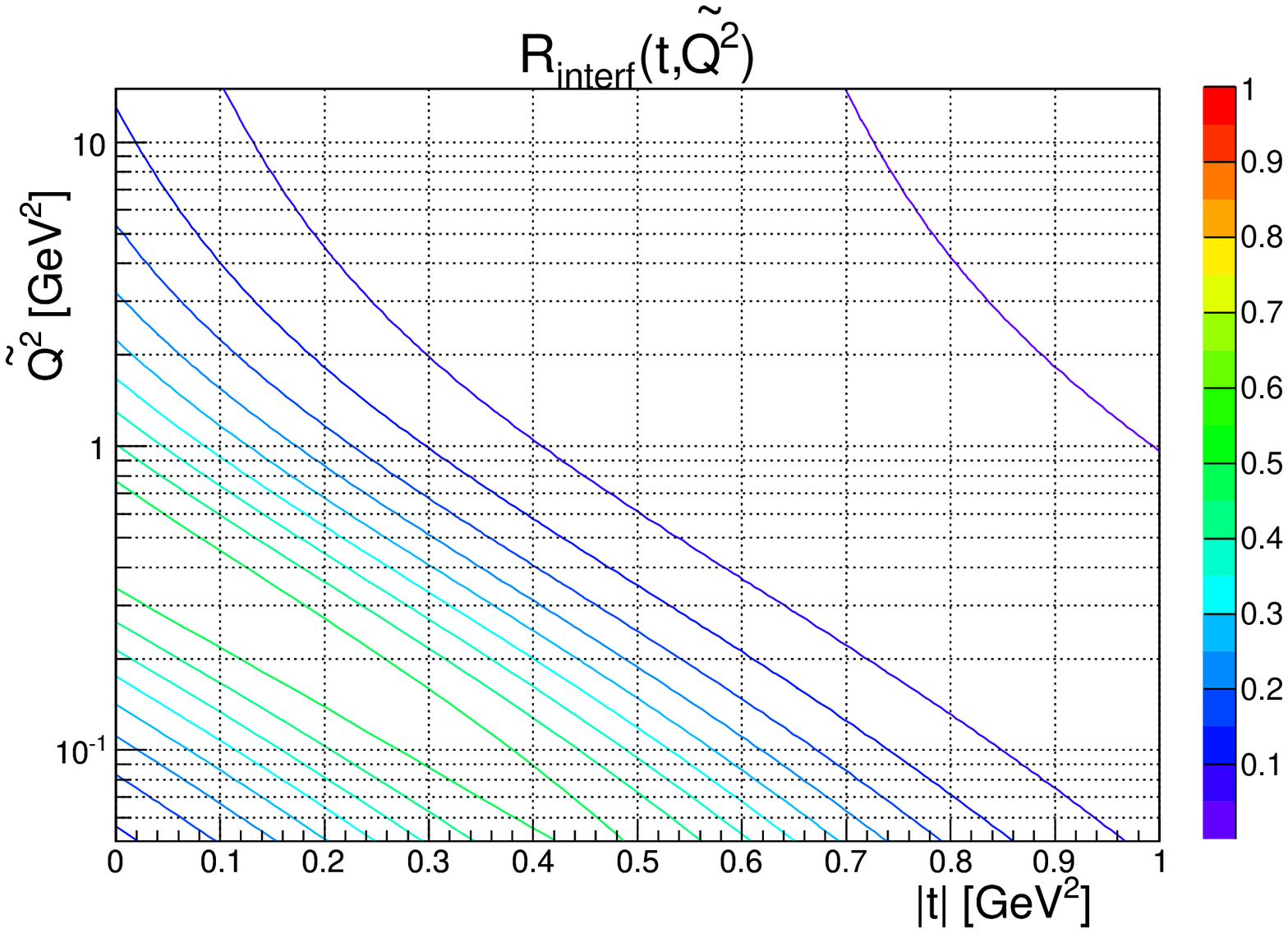}
  \caption{\label{fig:Rsh_surf} Left column: soft (upper surface), hard (middle surface) and interference (bottom surface) components of the ratio $R_i(\widetilde {Q^2}, W, t)$
  are shown  as functions of $\widetilde{Q^2}$ and $t$, for $W=70$~GeV. Right column: some representative curves of the surfaces projected on the ($t, \widetilde {Q^2}$  ) plane.}
\end{figure*}



\newpage
\section{ Hadron-induced reactions: high-energy $pp$ scattering}\label{sec:pp}
Hadron-induced reactions differ from those induced by photons at least in two aspects. First, hadrons are on the mass shell and hence the relevant processes are typically ``soft''. Secondly, the mass of incoming hadrons is positive, while the virtual photon has negative squared ``mass". Our attempt to include hadron-hadron scattering into the analysis with our model has the following motivations: a)
by vector meson dominance (VMD) the photon behaves partly as a meson, therefore meson-baryon (and more generally, hadron-hadron) scattering has much in common with photon-induced reactions. Deviations from VMD may be accounted for the proper $Q^2$ dependence of the amplitude (as we do hope is in our case!); b) of interest is the connection between
space- and time-like reactions;  c) according to recent claims (see e.g. Ref. \cite{L,DL_tr}) the highest-energy (LHC) proton-proton scattering data indicate the need for a ``hard" component in the Pomeron (to anticipate, our fits do not found support the need of any noticeable ``hard" component in $pp$ scattering).

We did not intend to
a high-quality fit to the $pp$ data; that would be impossible without the inclusion of subleading contributions and/or the Odderon. Instead we normalized the parameters of our leading Pomeron term according to recent fits by Donnachie and Landshoff \cite{L} including, apart from a soft term, also a hard one.

 The $pp$ scattering amplitude is written in the form similar to the amplitude  (\ref{eq:Amplitude_hs}) for VMP or DVCS, the only difference being that the normalization factor is constant since the $pp$ scattering amplitude does not depend on $Q^2$:
\begin{equation}\label{eq:Amplitude2_pp}
 A^{pp}(s,t)=
  A^{pp}_s\, e^{-i\frac{\pi}{2}\alpha_s(t)} \left(\frac{s}{s_{0}}\right)^{\alpha_s(t)} e^{b_st}
 +A^{pp}_h\, e^{-i\frac{\pi}{2}\alpha_h(t)} \left(\frac{s}{s_{0}}\right)^{\alpha_h(t)} e^{b_ht}.
\end{equation}


We fixed the parameters of Pomeron trajectories in accord with those of Refs. \cite{Lpp,L})
 $$\alpha_{s}(t)=1.084+0.35t,\qquad \alpha_{h}(t)=1.30+0.10t.$$

With these trajectories the total cross section
\begin{equation}\label{eq:cstot_2}
     \sigma_{tot}=\frac{4\pi}{s} Im\;A(s,{t=0})
\end{equation}
was found compatible with the LHC data, as seen in the left plot of Fig.~\ref{fig:cs_pp}. From the comparison of Eq.~(\ref{eq:cstot_2}) to the LHC data we get
$$A^{pp}_s=-1.73 \text{\,mb}\cdot\text{GeV}^2,\,\,\quad\quad
A^{pp}_h=-0.0012 \text{\,mb}\cdot\text{GeV}^2.$$
The parameter $b_s$ was determided by fitting the differential and integrated elastic cross sections to the data taken from Refs.~\cite{LHCpp,ppPDG,pp_el}. To this aim, we used  Eqs.~(\ref{eq:dcsdt(h+s)})
and (\ref{eq:cs(h+s)}), the normalization factors $H_s$ and $H_h$ replaced with $A^{pp}_s$ and $A^{pp}_h$, respectively, according to Eq.~(\ref{eq:Amplitude2_pp}).
The parameter $b_h$, was set to be equal to $b_s$, since for DVCS and VMP these parameters assume similar values, as seen from  in Tables~\ref{tab:fit1(s+h)} and \ref{tab:fixed_trajec}.
By adjusting the theoretical curves to the data, we get
$b_s=b_h=1.8\text{\,GeV}^{-2}$. The comparison with the experimental data from
Refs.~\cite{LHCpp}-\cite{pp_el} is shown in Fig.~\ref{fig:cs_pp} (right plot) for the integrated cross section and in Fig.~\ref{fig:dcsdt_pp} for the differential elastic cross section.

 \begin{figure}[!ht]
  \centering
  \includegraphics[trim = 0mm 0mm 12mm 2mm,clip, scale=0.4]{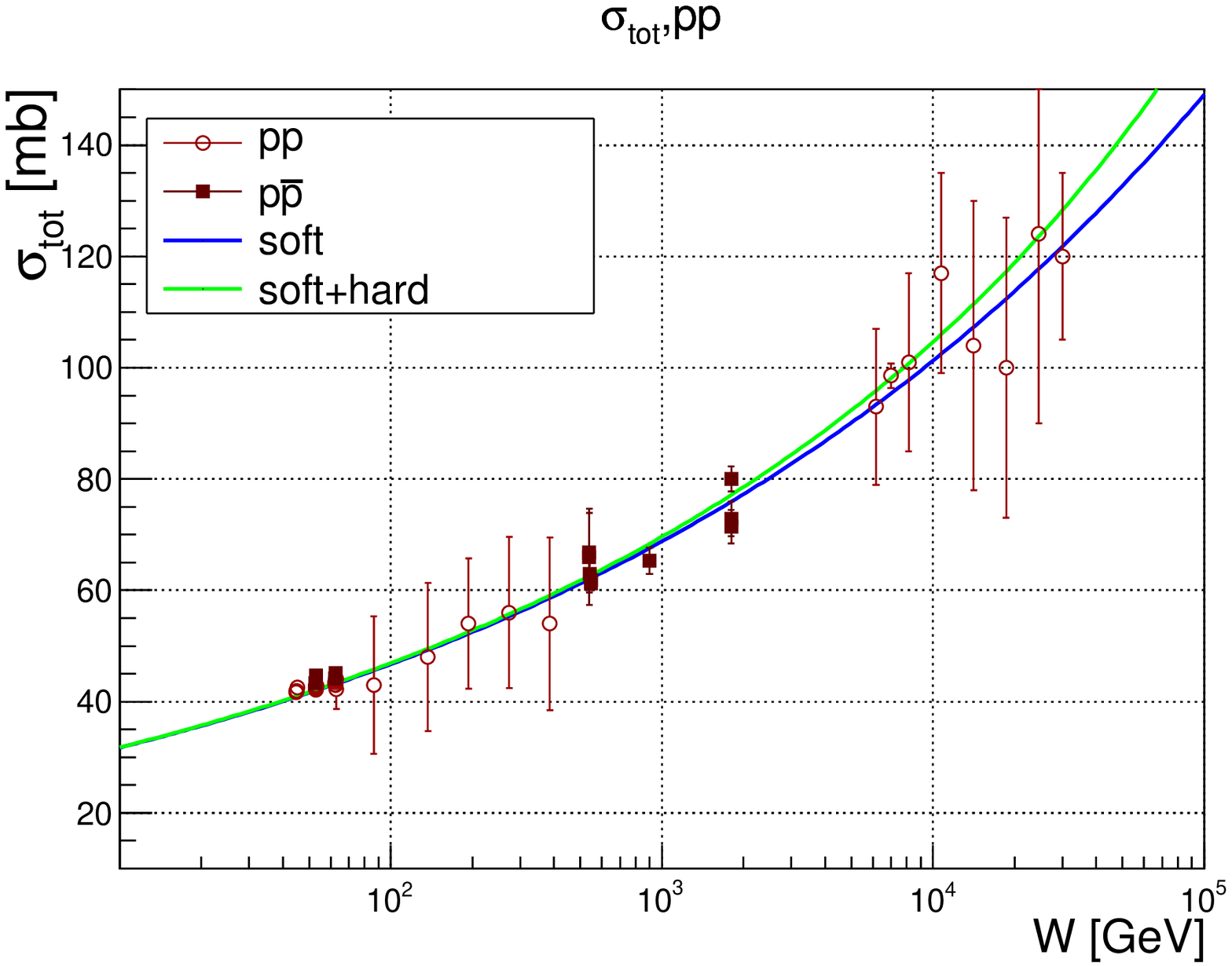}
  \includegraphics[trim = 0mm 0mm 12mm 2mm,clip, scale=0.4]{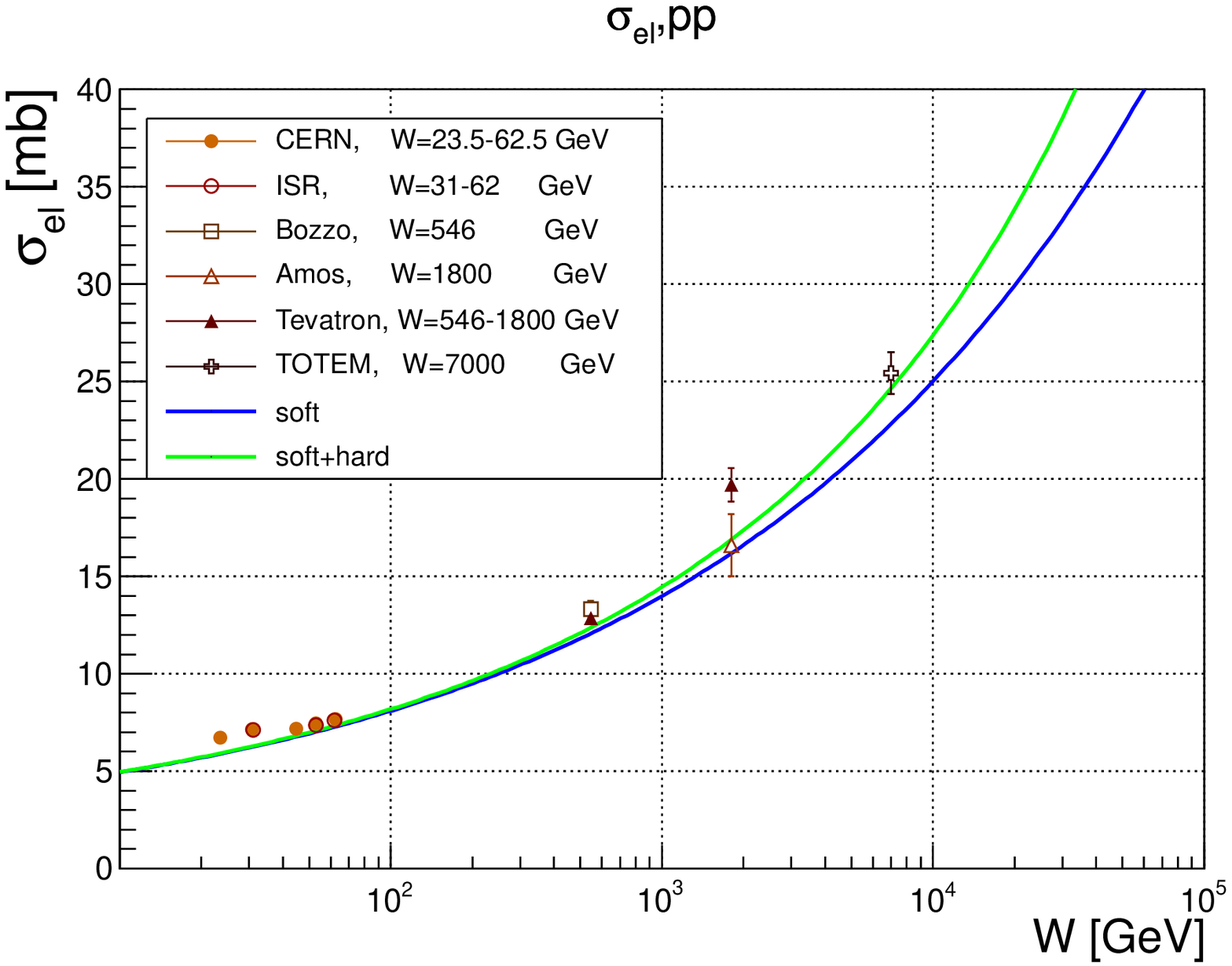}
  \caption{ \label{fig:cs_pp} Calculated total  $\sigma_{tot}(W)$ and the elastic cross sections $\sigma_{el}(W)$  for $pp$ scattering with the data from Refs.~\cite{LHCpp,ppPDG,pp_el}.}
 \end{figure}

Next, by using formula (\ref{eq:B(h+s)})
we calculated the forward slope $B$, shown in Fig.~\ref{fig:B_pp} together with data from Refs.~\cite{LHCpp} and \cite{Bpp}.

From these figures we conclude that, while the data on total cross section is compatible with a small ``hard" admixture in the amplitude, the slope parameter with a hard component included seems to manifest a wrong tendency, by slowing down with increasing energy, while the TOTEM measurements \cite{LHCpp} show the contrary.


\begin{figure}[!ht]
  \centering
  \includegraphics[trim = 0mm 0mm 12mm 2mm,clip, scale=0.6]{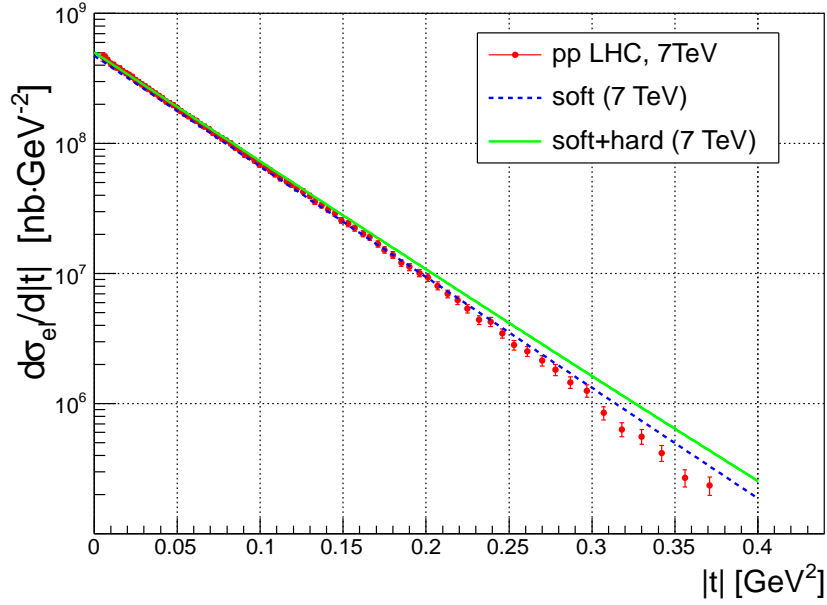}
  \caption{ \label{fig:dcsdt_pp} The differential elastic cross section $d\sigma_{el}/dt$ for $pp$-scattering. The data were taken from Ref.~\cite{LHCpp}.}
 \end{figure}

 \begin{figure}[!ht]
  \centering
  \includegraphics[trim = 0mm 0mm 12mm 2mm,clip, scale=0.6]{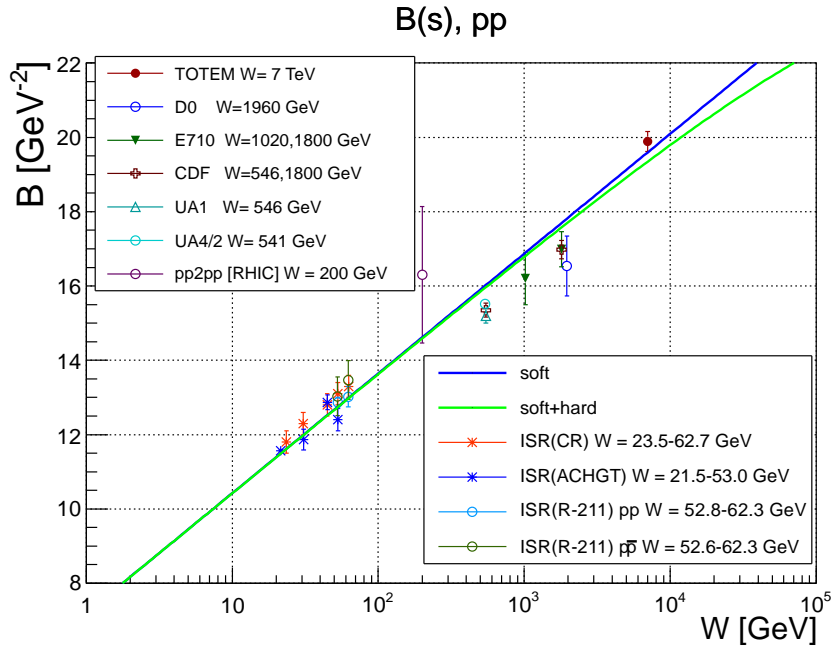}
  \caption{ \label{fig:B_pp}The $B$ of the differential elastic cross section $d\sigma_{el}/dt$ for $pp$-scattering as a function of $W$. The data are from Refs.~\cite{LHCpp,Bpp}.}
 \end{figure}

\newpage
\section{Discussion of the results and conclusions}
\label{sec:conclusion}
 In this paper we have proposed an economic model describing both ``soft" and ``hard" exclusive production of vector particles. It features a unique, ``universal" Pomeron, the same for all processes. This Pomeron is made of two terms, a ``soft" and a ``hard" components, their relative weight depending on the  ``softness" or ``hardness" of the given processes.

 The model incorporates some features of earlier models Refs.~\cite{Capua,FazioPhysRev,Acta}, such as the interplay between the  dependence of the scattering amplitude on the virtuality $Q^2$ and the squared momentum transfer $t$.

In the framework of the model we have analyzed all available data on vector meson ($\rho^0$, $\omega$, $\phi$, $J/\psi$, $\varUpsilon$, $\Psi$(2S)) production and DVCS
obtained at HERA by the H1 and ZEUS Collaborations.
A global fit was performed with a small number of free parameters, namly eight parameters: four parameters of Pomeron trajectories and five parameters for the normalization of the cross sections from six processes), universal for all reactions. By fixing the parameters of the Pomeron trajectories, their number reduces to six.
The results of the fit are presented in Figs.~\ref{fig:cs(Q2)} ($\sigma_{el}(\widetilde{Q^2})$), Figs.~\ref{fig:cs_rho1(W)}-\ref{fig:cs_Ups(W)} ($\sigma_{el}(W)$), Figs.~\ref{fig:dcsdt_rho}-\ref{fig:dcsdt_Jpsi.php1} ($d\sigma_{el}(t)/dt$) and Figs.~\ref{fig:B(Q2)}-\ref{fig:B(W)} (the slope $B$) for VMP and in Figs.~\ref{fig:dcsdt.DVCS}-\ref{fig:csQ2.DVCS} for DVCS. The values of the parameters are quoted in Table~\ref{tab:fit1(s+h)} and Table~\ref{tab:fixed_trajec}.

The resulting fit is reasonable, despite the following minor problems:
 \begin{itemize}
  \item
  in order to incorporate DVCS together with VMP we need to assign some non-zero value to the ``mass" of the real photon, that can be treated as an effective mass of quark-antiquark system into which the virtual photon fluctuates. From the fit we obtained $M_{DVCS}^{eff}=1.8$~GeV;
  \item
  there are some systematic shifts of theoretical curves with respect to the experimental data
 in the regions of low $Q^2$, $W$ and $t$ for the $J/\psi$ fit (see Figs.~\ref{fig:cs_Jpsi(W)},~\ref{fig:dcsdt_Jpsi} and \ref{fig:dcsdt_Jpsi.php1}). This effect may come both from the absence of the secondary Reggeons, and from the influence of the soft (and/or the interference) term of the elastic cross section $\sigma_{el}(Q^2, W, t)$;
 \end{itemize}


Among the remaining  open problems, to be treated in subsequent studies, are:

\begin{itemize}
\item
 in the present paper we have neglected sub-leading Regge contributions. They must must be included in any extension of the model to lower energies (below $30$ GeV);

\item
the $\widetilde{Q^2}$ dependence of the scattering amplitude introduced in the present paper empirically has to be compared with the results of unitarization and/or QCD evolution. We intend to come back to this point;

\item
as seen from Sec.~\ref{sec:Balance}, the ``soft" component of the Pomeron dominates in the region of small $t$ and small $\widetilde{Q^2}$. Hence, a parameter, responsible for the ``softness'' and/or ``hardness'' of processes, should be a combination of $t$ and $Q^2$. A simple solution was suggested in Ref. \cite{Capua} with the introduction of the variable $z=t-Q^2$. The interplay of these two variables remains  an important open problem that requires further investigation.
\end{itemize}

The extension of our formalism to hadronic reactions ($pp$ scattering) shows that
the available data can be will described by 
a single - soft - component.

\section*{Acknowledgements}
\label{sec:Acknowledgements}
L.J. thanks the Dipartimento di Fisica dell'Universit`a della Calabria and the Istituto Nazionale di Fisica Nucleare - Gruppo Collegato di Cosenza, where part of this work was done, for their hospitality and support. 
He was supported partly also by the grant "Matter under extreme conditions" of the
National Academy of Sciences of Ukraine, Dept. of Astronomy and
Physics.

\end{document}